\documentclass[prd,aps,reprint,floatfix,preprintnumbers,showpacs,nofootinbib,notitlepage,superscriptaddress]{revtex4-1}
\usepackage{amsthm}   
\usepackage{amsmath}  
\usepackage{amssymb}  
\usepackage{mathrsfs}
\usepackage{color}
\usepackage[all]{xy}
\usepackage[pdftex]{graphicx}
\usepackage{indentfirst}
\usepackage[pdftex]{hyperref}
\usepackage{subfigure}
\usepackage{bbold}

\newcommand{\pa}{\partial}

\newcommand{\be}{\begin{equation}}
\newcommand{\ee}{\end{equation}}
\newcommand{\ba}{\begin{eqnarray}}
\newcommand{\ea}{\end{eqnarray}}

\newcommand{\en}{\nonumber\\}

\newcommand{\uu}{\mathbf{u}}
\newcommand{\de}{\delta}

\newcommand{\da}{\dagger}
\newcommand{\kk}{\mathbf{k}}
\newcommand{\xx}{\mathbf{x}}
\newcommand{\rr}{\mathbf{r}}

\newcommand{\yy}{\mathbf{y}}

\newcommand{\pp}[1]{\mathbf{p}_{#1}}
\newcommand{\qq}{\mathbf{q}}

\definecolor{darkred}{RGB}{175,0,0}
\definecolor{darkblue}{RGB}{0,0,175}

\newcommand{\qgal}{{\bf q}_{\rm gal} }
\newcommand{\qsub}{{\bf q}_{\rm sub} }
\newcommand{\qenv}{{\bf q}_{\rm env} }

\newcommand{\qsrc}{{\bf q}_{\rm src } }
\newcommand{\qinst}{{\bf q}_{\rm inst } }
\newcommand{\qfore}{{\bf q}_{\rm fore } }

\newcommand{\psub}{\phi_{\rm sub } }
\newcommand{\ksub}{\kappa_{\rm sub}}

\newcommand\sub{{\rm sub}}
\newcommand{\src}{{\rm src}}
\newcommand\vs{\mathbf{s}}
\newcommand\img{{\rm img}}

\newcommand\Cmat{{\bf C}}
\newcommand\Gmat{{\bf G}}
\newcommand\Fmat{{\bf F}}
\newcommand\Mmat{{\bf M}}
\newcommand\Qmat{{\bf Q}}

\newcommand\Gammat{\boldsymbol{\Gamma}}

\newcommand{\mnras}{Mon. Not. R. Astron. Soc.}
\newcommand{\pasp}{Publ. Astron. Soc. Pac.}

\newcommand{\jcap}{Journal of Cosmology and Astroparticle Physics}
\newcommand{\apjl}{The Astrophysical Journal Letters}

\begin{document}
\title{Beyond subhalos: Probing the collective effect of the Universe's small-scale structure with gravitational lensing}
\author{Francis-Yan Cyr-Racine}\email{fycr@unm.edu}
\affiliation{Department of Physics, Harvard University, Cambridge, Massachusetts 02138, USA}
\affiliation{Department of Physics and Astronomy, University of New Mexico, Albuquerque, New Mexico 87131, USA}

\author{Charles R. Keeton}
\affiliation{Department of Physics and Astronomy, Rutgers, The State University of New Jersey, Piscataway, New Jersey 08854, USA }

\author{Leonidas A. Moustakas}
\affiliation{Jet Propulsion Laboratory, California Institute of Technology, Pasadena, California 91109, USA}

\date{\today}
\begin{abstract}
Gravitational lensing has emerged as a powerful probe of the matter distribution on subgalactic scales, which itself may contain important clues about the fundamental origins and properties of dark matter. Broadly speaking, two different approaches have been taken in the literature to map the small-scale structure of the Universe using strong lensing, with one focused on measuring the position and mass of a small number of discrete massive subhalos appearing close in projection to lensed images, and the other focused on detecting the collective effect of all the small-scale structures between the lensed source and the observer. In this paper, we follow the latter approach and perform a detailed study of the sensitivity of galaxy-scale gravitational lenses to the ensemble properties of small-scale structure. As in some previous studies, we adopt the language of the substructure power spectrum to characterize the statistical properties of the small-scale density field. We present a comprehensive theory that treats lenses with extended sources as well as those with time-dependent compact sources (such as quasars) in a unified framework for the first time. Our approach uses mode functions to provide both computational advantages and insights about couplings between the lens and source. The goal of this paper is to develop the theory and gain the intuition necessary to understand how the sensitivity to the substructure power spectrum depends on the source and lens properties, with the eventual aim of identifying the most promising targets for such studies. 
\end{abstract}

\maketitle

\section{Introduction}
The distribution of matter on kiloparsec scales and smaller may hold important clues about the fundamental nature of dark matter. Within the standard cold dark matter (CDM) paradigm, for instance, we expect dark matter at these scales to be distributed among a large number of low-mass subhalos that are largely devoid of gas and stars \cite{Springel:2008cc}. On the other hand, theories in which dark matter has a significant free-streaming length \cite{Bond:1983hb,Bode:2000gq,Dalcanton:2000hn,Boyanovsky:2008ab,Boyanovsky:2011aa} or interacts with a relativistic species at early times (see, e.g.,~Refs.~\cite{1992ApJ...398...43C,Boehm:2001hm,Ackerman:2008gi,Feng:2009mn,Kaplan:2009de,Aarssen:2012fx,Cyr-Racine:2013ab,Cyr-Racine:2014aa,2014PhRvD..90d3524B,2014MNRAS.445L..31B,Schewtschenko:2014fca,Chu:2014lja,Buen-Abad:2015ova,2016PhRvD..93l3527C,Chacko:2016kgg,Ko:2016uft}) predict a dearth of low-mass dark matter subhalos. Moreover, the presence of dark matter self-interaction \cite{2012MNRAS.423.3740V,Rocha:2012jg,2013MNRAS.430..105P,2013MNRAS.431L..20Z,Kaplinghat:2013xca,Kaplinghat:2015aga,Vogelsberger_2015} or dissipative dynamics \cite{Fan:2013tia,Fan:2013yva,Buckley:2017ttd,Agrawal:2017rvu,Agrawal:2017pnb} can modify the inner structure of subhalos, hence providing another potential probe of dark matter microphysics.  Ultralight axions \cite{Hu_2000,Hui_2016} might also lead to interesting phenomenology on small scales (see, e.g.,~Refs.~\cite{Du_2016,Mocz:2017wlg}). 

Several observational techniques can be used to probe the dark matter distribution on subgalactic scales. In addition to kinematical studies of ultrafaint dwarf galaxies within the Local Group \cite{Walker_Penarrubia_2011,Salucci:2011ee,Laporte:2013fwa,Bonnivard:2015xpq,Caldwell:2016hrl,Banik:2018pjp}, detailed analyses of potential perturbations to local stellar streams \cite{2014ApJ...788..181N,2016ApJ...820...45C,2016PhRvL.116l1301B,2016MNRAS.463..102E,2017MNRAS.466..628B}, to the Milky Way's galactic disk  \cite{2012ApJ...750L..41W,Feldmann:2013hqa}, or to the timing of distant pulsars \cite{Clark:2015tha,Clark:2015sha} have been proposed as methods to put constraints on the abundance of low-mass subhalos. Analyses of stellar wakes \cite{Buschmann:2017ams} could also be used to detect starless dark matter subhalos locally.

Both within and beyond the Local Group, gravitational lensing has emerged as a powerful probe of the dark matter distribution on the smallest scales. Indeed, nearby dark matter subhalos devoid of stars could be detected through lensing using subtle time-dependent astrometric perturbations of more distant sources \cite{Erickcek:2010fc,Li:2012qha,VanTilburg:2018ykj}. Further out, galaxy-scale strong lensing systems in which a background high-redshift source is multiply imaged by a massive foreground galaxy are promising laboratories to study the dark matter distribution on the smallest scales. For example, the presence of flux-ratio anomalies in quasar lenses \cite{Mao:1998aa,Metcalf:ad,Chiba:aa,Metcalf:ac,Dalal:aa,Keeton:2003ab,Metcalf:2010aa} has led to a measurement of the typical abundance of mass substructures within lens galaxies \cite{Dalal:2002aa,Hsueh:2019ynk}, with future measurements appearing promising (see, e.g.,~Refs.~\cite{Gilman:2017voy,Gilman:2019vca}) if potential systematics can be properly accounted for \cite{Xu:2011aa,Xu:2010aa,2015MNRAS.447.3189X,Hsueh:2016aih,Hsueh:2017nlk,Hsueh:2017zfs}. Quasar flux ratio measurements have also been used to constrain the position and mass of potential individual subhalos within the lens galaxies \cite{Fadely:2009aa,Nierenberg:2014aa,Nierenberg:2017}. In a different regime, the strong magnification near the caustic of galaxy clusters acting as strong lenses could also be used to probe subhalos \cite{Dai:2018mxx}.

Using images of lensed extended sources, gravitational imaging \cite{Koopmans_2005,Vegetti:2008aa} has led to the statistically significant detection of a few subhalos with masses above $\sim10^8 M_\odot$ \cite{Vegetti_2010_1,Vegetti_2010_2,Vegetti_2012}. A somewhat similar technique using spatially resolved spectroscopic observations of gravitational lenses \cite{Moustakas:2003aa,Hezaveh:2012ai} has also led to the direct detection of a $\sim 10^9 M_\odot$ subhalo \cite{Hezaveh_2016_2}. Taken together, these measurements can in principle be used to put constraints on the subhalo mass function (see e.g.~Refs.~\cite{Vegetti:2009aa,2014MNRAS.442.2017V,Li:2015xpc,Ritondale:2018cvp}), although interpreting these detections in terms of physical subhalo masses might be more subtle than initially thought \cite{Minor:2016jou}. In the near future, transdimensional techniques \cite{Brewer:2015yya,Daylan:2017kfh} will allow a better understanding of various lensing degeneracies associated with subhalo detection within gravitational lenses.

Within the CDM paradigm, typical lens galaxies are expected to contain a large amount of mass substructure, with the subhalo mass function rapidly rising toward small halo mass \cite{Springel:2008cc,Despali:2016meh}. While the smallest subhalos are likely to be individually too light to be detected with standard gravitational imaging techniques, their collective effect might be detectable, in the spirit of the original work by Dalal and Kochanek \cite{Dalal:2002aa}. Such a collective approach based on approximate Bayesian computation was used in Ref.~\cite{Birrer2017} to put constraints on the abundance of substructure within the lens RX J1131-1231. 

A different approach brought forward by Ref.~\cite{Hezaveh_2014} (see also \cite{2018MNRAS.474.1762C}) aims at characterizing the mass substructure through their power spectrum. Interestingly, the substructure power spectrum contains information about the inner profile, mass function, and abundance of subhalos within lens galaxies \cite{Rivero:2017mao,Brennan:2018jhq,Rivero:2018bcd}, hence packing a lot of information within a single function. More generally, the power spectrum allows one to capture the effects of overdensities that cannot be easily described within the traditional language of the halo model (i.e. pancakes, streams, etc.) \cite{Angulo:2013sza}. Since the density field on such small scales is expected to be non-Gaussian, the power spectrum does not in general capture all its properties. The power spectrum can nonetheless contain important clues about the behavior of dark matter on subkiloparsec scales, just like measuring the nonlinear matter power spectrum of the large-scale structure of the Universe can lead to important information about the sum of neutrino masses. Recently, Ref.~\cite{Bayer:2018vhy} provided the first upper limits on the amplitude of the substructure power spectrum in the lens SDSS J0252+0039. 

In this work, we perform an in-depth analysis of the sensitivity of gravitational lens images to the substructure power spectrum, focusing on data taken with charged coupled devices, which are common in optical astronomy. We present here a likelihood-based mathematical framework necessary to extract the substructure power spectrum directly from pixel-based images. While our general approach follows a similar philosophy to that of Ref.~\cite{Hezaveh_2014}, our computational technique differs at several levels, especially in our use of a mode function-based approach. Our approach also differs significantly from Ref.~\cite{2018MNRAS.474.1762C} since we make no \emph{a priori} assumption about the statistical symmetries of the lensed source. Importantly, we extend our power spectrum mathematical framework to include compact time-varying sources such as quasars, hence opening substructure power spectrum measurements to a broader range of gravitational lenses. Since the goal of this paper is to present the framework necessary to extract measurements of the substructure power spectrum from lensed images and develop some intuition about their sensitivity to this latter quantity, we focus here on simple parametric source and lens models. Also, to avoid the complexity related to multiplane lensing (see e.g.~Refs.~\cite{Keeton:2003aa,McCully_2014,Despali:2017ksx}), we shall concentrate in this work on substructure within the lens galaxy. We  note however that most of the machinery developed here is likely applicable in the presence of line-of-sight structure with only minor modifications, as long as the effects of the substructure (both subhalos and those along the line of sight) on the lensed image are small. 

Given the unique potential of this technique in probing subkiloparsec structure within galaxies and along the line of sight at cosmological distances from the Milky Way, we aim this paper at an audience that is not necessarily familiar with galaxy-scale strong lensing. As such, we carefully review the different ingredients and assumptions entering our analysis. Readers with expertise in cosmic microwave background or large-scale structure analyses will find several familiar concepts and techniques throughout this paper. Expert readers could skip directly to Sec.~\ref{sec:likelihood_ext_src} for details about our method to extract the substructure power spectrum from images of gravitationally lensed sources.

This paper is organized as follows. In Sec.~\ref{sec:small_scale_struc}, we review the mass decomposition of the lens galaxy into macro lens and substructure, and then introduce the substructure convergence power spectrum. In Sec.~\ref{sec:sub_lensing}, we review the impact of mass substructure on observed images of galaxy-scale gravitational lenses, focusing on extended sources. In Sec.~\ref{sec:likelihood_ext_src}, we present the derivation of our likelihood for the substructure power spectrum in the case of extended lensed images. The numerical implementation of this likelihood is discussed in Sec.~\ref{sec:fourier_impl}, and simple Fisher forecasts are presented in Sec.~\ref{sec:Fisher}. We present in Sec.~\ref{sec:numerical_results} complete Markov Chain Monte Carlo analyses of mock lensed images of extended sources to assess sensitivity to the substructure power spectrum. In Sec.~\ref{sec:likelihood_comp_src}, we generalize our likelihood computation to include time-dependent compact sources such as quasars and present simple Fisher forecasts. We finally conclude in Sec.~\ref{sec:conclusions}.

Throughout this paper, we assume a Planck 2015 cosmology \cite{Ade:2015xua}. We also take the redshift of the source to $z_{\rm src} =0.6$ and that of the lens to be $z_{\rm lens} = 0.25$, which results in a critical density for lensing $\Sigma_{\rm crit} = 5.998\times10^{10} M_\odot/{\rm arcsec}^2 = 3.686\times10^9 M_\odot/{\rm kpc}^2$ in the lens plane. A useful number to keep in mind is that for these choices of cosmology and redshifts, 1 arcsec $\approx4$ kpc in the lens plane.

\section{Small-scale structure and lens galaxies}\label{sec:small_scale_struc}
We begin this paper by reviewing the distinction between the so-called macro lens mass model and the small-scale  substructure contained within the lens galaxies or along the line of sight. We then review the relevant statistical properties of the substructure that are most interesting from a gravitational lensing point of view.
\subsection{Mass decomposition for galaxy-scale lenses}
In this work, we specialize to the case of galaxy-scale strong gravitational lenses, in which multiple images of a background source are generated. In general, the exact structure of the gravitational potential $\phi_{\rm lens}$ responsible for the lensing is the result of the complex assembly history of the lens galaxy as well as its subsequent dynamical evolution. In addition, structures along the line of sight can also contribute to the richness of the projected gravitational potential. Despite this apparent complexity, many observed galaxy-scale gravitational lenses can be reasonably fitted with relatively simple mass models, such as power-law ellipsoids (see however Ref.~\cite{Woldesenbet:2011aa}).

A typical lens galaxy contains structure on a variety of scales, with the larger scale features responsible for the broad morphology of the observed lensed images, while the small-scale structures (e.g.~satellite galaxies, giant molecular clouds, globular clusters, etc.) give rise to small corrections to the lensed observables. This suggests that we can decompose the lensing convergence [i.e.~the two-dimensional (2D) projected mass density divided by the critical surface density for lensing] into a dominant macro component $\kappa_0(\yy)$, and a small contribution $\ksub(\yy)$ parametrizing the difference between the actual projected mass distribution and the dominant component $\kappa_0$, that is,
\be\label{eq:smooth+subs}
\kappa_{\rm lens}(\yy) = \kappa_0(\yy)+\ksub(\yy).
\ee
Note that we have absorbed the mean convergence in the substructure (denoted $\bar{\kappa}_{\rm sub}$) within $\kappa_0$ such that the $\ksub$ field as defined above has zero expectation value, $\langle \ksub\rangle =0$.  We note that in the absence of lensing time-delay observations, stellar kinematic measurements, or strong priors on the brightness and size of the source, it is difficult to constrain $\bar{\kappa}_{\rm sub}$ due to the mass-sheet degeneracy \cite{1985ApJ...289L...1F}. We shall refer to $\kappa_0$ (and $\phi_0$) as the macro lens (or component) since it is responsible for determining the broad configuration of the lens. In general, it contains the contributions from the smooth dark matter halo, the dominant baryonic structure (disk, bulge, or otherwise), and possibly from single massive subhalos significantly affecting the configuration of the lens (such as those identified in Refs.~\cite{2014MNRAS.442.2017V,Vegetti_2012,Hezaveh_2016_2}). 

On the other hand, the substructure convergence $\ksub$ (and its related lensing potential $\psub$) contains contributions from the usual dark matter subhalos and satellite galaxies orbiting the main lens galaxy, as well as from other astrophysical structures such as tidal streams, debris, dense gas clouds, and globular clusters. The structure of the line-of-sight density field also contributes to $\ksub$, although we do not explicitly take it into account here since this would require multiplane lensing. The crucial point is that the perturbations encoded in $\ksub$ are subdominant\footnote{By construction, if the $\ksub $ perturbations were large, they would lead to easily detectable effects, implying that they should have been absorbed in $\kappa_0$.} compared to $\kappa_0$, hence resulting in subtle disturbances in the lensed images. Detecting the presence of these small perturbations and characterizing their statistical properties could yield important clues about the nature of dark matter. 
\subsection{Two-point statistics for substructure convergence}\label{sec:power_spectrum}
The typical length scales probed by strong lensing observations are deep into the nonlinear regime of cosmological structure formation where both baryon- and dark matter-dominated objects contribute significantly to the overall projected mass density. Taken at face value, this should imply that the statistics of the projected mass density fluctuations encoded in $\ksub$ are highly non-Gaussian and difficult to compute from first principles without resorting to expensive numerical simulations. However, in the cases where $\ksub$ and $\psub$ receive contributions from a large number of dark matter and baryonic mass substructures in the lens galaxy, it is possible to invoke the central limit theorem to argue that the statistics of $\ksub$ and $\psub$ are approximately Gaussian \cite{Keeton:2009aa,Hezaveh_2014,Cyr-Racine_2015}. This occurs for instance in cold dark matter models where the subhalo mass function rises steeply toward small halo mass. For such approximately Gaussian scenarios, the statistical properties of the substructure convergence field are almost entirely captured by its two-point correlation function, or its Fourier transform, the power spectrum.

In a realistic lens galaxy, the mass substructure convergence and potential will never be exactly Gaussian. As was argued in Ref.~\cite{Hezaveh_2014}, the non-Gaussian signatures are dominated by massive subhalos or line-of-sight structure appearing close in projection to lensed images. In high signal-to-noise images, these massive halos could in principle be directly detected through gravitational imaging \cite{Koopmans:aa,Vegetti:2008aa,Hezaveh:2012ai} and incorporated into the main lens model described by $\kappa_0$, hence leaving the statistics of the mass substructures contributing to $\ksub$ roughly Gaussian. But even if the statistics of the substructure convergence are not entirely Gaussian, it is nonetheless interesting to measure the substructure convergence power spectrum since it still contains important information about small-scale structure within lens galaxies.

In contrast to the more familiar case of cosmological large-scale structure, we do not expect the substructure density field within a single lens galaxy to be either homogeneous or isotropic. However, given the relatively small projected area probed by the strong lensing region, neglecting the spatial variation of the substructure power spectrum across the lensed image is likely a good approximation. The relevant quantity is then the lens plane-averaged substructure power spectrum, which for a single lens plane is defined as \cite{Rivero:2017mao}
\be\label{eq:power_spectrum}
P_{\rm sub}(\kk) =\int d^2\rr \, e^{-i \kk\cdot\rr} \int_A d^2\vs\,\frac{\big\langle\ksub(\vs) \ksub(\vs+\rr)\big\rangle}{A},
\ee
where $\kk$ is the Fourier wave number and $A$ is the area of the strong lensing region where we perform the spatial average. This power spectrum should be interpreted as an ensemble average over multiple realization of a given lens, or for practical purposes, as an ensemble average over multiple galaxy-scale lenses. As such, we expect that it will be on average statistically isotropic, and the monopole power spectrum should carry thus most of the signal. The latter is simply given by 
\be\label{eq:monopople_power_spectrum}
P_{\rm sub}^{(0)}(k)= \frac{1}{2\pi}\int_{0}^{2\pi} d\theta_k \,P_{\rm sub}(\kk),
\ee
where $\theta_k$ is the polar angle of the Fourier wavenumber $\kk$, and $k = |\kk|$. In this work, we shall focus our attention on the monopole contribution to the substructure convergence power spectrum but we note that the formalism developed below can easily accommodate anisotropic contributions\footnote{Since lens galaxies are generally not spherically symmetric due, e.g., to the presence of a disk or an elliptical baryonic distribution, it would be interesting, though likely difficult, to look at the possible dependence of the convergence power spectrum on the angle between Fourier modes and the baryonic major axis.} to the power spectrum.

\begin{figure}[t!]
\includegraphics[width=0.482\textwidth]{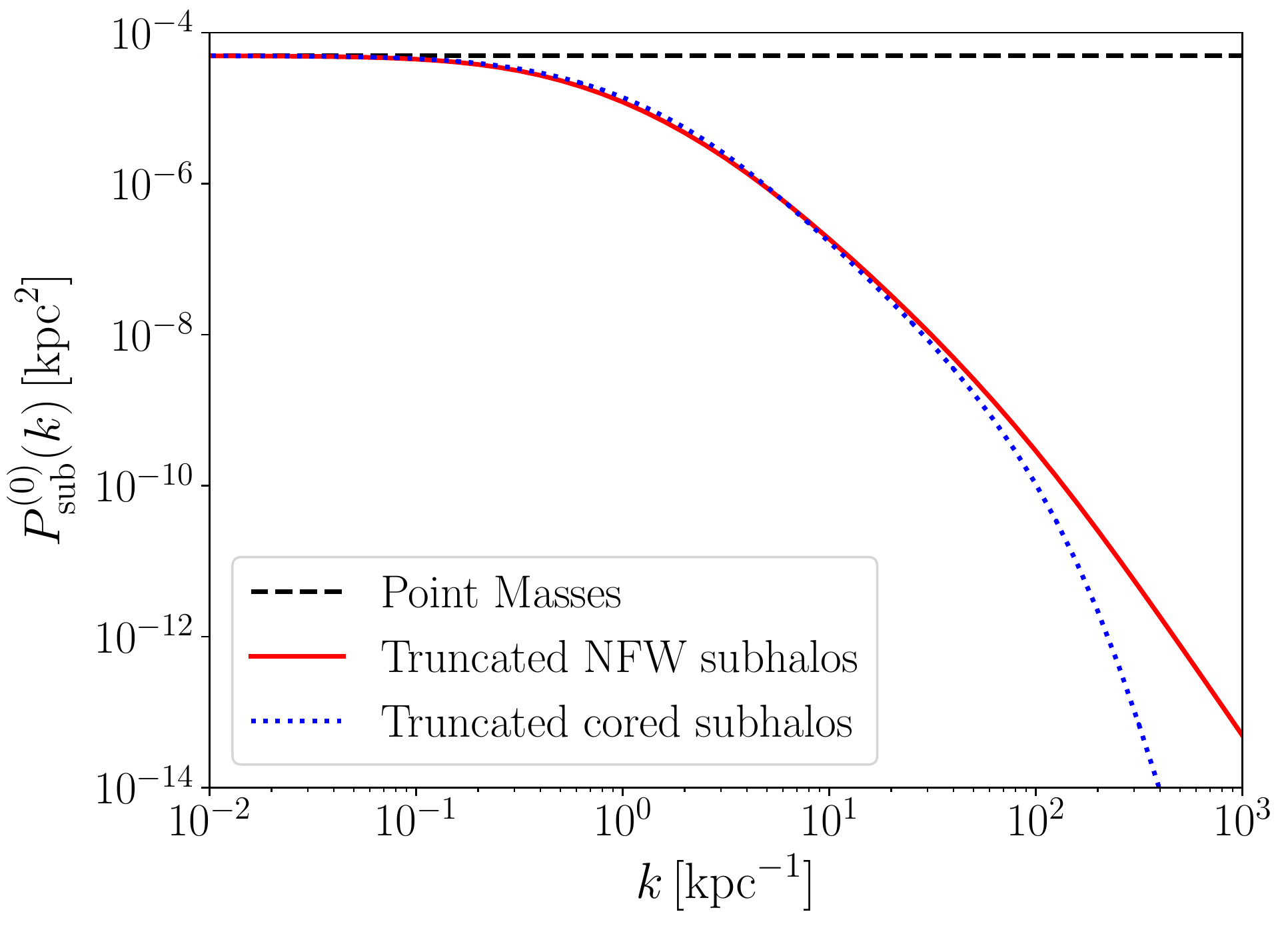}\\
\begin{flushright}
\includegraphics[width=0.455\textwidth]{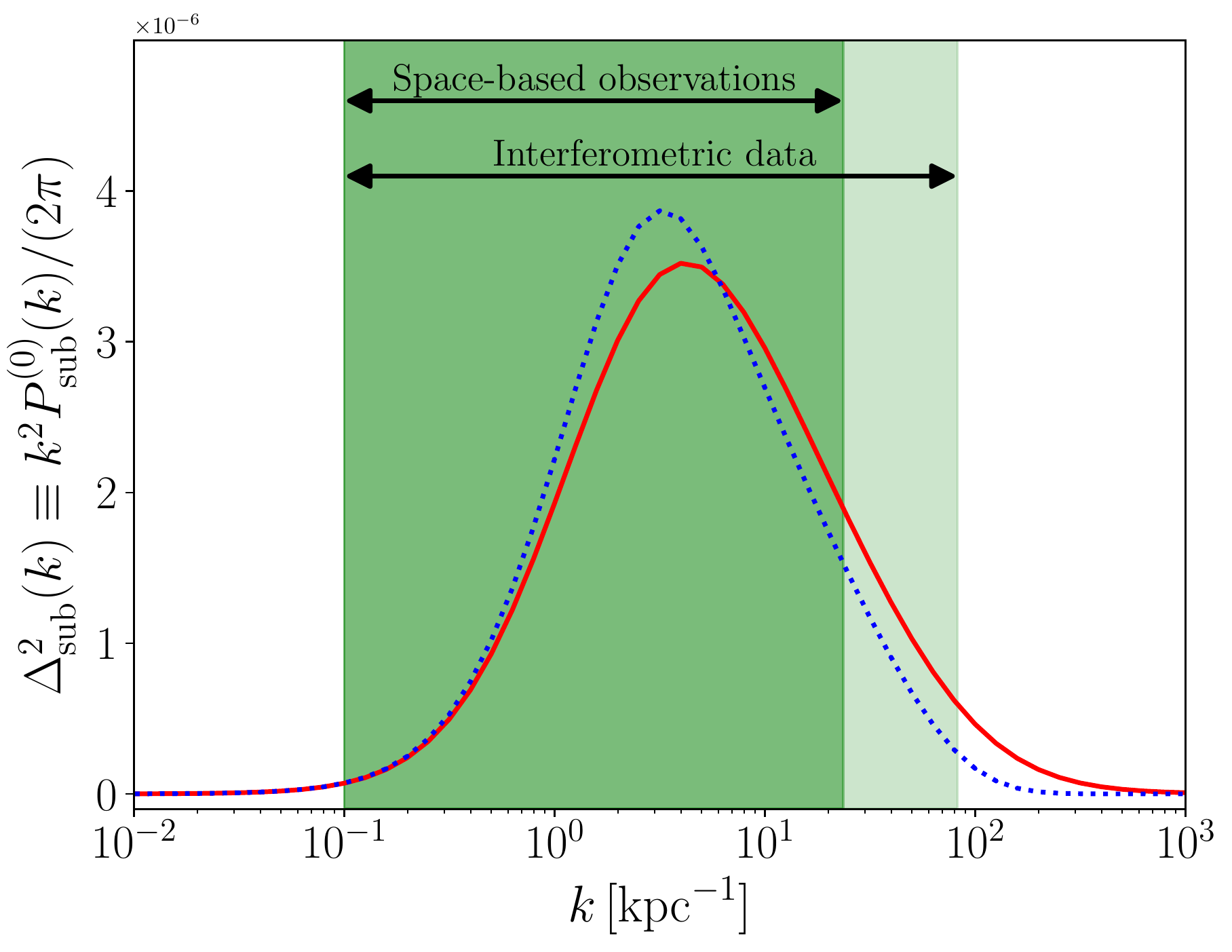}
\end{flushright}
\caption{The upper panel shows the substructure convergence power spectra for a population of point masses (dashed black), of truncated Navarro-Frenk-White (NFW) subhalos (solid red), and of truncated cored (Burkert) subhalos (dotted blue). Here, we have assumed a mean substructure abundance of $\bar{\kappa}_{\rm sub}=0.01$, a power-law subhalo mass function $dN/dm\propto m^\beta$ with $\beta=-1.9$ and $10^5 M_\odot < m < 10^8 M_\odot$. For the truncated NFW, we take the scale radius to scale with the subhalo mass as $r_{\rm s} =  0.11\,(m/10^6 M_\odot)^{1/2}$ kpc. The subhalos are tidally truncated at a radius $r_{\rm t} =  (m/10^6 M_\odot)^{1/3}(r_{\rm 3D}/100\, {\rm kpc})^{2/3}$ kpc \cite{Binney2008}, where $r_{\rm 3D}$ is the three-dimensional distance of the subhalo from the center of the host. We average over all possible $r_{\rm 3D}$ locations of a subhalo up to a radius of 409 kpc, as described in Ref.~\cite{Rivero:2017mao}. We adopt similar relations for the truncated cored subhalos but with a core radius $r_{\rm c} = 0.7\,r_{\rm s}$. The lower panel shows the dimensionless power per log interval in wave number, $\Delta_{\rm sub}^2(k)\equiv k^2P_{\rm sub}^{(0)}/(2\pi)$, for both the truncated NFW and cored subhalo population. The two green bands display the range of scales that could be probed by space-based observations (dark green) and interferometric data (light green).}\label{fig:Psub_example}
\end{figure}

A detailed study of the substructure convergence power spectrum for a dark matter-only population of subhalos was presented in Ref.~\cite{Rivero:2017mao}. We illustrate in the upper panel of Fig.~\ref{fig:Psub_example} examples of the monopole convergence power spectrum for this idealized scenario. We note that we only display the one-subhalo term and we thus neglect here subhalo-subhalo clustering. The black dashed line shows the convergence power spectrum for an unclustered distribution of point masses, which is essentially a white noise (flat) power spectrum. The amplitude of this power spectrum is simply given by \cite{Keeton:2009aa,Hezaveh_2014,Cyr-Racine_2015,Rivero:2017mao}
\be
P_{\rm sub}^{(0)}(k) = \frac{\bar{\kappa}_{\rm sub} \langle m^2\rangle}{\Sigma_{\rm crit} \langle m\rangle},  \qquad \text{(point masses)}
\ee
where $\langle m^n\rangle$ is the $n$th moment of the subhalo mass function \cite{Cyr-Racine_2015}. 
We also display the convergence power spectrum for two other populations of subhalos, a truncated Navarro-Frenk-White (NFW) \cite{Navarro_1996} subhalo population inspired by standard CDM, and a cored (modeled with a Burkert \cite{Burkert_1995} profile) subhalo population inspired by self-interacting dark matter. In both cases, we see that the convergence power spectrum follows that of the point masses on scales larger than the size of the largest subhalos. Once that scale is reached, however, the convergence power spectrum for both the NFW and cored population begins to rapidly decay away from the point mass case. On small enough scales, the convergence power spectrum starts to probe the inner slope of the subhalos' density profile, with the cored subhalo case displaying less power on these scales than the more cuspy NFW subhalos. We refer the reader to the figure caption for more details on the parameter used to generate these power spectra.

To understand on which scales the mass substructures matter the most, it is instructive to consider the dimensionless power spectrum
\be
\Delta_{\rm sub}^2(k) \equiv \frac{k^2 P_{\rm sub}^{(0)}(k)}{2\pi},
\ee
which describes the amount of substructure convergence power per logarithmic interval in wave number. We illustrate this dimensionless power spectrum in the lower panel of Fig.~\ref{fig:Psub_example} for both the truncated NFW and truncated  cored subhalo population. Since we observe that most of the power lies on scales $0.03-10$ kpc for these plausible models, we naturally expect this range of scales to dominate the constraints that we can put on substructure using lensing data. Remarkably, such length scales correspond to the typical scales probed by galaxy-size strong lens observations. Assuming Einstein radii of $1$--$2''$ and redshifts of $z_{\rm lens}=0.1$--$1$ which are typical for galaxy-scale lenses (see e.g.~Ref.~\cite{Bolton2008}), the range of scales that can in principle\footnote{We say ``in principle'' since the exact length scales that can actually be probed depend on the structure of the lensed source.} be probed with space-based lensing observations (assuming a 70 mas image resolution) is indicated by the dark green band in the lower panel of Fig.~\ref{fig:Psub_example}. The higher image resolution achievable with interferometric data (such as those from the Atacama Large Millimeter/submillimeter Array) can significantly extend this range to higher wave numbers (pale green band), hence allowing measurements on most of the length scales where mass substructure matters the most. Thus, galaxy-scale strong lenses constitute ideal laboratory to study small-scale dark matter structures. 

In the remainder of this paper, we shall use both the point mass and the truncated NFW power spectra shown in Fig.~\ref{fig:Psub_example} as benchmark fiducial models to generate simulated data and test their sensitivity to the amplitude and shape of the convergence power spectrum.

\section{Gravitational lensing in the presence of mass substructures}\label{sec:sub_lensing}
In this section, we review the impact of small-scale structure on lensing residuals for the case of extended, time-independent sources. Time-dependent compact sources such as quasars will be treated separately in Sec. \ref{sec:likelihood_comp_src}.
\subsection{Preliminaries}
Let us take a source at redshift $z_{\rm src}$ with a surface brightness profile at wavelength $\lambda_{\rm src} \equiv \lambda/(1+z_{\rm src})$ and time $t_{\rm s}$ described by $\tilde{S}(\uu,t_{\rm s},\lambda_{\rm src})$, where $\lambda$ is the wavelength of the observation. Here, $\uu$ stands for two-dimensional angular coordinates in the source plane. We also take a gravitational lens at redshift $z_{\rm lens}$ specified by the projected mass density (convergence) $\kappa_{\rm lens}(\yy)$ and its related lensing potential $\phi_{\rm lens}(\yy)$, which we assume to be static in time.\footnote{On the typical timescale of a single astronomical observation, this is indeed a very good approximation. However, for a multi-epoch observational campaign of a given lens system, it is possible to notice changes in the gravitational potential due to stellar microlensing. In this work, we treat microlensing separately from $\phi_{\rm lens}$ (see Sec.~\ref{sec:TD_compact_source}) and we can therefore take the latter to be static.} Again, $\yy$ stands for two-dimensional angular coordinates, but in the image plane. The lensing convergence and potential are related to each other via the usual Poisson equation $\nabla^2\phi_{\rm lens} = 2\kappa_{\rm lens}$. Let us denote the lensed image of the source detected at time $t$ and wavelength $\lambda$ by $\tilde{O}(\yy,t,\lambda)$. Formally, the relation between the source and the lensed image is \cite{schneider1999gravitational}
\begin{align}\label{eq:formal_sol}
\tilde{O}(\yy,t,\lambda) &= \int d\uu\, \int dt_{\rm s} \, \tilde{S}(\uu,t_{\rm s},\lambda_{\rm src} )\,\de\left(t-t_{\rm s} -\tau(\yy,\uu) \right)\en
&\qquad \qquad \times \de\left(\uu-\yy+\vec{\nabla}\phi_{\rm lens}(\yy)\right) ,
\end{align}
where $\vec{\nabla}$ denotes the gradient with respect to $\yy$, $\de$ is the Dirac delta function, and where the excess time delay is
\be\label{eq:time_delay_def}
\tau(\yy,\uu) =t_0\left[ \frac{1}{2}|\yy-\uu|^2-\phi_{\rm lens}(\yy)\right],
\ee
with
\be\label{eq:time_constant}
t_0\equiv \frac{1+z_{\rm lens}}{c}\frac{D_{\rm l} D_{\rm s}}{D_{\rm ls}},
\ee
where $c$ is the speed of light, $D_{\rm l}$, $D_{\rm s}$, and $D_{\rm ls}$ are the angular diameter distance between the observer and the lens, the observer and the source, and the lens and the source, respectively. Performing the integrals in Eq.~\eqref{eq:formal_sol}, we obtain
\be
\tilde{O}(\yy,t,\lambda) =  \tilde{S} \left(\yy-\vec{\nabla}\phi_{\rm lens}(\yy) ,t -\tau(\yy),\lambda_{\rm src}\right),
\ee
where we use the shorthand notation $\tau(\yy) = \tau\left(\yy,\yy-\vec{\nabla}\phi_{\rm lens}(\yy)\right)$. In a typical observational scenario, the lens galaxy itself, the sky background, and possibly other objects along or close to the line of sight can also contribute to the observed photon flux in addition to the image of the lensed source. We take this into account by adding an external surface brightness contribution $\tilde{L}(\yy,t,\lambda)$ to the lensed image $\tilde{O}(\yy,t,\lambda)$.  Furthermore, the image is usually observed through a filter $F_{\lambda_i}(\lambda)$ centered at a characteristic wavelength $\lambda_i$. We thus define the two following quantities:
\begin{align}
S_{\lambda_i}(\uu,t_{\rm s}) &= \int d\lambda \,F_{\lambda_i}(\lambda) \tilde{S}(\uu,t_{\rm s},\lambda_{\rm src}),\\
L_{\lambda_i}(\yy,t) &= \int d\lambda \,F_{\lambda_i}(\lambda) \tilde{L}(\yy,t,\lambda),
\end{align}
which are the wavelength-integrated lensed source and external surface brightness, respectively. In addition, the light from all sources will be processed by the optics of the instrument used to observe it, as well by potential atmospheric disturbances. We take this effect into account by convolving the image with a point-spread function (PSF) $W_{\lambda_i}(\yy,t)$, which in general depends on the wavelength $\lambda_i$ and time of the observation. The actual model for the observed surface brightness $\hat{O}_{\lambda_i}(\xx,t)$ is thus given by
\begin{align}\label{eq:full_gen_sol}
\hat{O}_{\lambda_i}(\xx,t) &= \int d\yy\, \Big[S_{\lambda_i} \left(\yy-\vec{\nabla}\phi_{\rm lens}(\yy) ,t -\tau(\yy)\right)\\
&\qquad \qquad \qquad+ L_{\lambda_i}(\yy,t)\Big] W_{\lambda_i}(\xx - \yy,t).\nonumber
\end{align}

Finally, the light is usually collected for an exposure of length $T_{\rm exp}$ on a detector made of an array of two-dimensional pixels, and then converted to counts per pixel. Taking $P_j(\xx)$ to be the pixel response function of the $j$th pixel\footnote{For an ideal pixel, this function should be unity within the area spanned by the pixel, and zero elsewhere.}, and $\mathcal{S}_{\rm inv}^{(\lambda_i)}$ to be the inverse sensitivity of the detector for the filter labeled by $\lambda_i$, the number of counts in the $j$th pixel for the $k$th exposure is
\begin{align}\label{eq:brightness_to_count}
O_{\lambda_i}(\xx_j,t_k) &= \frac{1}{\mathcal{S}_{\rm inv}^{(\lambda_i)}} \int_{t_k}^{t_k+T_{\rm exp}} dt \int d^2\xx \,P_j(\xx) \hat{O}_{\lambda_i}(\xx,t)\en
&\approx \frac{A_{\rm pix} T_{\rm exp}}{\mathcal{S}_{\rm inv}^{(\lambda_i)}} \hat{O}_{\lambda_i}(\xx_j,t_k),
\end{align}
where we have assumed in going from the first to the second line that the lensed source, foregrounds, and the PSF are static on the exposure timescale, and that the pixel response function is uniform across the detector and given by a 2D rectangular function. Here, $A_{\rm pix}$ is the area of a pixel, and $\xx_j$ is the position of the $j$th pixel. 
\subsection{Lensing residuals for extended sources}
In this section, we review (see e.g. Refs.~\cite{2001ASPC..237...65B,Koopmans:aa,Vegetti:2008aa}) the structure of the residuals between a strongly lensed image of an extended source (e.g.~a galaxy) created using only a macro lens  $\kappa_{\rm lens}=\kappa_0$ from an image generated by a mass model that includes substructure, $\kappa_{\rm lens} = \kappa_0 + \ksub$. By causality, an extended source can be considered static on the typical time scales associated with astronomical observations. We can thus neglect the time dependence of the source in Eq.~\eqref{eq:full_gen_sol}. However, the PSF and foreground light can vary from observations to observations, and we keep their time dependence explicit.  Since the $\psub$ potential causes only small distortions to the observed image, we can perform a perturbative analysis in $\psub$ and expand $\hat{O}_{\lambda_i}(\xx,t)$ from Eq.~\eqref{eq:full_gen_sol} as
\begin{align}\label{eq:O_lambda}
\hat{O}_{\lambda_i}(\xx, t) &\approx \int d\yy W_{\lambda_i}(\xx-\yy,t) \Big[S_{\lambda_i} \left(\yy-\vec{\nabla}\phi_0(\yy)\right)  \\
& -\vec{\nabla}_{\uu}S_{\lambda_i} \left(\uu \right)\big|_{\uu = \yy-\vec{\nabla}\phi_0(\yy)}\cdot \vec{\nabla}\psub(\yy) +L_{\lambda_i}(\yy,t)\Big],\nonumber
\end{align}
where $\vec{\nabla}_{\uu}$ denotes the source-plane gradient, and where $\vec{\nabla}\psub(\yy)$ is the deflection vector field created by the substructure.  We note that the source-plane gradient arising in Eq.~\eqref{eq:O_lambda} can be translated to the image plane via the relation
\be\label{eq:u2y_convertion}
\vec{\nabla}_{\uu}S_{\lambda_i} \left(\uu \right)\big|_{\uu = \yy-\vec{\nabla}\phi_0(\yy)} = \left(\frac{\pa \yy}{\pa\uu}\right)\cdot \vec{\nabla}S_{\lambda_i}\left(\yy-\vec{\nabla}\phi_0(\yy)\right),
\ee
where we recognize that the prefactor is nothing more than the magnification tensor ${\bf M}_0 \equiv \pa\yy/\pa\uu$ generated by the macro lens component. Denoting by $\hat{O}_\lambda^{(0)}(\xx,t)$ the image of the source lensed purely by the macro potential (as well as potential foregrounds), the residuals between the image of a source lensed by the total lens potential $\phi_{\rm lens}$, and that of the same source lensed only by the macro component $\phi_0$ is
\begin{align}\label{eq:delta_O}
\de \hat{O}_{\lambda_i,{\rm sub}}(\xx,t) &\equiv \hat{O}_{\lambda_i}(\xx,t) - \hat{O}_{\lambda_i}^{(0)}(\xx,t)\\
&\approx-\int d\yy\, W_{\lambda_i}(\xx-\yy,t) \en
&\qquad \times \Big[ \vec{\nabla}_{\uu}S_{\lambda_i} \left(\uu \right)\big|_{\uu = \yy-\vec{\nabla}\phi_0(\yy)}\cdot \vec{\nabla}\psub(\yy) \Big].\nonumber 
\end{align}
We thus obtain the well-known result \cite{2001ASPC..237...65B,Koopmans:aa,Vegetti:2008aa} that the lensed image residuals of an extended source are proportional to the gradient of that source evaluated in the image plane. Residuals are largest when this latter gradient is either aligned or antialigned with the deflection field created by substructures. The above expression automatically captures the well-known fact that large, smoothly varying sources lead to lensed images that are largely insensitive to short-scale variations in the substructure potential.
\begin{figure*}[t!]
\centering
\includegraphics[width=0.42\textwidth]{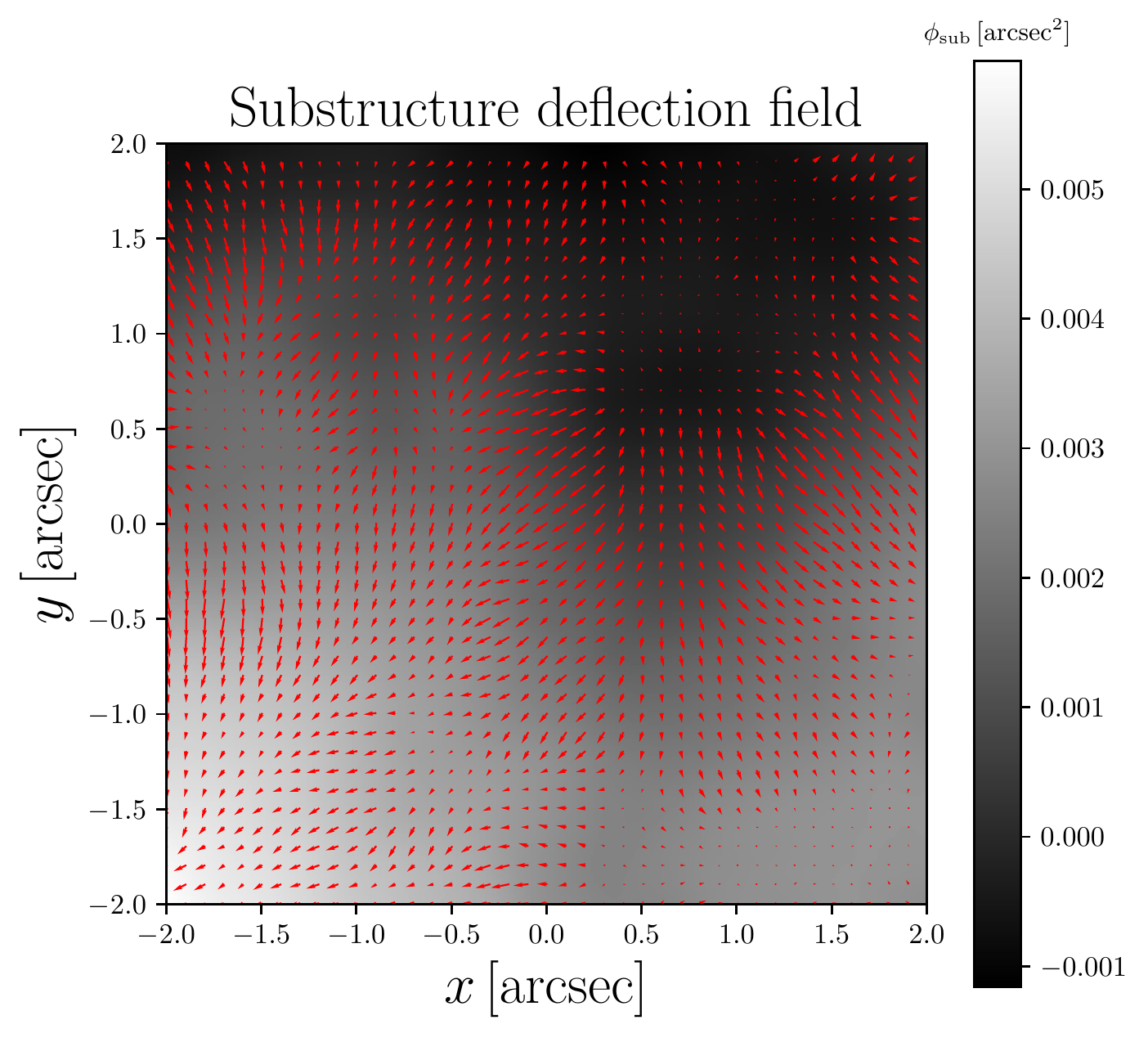}
\includegraphics[width=0.42\textwidth]{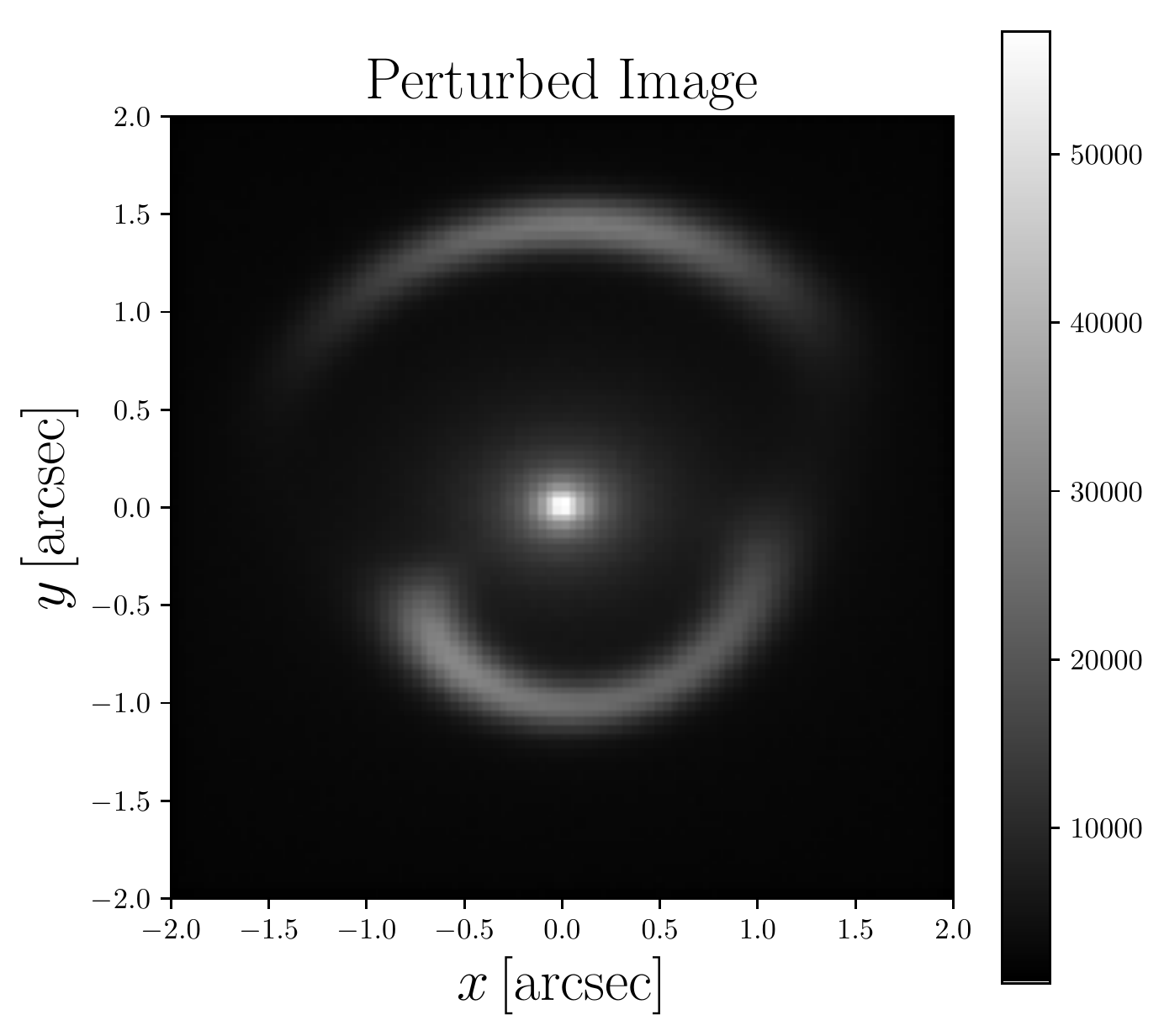}\\
\includegraphics[width=0.42\textwidth]{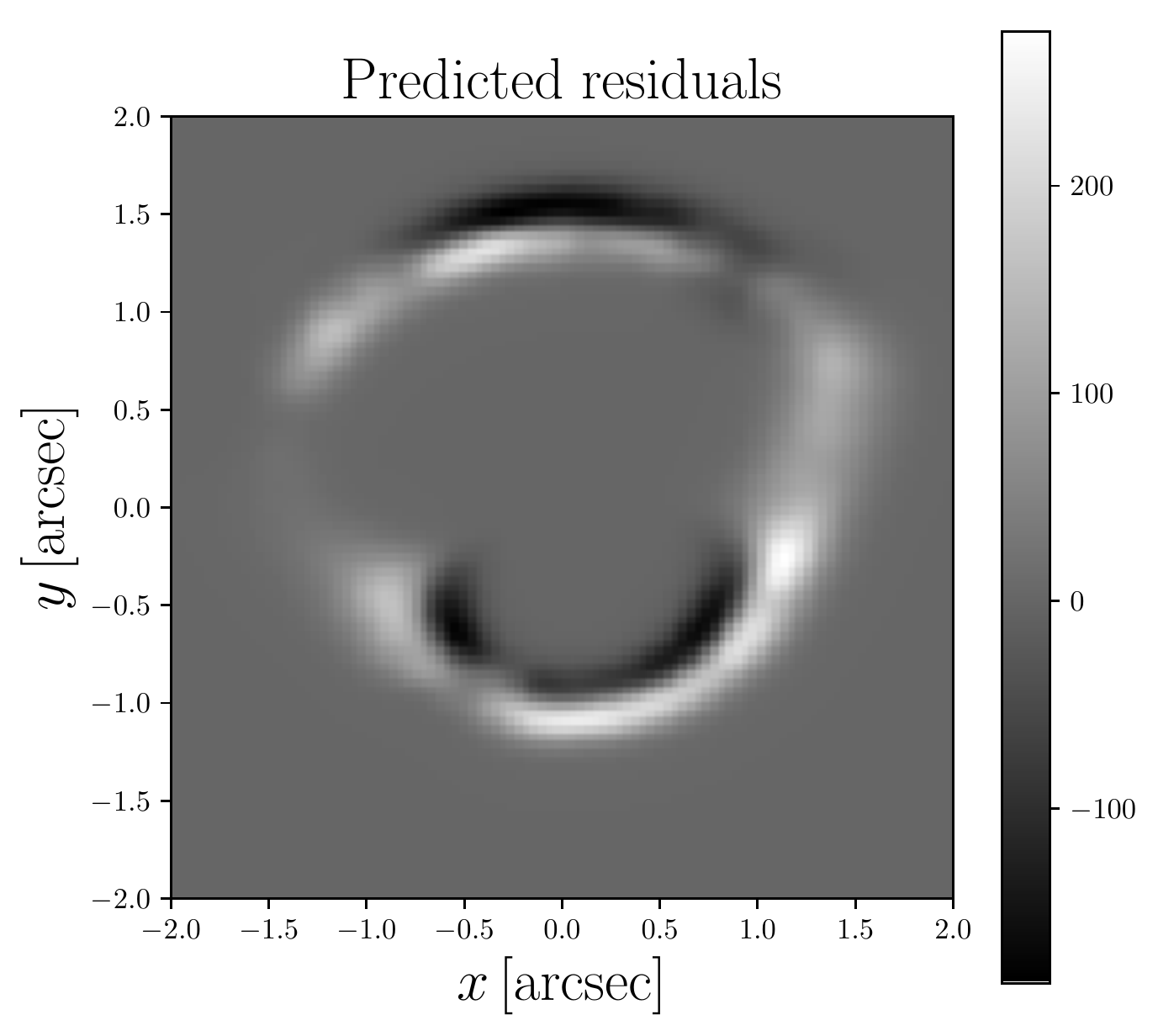}
\includegraphics[width=0.42\textwidth]{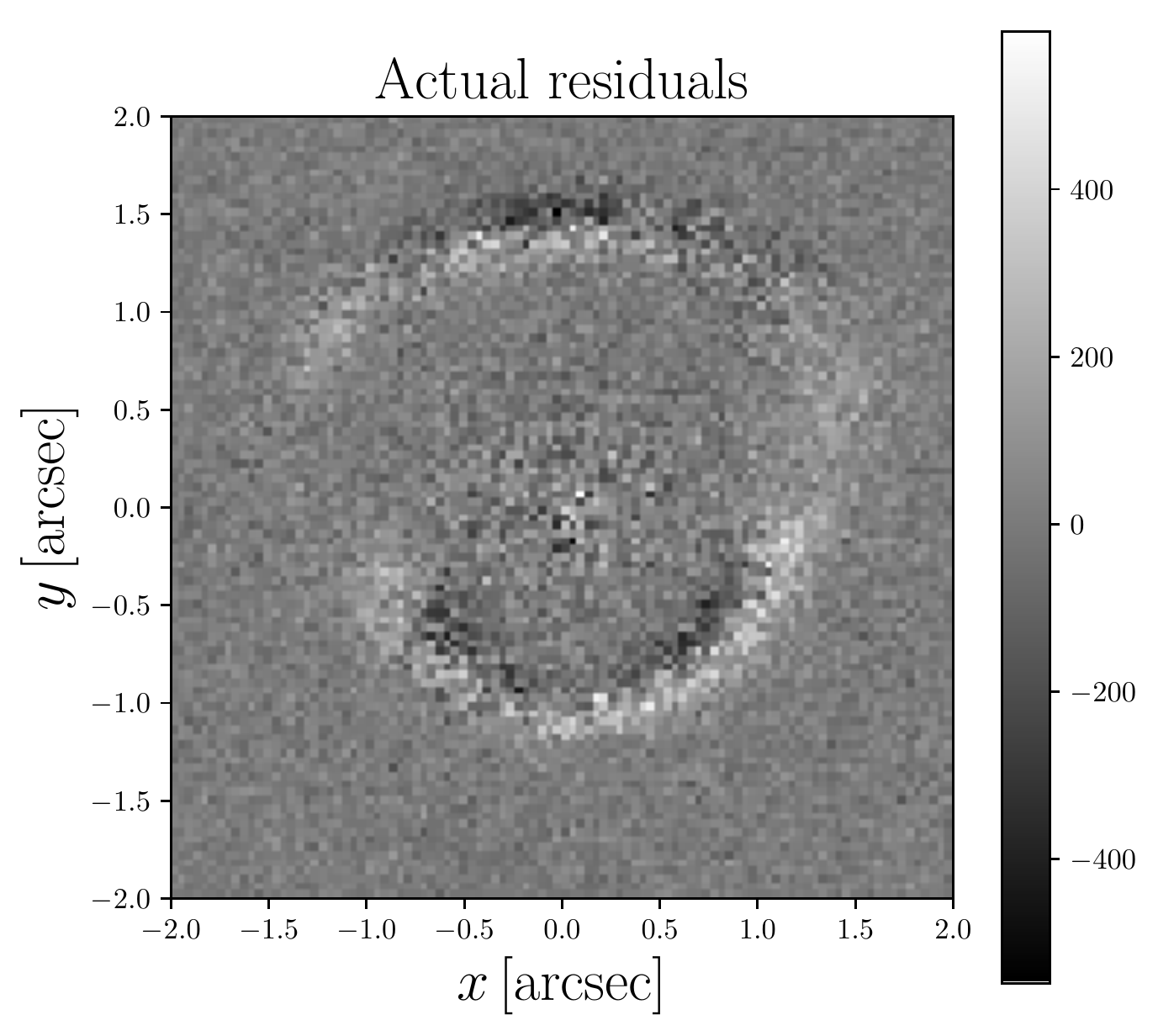}\\
\includegraphics[width=0.42\textwidth]{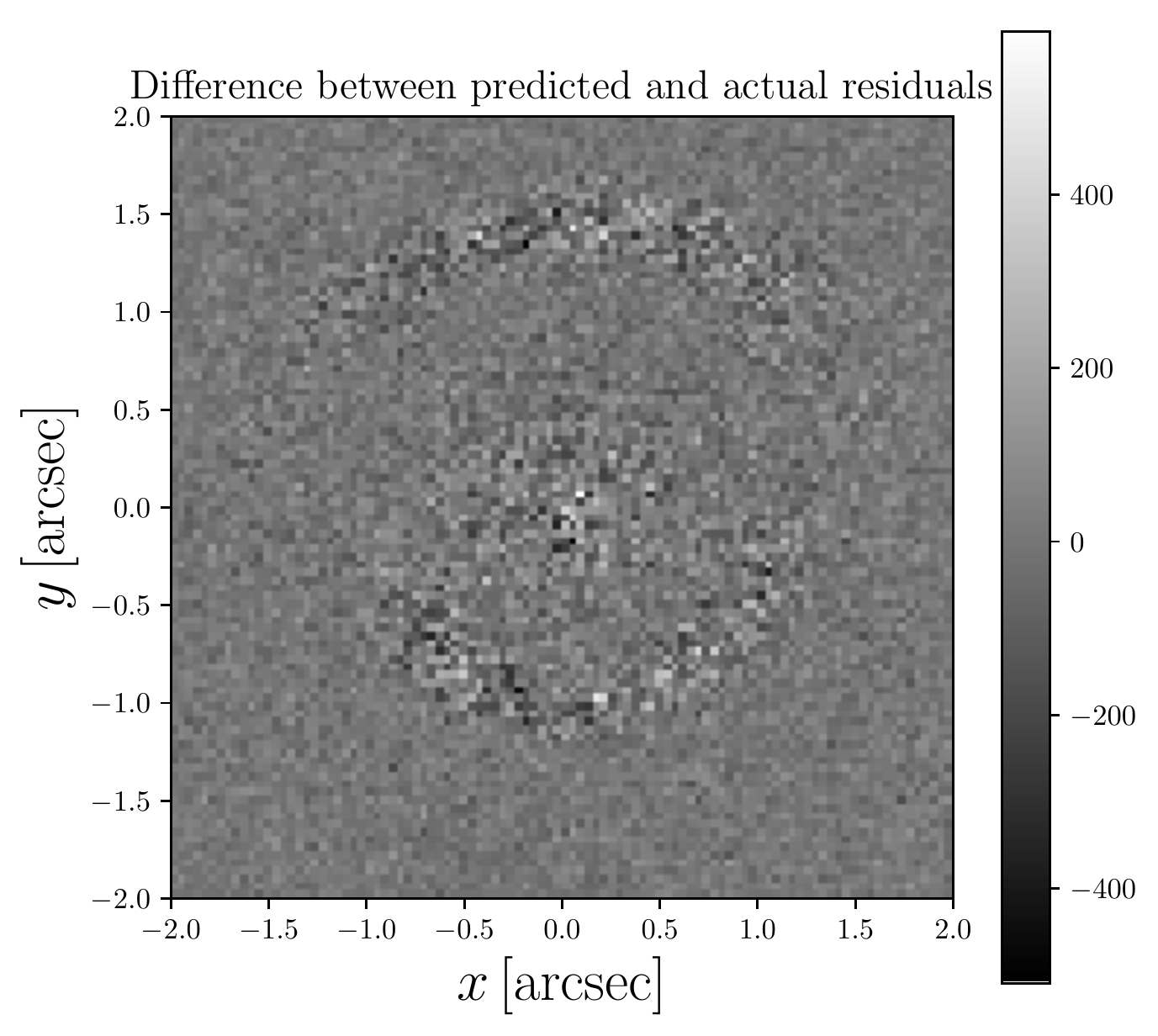}
\includegraphics[width=0.42\textwidth]{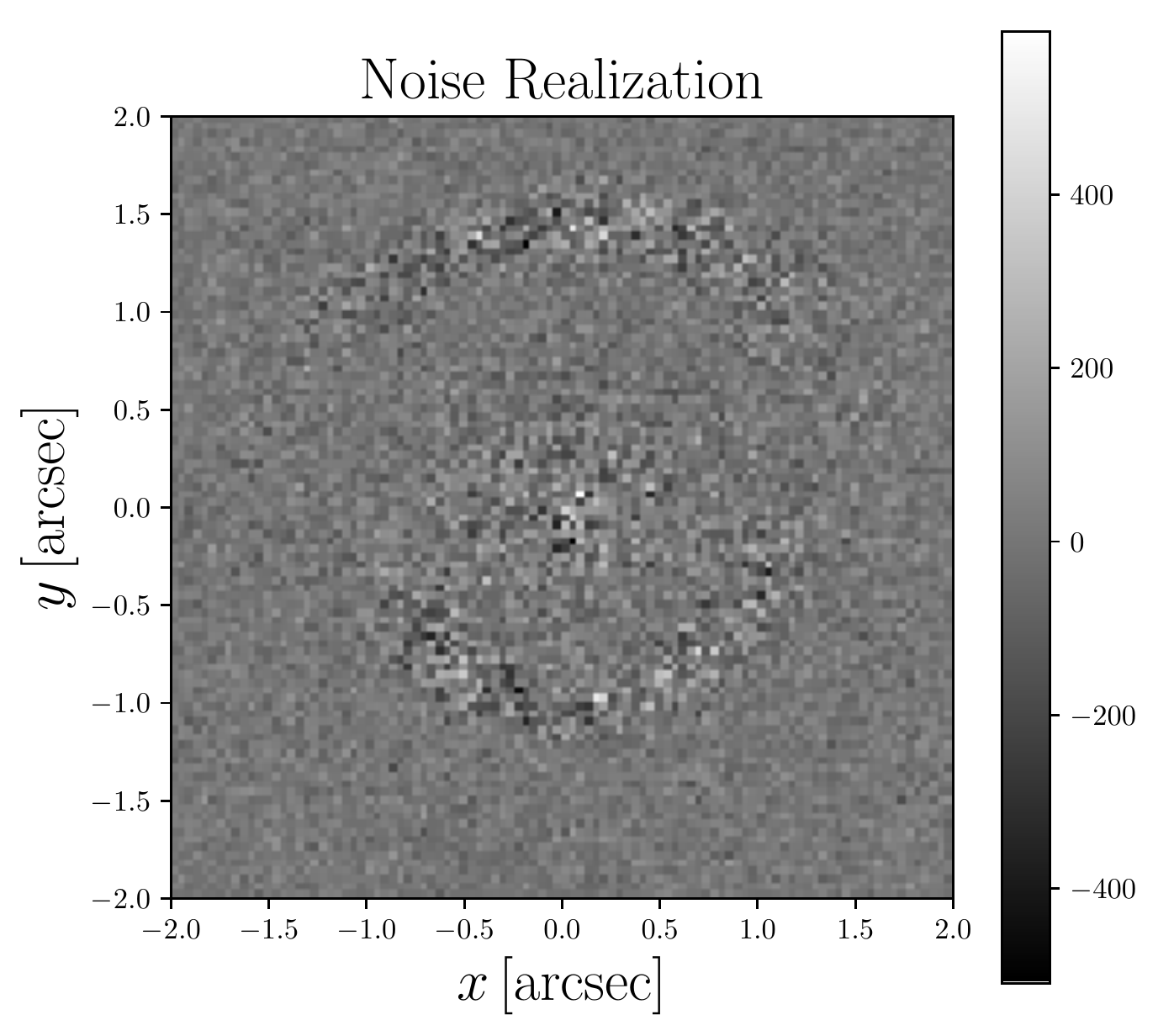}
\caption{Illustration of the lensing residuals in the presence of substructure. The upper left panel shows the substructure deflection field (red arrows) from a population of truncated NFW subhalos with a convergence power spectrum given by the red solid line of Fig.~\ref{fig:Psub_example}. The upper right panel shows a mock image of a background galaxy (modeled as a $n=1/2$ S\'ersic profile) that is strongly lensed by a foreground galaxy (modeled as an isothermal ellipsoid plus external shear). The light from the lens galaxy (modeled as a $n=4$ S\'ersic), sky background, and Poisson noise are added to the image.  The middle left panel shows the predicted image residuals as given by Eq.~\eqref{eq:delta_O} for the specific substructure realization used to generate the lensed image. The middle right panel shows the actual image residuals gotten by subtracting from the mock image a model image generated with the same lens galaxy, source parameters, environment, and observational conditions, but without the substructure. Finally, the lower left panel shows the difference between the residuals predicted by Eq.~\eqref{eq:delta_O} and the actual image residuals which is entirely consistent with the Poisson noise realization used to generate the mock image (lower right panel).}\label{fig:residuals_example}
\end{figure*}
%

\subsection{Validity of the linear approximation}\label{sec:valid_first_order}
We illustrate in Fig.~\ref{fig:residuals_example} how the image residuals predicted by Eq.~\eqref{eq:delta_O} (middle left panel) compare with the actual residuals (middle right panel) gotten by simply taking the difference $\hat{O}_{\lambda_i}(\xx,t) - \hat{O}_{\lambda_i}^{(0)}(\xx,t)$ between an actual mock image generated with a nonvanishing $\ksub$ and a model image generated using only a macro lens model $\kappa_0$. The details of the lens, source, foreground, noise, and PSF models used to generate these images can be found in Appendix \ref{sec:img_sims} below. Here, the substructures are modeled as truncated NFW subhalos with a convergence power spectrum given by the red solid line of Fig.~\ref{fig:Psub_example}. Despite the noise, we see that the predicted residuals match very closely the actual image residuals, implying that the most of the substructure information is captured by the first-order expansion performed in Eq.~\eqref{eq:delta_O}. This is also illustrated in the lower left panel of Fig.~\ref{fig:residuals_example} where we display the difference between the predicted and actual image residuals, which is entirely consistent with the Poisson noise realization (lower right panel) used to generate the mock image. 

Mathematically, once can assess the validity of the first-order approximation used to derive Eq.~\eqref{eq:delta_O} by looking at the next order contribution to the lensing residuals, which takes the form
\be
\frac{1}{2}\nabla\psub(\yy)\cdot \mathcal{H}_{S_{\lambda_i}}\cdot \nabla\psub(\yy),
\ee
where $\mathcal{H}_{S_{\lambda_i}}$ is the Hessian matrix containing the second derivatives of the source surface brightness profile. Let us estimate the magnitude of this second-order term compared to the leading-order contribution $\nabla_\uu S_{\lambda_i}\cdot \nabla\psub$. For a source of typical size $r_{\rm src}$, the ratio of the determinant of the Hessian matrix to that of the norm of the gradient of the source is $\sim 1/r_{\rm src}$, which implies that the second-order corrections to Eq.~\eqref{eq:delta_O} are negligible if
\be
\frac{| \mathcal{H}_{S_{\lambda_i}}|| \nabla\psub|}{|\nabla_\uu S_{\lambda_i}|}\sim \frac{|\nabla\psub|}{r_{\rm src}}\ll 1,
\ee
that is, the typical magnitude of the substructure deflection must be much smaller than the size of the source. Since the typical magnitude of the substructure deflection field is \cite{Keeton:2009aa,Cyr-Racine_2015}
\be
| \nabla\psub|\sim \left(\frac{\bar{\kappa}_{\rm sub} \langle m^2 \rangle}{\Sigma_{\rm crit} \langle m \rangle}\right)^{1/2}\sim 10^{-3} \,\text{arcsec},
\ee
the first-order approximation should be valid for a broad range of galaxy-scale lenses with extended sources where the source sizes are typically in the range $0.1-1$ arcsec (see e.g.~Ref.~\cite{Bolton2008}). However, the first-order calculation obviously breaks down for compact sources such as quasars for which a different treatment is necessary (see Sec.~\ref{sec:TD_compact_source}).

\subsection{Degeneracy with the source brightness profile}
When confronted with the image residuals given in Eq.~\eqref{eq:delta_O}, the immediate question that comes to mind is: \emph{can these residuals be reabsorbed by an appropriate modification to the source brightness profile?} To answer this question, let us imagine that we add a small contribution $\de S_{\lambda_i}(\uu)$ to the source surface brightness profile, that is,
\be
S_{\lambda_i}(\uu)\to S_{\lambda_i}(\uu) + \de S_{\lambda_i}(\uu).
\ee
We want $\de S_{\lambda_i}(\uu)$ to absorb the image residuals introduced by the presence of substructures within the lens. To do so, the small source correction once projected to the image plane must have the form
\be\label{eq:src_mod_image_plane}
\de S_{\lambda_i}(\yy - \nabla\phi_0(\yy)) = \nabla_\uu S_{\lambda_i}\big|_{\uu = \yy-\vec{\nabla}\phi_0(\yy)} \cdot \nabla \psub(\yy),
\ee
where we only kept terms that are first order in perturbation variables $\{\nabla\psub,\de S_{\lambda_i}\}$. To actually compute the source correction, we must project this expression back to the source plane. First, in the weak lensing regime, the lens equation $\uu = \yy - \nabla\phi_0(\yy)$ has a unique solution $\yy(\uu)$ and projecting Eq.~\eqref{eq:src_mod_image_plane} back to the source plane is a well-defined procedure. It is then always possible to add a source correction of the form
\be
\de S_{\lambda_i}(\uu) = \nabla_\uu S_{\lambda_i}(\uu)\cdot \nabla \psub(\yy(\uu)).
\ee
to ``gauge'' away the image residuals caused by substructures\footnote{We note that if the structure of the source is well-known---as is the case for the cosmic microwave background---it is not necessarily possible to just ``gauge'' away the effect of the substructure, even in the weak lensing regime. See Ref.~\cite{Nguyen:2017zqu} for details.}. We note that the same logic applies in the case where $\nabla\psub$ is a constant vector. 

However, in the strong lensing regime where the lens equation $\uu = \yy - \nabla\phi_0(\yy)$ has multiple solutions and $\nabla\psub$ is a random nonconstant vector, the mapping between lens and source planes is no longer one to one. It is then no longer possible to define a unique $\de S_{\lambda_i}(\uu)$ that can entirely absorb the image residuals given in Eq.~\eqref{eq:delta_O}. For instance, let us imagine the case where the lens equation has two distinct solutions $\yy_1(\uu)$ and $\yy_2(\uu)$. We can still define a source brightness correction $\de S^{(1)}_{\lambda_i}(\uu)$ using the first solution $\yy_1(\uu)$. Adding this correction to the overall source model will indeed nullify the image residuals in the neighborhood of the first image, but will also introduce extra residuals near the second image. This argument is easily generalized to a larger number of images. The main message here is that the redundancy introduced by having multiple images of the lensed source ensures that substructure effects can never be entirely absorbed by adding complexity to the source.  

Thus, in the strong lensing regime, the image residuals caused by the stochastic substructure in the lens (or along the line of sight) are in general not degenerate with the source surface brightness profile. Of course, for a realistic noisy image with imperfect knowledge of the PSF, it will be always possible to absorb some of the residuals by a modification of the source brightness profile (due to the specifics of the noise realization and the PSF side lobes), but the argument presented above shows that it is never possible to completely eliminate the substructure-caused residuals by changing the source. 
\subsection{Degeneracy with foregrounds}
One might also worry that the image residuals caused by the substructure could be degenerate with foreground light, either from the lens galaxy itself, or other faint objects along the line of sight. Unfortunately, this possibility appears difficult to eliminate since there is no a priori reason for why foregrounds could not mimic substructure-caused residuals. However, Eq.~\eqref{eq:delta_O} has a very specific functional form linking the structure of the source and of the macro lens to that of the residuals. Since the broad structure of the lens provides information about the general configuration of the source and of the macro lens, it should in principle be possible to distinguish image residuals caused by substructures from mismodeled foregrounds. In any case, it is unlikely in our opinion that foregrounds could exactly reproduce the mathematical structure given in Eq.~\eqref{eq:delta_O}, so they are unlikely to be completely degenerate with the impact of substructures. 

We thus conclude that it is in principle possible to extract information about mass substructure within gravitational lenses by examining image residuals such as those illustrated in Fig.~\ref{fig:residuals_example}. We now turn our attention to how exactly one could extract that information from lensed images of extended sources.

\section{Likelihood analysis for the substructure power spectrum: Extended source}\label{sec:likelihood_ext_src}
In this section, we analyze how gravitationally lensed images of an extended source can be used to extract constraints on the properties of the substructure inside lens galaxies. As noted in Ref.~\citep{Hezaveh_2014}, the fact that the leading-order image residuals for a lensed extended source are linearly proportional to the gradient of the substructure potential provides a straightforward way to write down a likelihood for the two-point correlation function of the substructure's projected mass density.  We derive a general expression for this likelihood below.

In the following, we denote the array of parameters describing the source as $\qsrc$, those describing the macro lens galaxy as $\qgal$, those describing its environment by $\qenv$, those describing the foreground light by $\qfore$, and those describing the properties of the instrument used to make the observations by $\qinst$. For notational convenience, we gather these different sets of parameters into a single array $\qq=\{\qsrc,\qgal,\qenv,\qfore,\qinst\}$. We also take the statistical properties of the mass substructures to be described by an array of parameters $\qsub$, which may include parameters describing the amplitude and shape of the substructure convergence power spectrum. In general, the data for a given lens system will consist of a time series of pixel counts taken with different filters centered at wavelength $\lambda_i$, that is,
\be
O_{\rm obs} = \big\{ O_{ {\rm obs},\lambda_i}(\xx_j,t_k)\big\},
\ee
where $t_k$ denotes the epoch of the $k$th observations, and $\xx_j$ the position of the $j$th pixel.

We first obtain the image residuals by subtracting from the data a model for the lensed image $O^{(0)}_\lambda(\xx_j,t_k;\qq)$ for a given choice of source, macro lens, environment, foreground light, and instrumental configuration. The observed residuals at the $j$th pixel are then given by 
\be
\de O_{{\rm obs},\lambda}(\xx_j,t_k;\qq) \equiv O_{{\rm obs},\lambda}(\xx_j,t_k) - O^{(0)}_\lambda(\xx_j,t_k;\qq).
\ee
For an appropriate choice of source structure and macro lens, the observed residuals should be caused by the effect of mass substructures and instrumental noise
\be
\de O_{{\rm obs},\lambda}(\xx_j,t_k;\qq)  = \de O_{{\rm sub},\lambda}(\xx_j,t_k;\qq,\qsub) +N_\lambda(\xx_j,t_k),
\ee
where $N_\lambda(\xx_j,t_k)$ is the instrumental noise in the $j$th pixel and where we have explicitly written that the substructure residuals depend on the choice of source, macro lens and its environment, as well as on parameters $\qsub$ describing the substructure population. 
\subsection{Basis function expansion}
To extract information from lensed images about the small-scale structure of the matter density field, it is useful to expand the substructure deflection field in terms of a set of orthonormal basis functions $\{\nabla\varphi_l\}$. In general, this basis should be chosen as to both maximally simplify the analysis and facilitate comparison between measurements and theoretical predictions. There are many possible choices of basis function, including a Fourier basis, a polar harmonic expansion, or a shapelet basis \cite{Refregier:2001fd}. To retain generality, we refrain at this point from specifying an actual orthonormal basis and write the deflection field as
\be\label{eq:expansion_phi_sub}
\nabla\psub(\xx) =\sum_{l = 1}^{N_{\rm modes}}  \mathcal{A}_{l} \nabla\varphi_l(\xx)
\ee
where $\mathcal{A}_{l}$ is the (usually complex) amplitude of the $l$th mode of the substructure deflection field. The basis functions satisfy the following orthonormality condition
\be\label{eq:orthonormality_cond}
\frac{1}{A_{\rm img}}\int_{A_{\rm img}} d^2\xx\, \nabla\varphi_l(\xx)\cdot\nabla\varphi^*_{l'}(\xx)=\de_{ll'},
\ee
where $A_{\rm img}$ is the area of the sky spanned by the data. This orthogonality condition can be used to invert Eq.~\eqref{eq:expansion_phi_sub} for the mode amplitudes
\begin{align}\label{eq:A_l_for_ksub}
\mathcal{A}_l &= \frac{1}{A_{\rm img}}\int_{A_{\rm img}} d^2\xx\, \nabla\varphi_l^*(\xx)\cdot\nabla\psub(\xx)\en
&=-\frac{2}{A_{\rm img}}\int_{A_{\rm img}} d^2\xx \,\varphi_l^*(\xx) \ksub(\xx),
\end{align}
where we have used integration by parts\footnote{We define our mode functions $\varphi_l$ such that they vanish on the boundary of the integration domain, ensuring that the surface term does not contribute to Eq.~\eqref{eq:A_l_for_ksub}.} and the Poisson equation $\nabla^2\psub = 2\ksub$ to write the last line. The residuals can then be written as
\begin{align}
\de O_{{\rm sub},\lambda}(\xx_j,t_k) &=\sum_{l = 1}^{N_{\rm modes}}  \mathcal{A}_{l} \mathcal{W}^\lambda_{l}(\xx_j,t_k),
\end{align}
where $\mathcal{W}_{l}^\lambda(\xx,t_k)$ is the gradient of the source projected into the $l$th mode and convolved with the PSF. For an extended source, it takes the form
\begin{align}
\mathcal{W}_{l}^\lambda(\xx,t_k) &= -\frac{A_{\rm pix} T_{\rm exp}}{\mathcal{S}_{\rm inv}^{(\lambda)}}\int d\yy\, W_\lambda(\xx-\yy,t_k) \\
&\qquad\qquad \times \nabla\varphi_l(\yy) \cdot\nabla_{\uu}S_\lambda\big|_{\uu = \yy-\vec{\nabla}\phi_0(\yy)},\nonumber
\end{align}
where we have used Eq.~\eqref{eq:brightness_to_count} and assumed a uniform pixel response function (the above kernel can easily be generalized for a more complex pixel response). While the $\nabla\varphi_l(\xx)$ modes are orthogonal by construction, we note that the kernels $\mathcal{W}_{l}^\lambda(\xx,t_k)$ do not generally form an orthogonal basis of the lensing residuals due to the presence of the gradient of the source. 
\subsection{Likelihood}\label{sec:single_obs}
It is useful at this point to introduce a matrix and vector notation that will streamline the likelihood derivation. We shall assume here a single observation at a given wavelength $\lambda_i$ and will therefore drop the time and wavelength indices for now. We will restore them in the next subsection.  Let us gather the observed lensing residuals into a vector $\de{\bf O}_{\rm obs}$ of length $N_{\rm pix}$. Similarly, we denote the image residuals caused by substructure as $\de{\bf O}_{\rm sub}$. We also gather the mode basis $\mathcal{W}_{l}(\xx_j)$ into an $N_{\rm pix} \times N_{\rm modes}$ matrix ${\bf W}_{\rm E}$, such that $({\bf W}_{\rm E})_{jl} = \mathcal{W}_{l}(\xx_j)$. We note that ${\bf W}_{\rm E}$ is a nothing more than a change-of-basis matrix from the mode space to the pixel space. We finally gather the mode amplitudes $\mathcal{A}_l$ into a vector ${\bf a}\equiv\{\mathcal{A}_l\}$ of length $N_{\rm modes}$. With these definitions, the image residuals caused by substructure can be simply written as
\be
\de {\bf O}_{\rm sub} = {\bf W}_{\rm E}\, {\bf a}.
\ee

Since our data consist of photon counts on pixels, we expect the noise in each pixel to have a Poisson contribution in addition to other instrumental sources such as readout noise. Since our aim is to detect the subtle effects of mass substructures, we are primarily interested in high signal-to-noise images in which the photon counts per pixel will be large, implying that we can approximate the Poisson shot noise by a Gaussian contribution. We thus assume that the noise has statistical properties entirely given by
\be
\big\langle N(\xx_i)N(\xx_j)\big\rangle_N = {\bf C}_{N,ij},
\ee
where $\langle\ldots\rangle_N$ denotes ensemble averaging over noise realizations. The likelihood for the parameters $\qq$ and $\qsub$ marginalized over the unknown amplitudes $\mathcal{A}_l$ is then
\begin{align}\label{eq:raw_likelihood}
\mathcal{L}(\qq,\qsub) &\propto \int  d{\bf a}\, d {\bf a}^\dagger \mathcal{P}_{\rm sub}( {\bf a}|\qsub) \\
& \qquad\times \frac{e^{-\frac{1}{2}(\de{\bf O}_{\rm obs} -  {\bf W}_{\rm E}\, {\bf a})^\dagger{\bf C}^{-1}_{N}(\de{\bf O}_{\rm obs} -  {\bf W}_{\rm E}\, {\bf a})}}{\sqrt{|{\bf C}_{N}|}}.\nonumber
\end{align}
In general, the conditional probability distribution for the $\mathcal{A}_l$ coefficients given a choice of substructure parameters $\qsub$, $\mathcal{P}_{\rm sub}({\bf a}|\qsub)$, is difficult to determine since it depends on rather complex galaxy formation physics. However, as we argued in Sec.~\ref{sec:power_spectrum} the statistics of the substructure convergence field could be approximated as Gaussian, implying that the statistics of the $\mathcal{A}_l$ coefficients can be approximately captured by their two-point functions. In this approximation, we have
 \be\label{eq:Gaussian_A_l}
 \mathcal{P}_{\rm sub}({\bf a}|\qsub)\simeq \frac{ e^{-\frac{1}{2}{\bf a}^\dagger {\bf C}_{\rm sub}^{-1}{\bf a}}}{\sqrt{(2\pi)^{N_{\rm modes}}|{\bf C}_{\rm sub}|}},
 \ee
where we have written the variance of the $\mathcal{A}_{l}$ coefficients as
\be
({\bf C}_{\rm sub})_{ll'}\equiv\langle \mathcal{A}_{l}\mathcal{A}^*_{l'}\rangle ,
\ee
 and where $|{\bf C}_{\rm sub}| = |\det{{\bf C}_{\rm sub}}|$. In the presence of small non-Gaussianities, we note that Eq.~\eqref{eq:Gaussian_A_l} could be generalized by performing an Edgeworth expansion similar to that performed in Ref.~\cite{Cyr-Racine_2015}. Given the basis functions $\nabla\varphi_l(\xx)$, the ${\bf C}_{\rm sub}$ takes the form
\be\label{eq:C_sub_from_P_sub}
({\bf C}_{\rm sub})_{ll'} = \frac{4}{A_{\rm img}^2}\int \frac{d^2\kk}{(2\pi)^2} P_{\rm sub}(\kk)\tilde{\varphi}_l(\kk) \tilde{\varphi}^*_{l'}(\kk),
 \ee
 where $\tilde{\varphi}_l(\kk)$ is the Fourier transform of $\varphi_l(\xx)$. It is then useful to define the following matrix
\be\label{eq:G_matrix}
{\bf G}\equiv {\bf W}_{\rm E}^\dagger {\bf C}_{N}^{-1} {\bf W}_{\rm E}, 
\ee
which has the convenient property of being Hermitian, ${\bf G}^\dagger = {\bf G}$. We note that the  matrix ${\bf G}_{ll'}$ is essentially the noise covariance matrix projected into the $l$ and $l'$ modes.  We also introduce ${\bf g}$, the noise-weighted data vector projected into the mode space
\be\label{eq:g_vector}
{\bf g} \equiv   {\bf W}_{\rm E}^\dagger {\bf C}_{N}^{-1}\de {\bf O}_{\rm obs},
\ee
as well as the standard $\chi^2$ in the absence of substructure 
\be
\chi^2 \equiv \de {\bf O}_{\rm obs}^{\rm T} {\bf C}_{N}^{-1} \de {\bf O}_{\rm obs}.
\ee
With the simplifying choice given in Eq.~\eqref{eq:Gaussian_A_l} the likelihood is Gaussian in the ${\bf a}$ variables, and we can thus analytically marginalize over these coefficients. Equation \eqref{eq:raw_likelihood} then becomes
\begin{align}\label{eq:like_one_image}
\mathcal{L}(\qq,\qsub) &\propto \frac{e^{-\frac{1}{2}[\chi^2-{\bf g}^{\dagger}{\bf D}^{-1}{\bf g}]}}{\sqrt{|{\bf C}_{N}||{\bf C}_{\rm sub}||{\bf D}|}},
\end{align}
where
\be
{\bf D}= {\bf G} + {\bf C}^{-1}_{\rm sub}.
\ee
The matrix ${\bf D}$ and the vector ${\bf g}$ depend on both the source and smooth lens parameters $\qq$, while the data enter through $\chi^2$ and ${\bf g}$. The substructure parameters $\qsub$ only enter through the covariance matrix ${\bf C}_{\rm sub}$. 

Equation \eqref{eq:like_one_image} is the likelihood written in the ``mode'' basis. It could be cast into the perhaps more familiar pixel basis by using the Woodbury matrix identity to write 
\begin{align}
\chi^2-{\bf g}^{\dagger}{\bf D}^{-1}{\bf g} & = \de {\bf O}_{\rm obs}^{\rm T} {\bf V}^{-1} \de {\bf O}_{\rm obs},
\end{align}
and
\be
|{\bf C}_{N}||{\bf C}_{\rm sub}||{\bf D}| = | {\bf V}|, 
\ee
where 
\be
{\bf V} \equiv {\bf C}_N + {\bf W}_{\rm E} {\bf C}_{\rm sub} {\bf W}_{\rm E}^\dagger. 
\ee
With this definition, the likelihood admits the simple form
\be\label{eq:likelihood_pixel_space}
\mathcal{L}(\qq,\qsub) \propto  \frac{e^{-\frac{1}{2}\de {\bf O}_{\rm obs}^{\rm T} {\bf V}^{-1} \de {\bf O}_{\rm obs}}}{\sqrt{|{\bf V}|}}.
\ee
While the matrix ${\bf V}$ has a simple interpretation as the ``noise + signal'' covariance matrix and the form of the likelihood given in Eq.~\eqref{eq:likelihood_pixel_space} is rather intuitive, we note that it is often computationally advantageous to use the mode basis likelihood (Eq.~\eqref{eq:like_one_image}) since it usually involves lower dimensional matrices and vectors. Indeed, due to noise, the number of measurable modes with nonvanishing signal-to-noise ratio is usually much smaller than the number of image pixels and it is more efficient to first perform the projection into the mode basis [Eqs.~\eqref{eq:G_matrix} and \eqref{eq:g_vector}] before computing the likelihood. Furthermore, depending on the exact choice of mode basis, the matrix ${\bf G}$ and vector ${\bf g}$ can have important symmetries that significantly simplify their computation (see e.g. Sec.~\ref{sec:num_impl} below). 
\subsection{Generalization to an ensemble of observations}\label{sec:multi_obs}
Let us now turn our attention to the case where we have a sequence of observations of the same gravitational lens taken at different time stamps $t_k$ and/or with different filters centered at wavelength $\lambda$. Generalizing Eq.~\eqref{eq:raw_likelihood} to the case of a time series of independent observations, the likelihood marginalized over the coefficients $\mathcal{A}_l$ for an extended source takes the form
\begin{align}
\mathcal{L}(\qq,\qsub) &\propto \int  d{\bf a}\, d {\bf a}^\dagger \mathcal{P}_{\rm sub}( {\bf a}|\qsub) \\
&\,\,\times \frac{e^{-\frac{1}{2}\sum_{k,\lambda}\Delta_\lambda(t_k)^\dagger{\bf C}^{-1}_{N,\lambda}(t_k)\Delta_\lambda(t_k)}}{\sqrt{|\tilde{{\bf C}}_{N}|}},\nonumber
\end{align}
where
\be
\Delta_\lambda(t_k) \equiv \de{\bf O}_{{\rm obs},\lambda}(t_k) -  {\bf W}_{\rm E,\lambda}(t_k)\, {\bf a},
\ee
and where the exact form of the $\tilde{\bf C}_N$ matrix will be given below. Note that the argument of the exponent is now summed over epochs and wavelengths, which is valid if the noise of observations taken at different epoch is uncorrelated. It is understood that each image could have its own source parameters (if the source has a different morphology at different wavelength, for instance), foreground parameters, pixelization, and PSF.  On the other hand, the macro lens, its environment, and the substructure contribution to the lensing deflection are taken to be the same across all images. Taking $\mathcal{P}_{\rm sub}({\bf a}|\qsub)$ as given in Eq.~\eqref{eq:Gaussian_A_l}, we can marginalize over the $\mathcal{A}_l$ coefficients to obtain
\begin{align}\label{eq:like_multi_image}
\mathcal{L}(\qq,\qsub) &\propto \frac{e^{-\frac{1}{2}[\tilde{\chi}^2-{\bf \tilde{g}}^{\dagger}{\bf \tilde{D}}^{-1}{\bf \tilde{g}}]}}{\sqrt{|\tilde{\bf C}_N| |{\bf C}_{\rm sub}||{\bf \tilde{D}}|}},
\end{align}
where
\be\label{eq:chi_and_g_def}
\tilde{\chi}^2 =\sum_\lambda \sum_{k = 1}^{N_{\rm obs}^\lambda}\chi_\lambda^2(t_k),\quad {\bf \tilde{g}}= \sum_\lambda \sum_{k = 1}^{N_{\rm obs}^\lambda} {\bf g}_\lambda(t_k),
\ee
\be\label{eq:D_def}
\quad {\bf \tilde{D}} = \sum_\lambda \sum_{k = 1}^{N_{\rm obs}^\lambda} {\bf G}_\lambda(t_k) + {\bf C}^{-1}_{\rm sub},
\ee
\be\label{eq:detCN}
|\tilde{\bf C}_N| = \prod_\lambda \prod_{k=1}^{N_{\rm obs}^\lambda} |{\bf C}_{N_\lambda}(t_k)|,
\ee
and where $N_{\rm obs}^\lambda$ is the number of exposures with a filter centered at wavelength $\lambda$. We note that these expressions are very similar to the single-observation case, except that the relevant quantities are now summed over all available observations. 

\section{Numerical implementation in a discrete Fourier basis}\label{sec:fourier_impl}
In this section, we specialize the general framework presented in the previous section to a discrete Fourier basis. We first present the specifics of the Fourier case before discussing the details of our numerical implementation. 
\subsection{Discrete Fourier basis}
We define our discrete orthonormal Fourier basis functions such that
\be\label{eq:mode_Fourier}
\varphi_l(\xx) =
\begin{cases}
 \frac{e^{i \kk_l\cdot \xx}}{k_l} & \text{if } \xx \in A_{\rm img}\\
0 & \text{otherwise,}
\end{cases}
\ee
where $k_l = |\kk_l|$, and $A_{\rm img}$ is the sky area spanned by the data. For a $N_{\rm pix} \times N_{\rm pix}$ image with side length $R$ ($A_{\rm img} = R^2$), the orthonormality condition given in Eq.~\eqref{eq:orthonormality_cond} implies that the wave number $\kk_l$ must take discrete values
\be\label{eq:ortho_modes}
\kk_l = \left(\frac{2\pi \,l_x }{R},\frac{2\pi\, l_y }{R}\right),
\ee
with
\be\label{eq:full_Fourier_mode}
l_x,l_y=
\begin{cases}
-\frac{N_{\rm pix}-1}{2},\ldots, \frac{N_{\rm pix}-1}{2} & (N_{\rm pix}\,\text{odd}) \\
-\frac{N_{\rm pix}}{2}-1,\ldots, \frac{N_{\rm pix}}{2}-1 & (N_{\rm pix}\,\text{even}),
\end{cases}
\ee
but where the zero mode with $l_x=l_y=0$ is not included. Here, it is understood that the mode index $l$ is a shorthand notation for the doublet $\{l_x,l_y\}$ characterizing the $x$ and $y$ coordinates of the Fourier mode.  In this basis, the $\mathcal{W}_{l}^\lambda(\xx,t_k)$ kernel takes the form
\begin{align}\label{eq:W_kernel_Fourier}
\mathcal{W}_{l}^\lambda(\xx,t_k)& = -\frac{i  A_{\rm pix} T_{\rm exp}}{\mathcal{S}_{\rm inv}^{(\lambda)}}\int d\yy\, W_\lambda(\xx-\yy,t_k) \,e^{i \kk_l\cdot \yy}\\
&\qquad\qquad \times \hat{\kk}_l \cdot\nabla_{\uu}S_\lambda\big|_{\uu = \yy-\vec{\nabla}\phi_0(\yy)},\nonumber
\end{align}
which has the symmetry property $\mathcal{W}_{l}^{\lambda*} = \mathcal{W}_{-l}^\lambda$, a consequence of the image residuals being real. This implies that the total number of independent Fourier modes for an $N_{\rm pix} \times N_{\rm pix}$ image is
\be
N_{\rm modes, ind} = 
\begin{cases}
\frac{N_{\rm pix}^2-1}{2} & N_{\rm pix}\,\text{odd}\\
\frac{N_{\rm pix}(N_{\rm pix}-1)}{2}-1 & N_{\rm pix}\,\text{even}.
\end{cases}
\ee
\begin{figure*}[t!]
\centering
\includegraphics[width=0.245\textwidth]{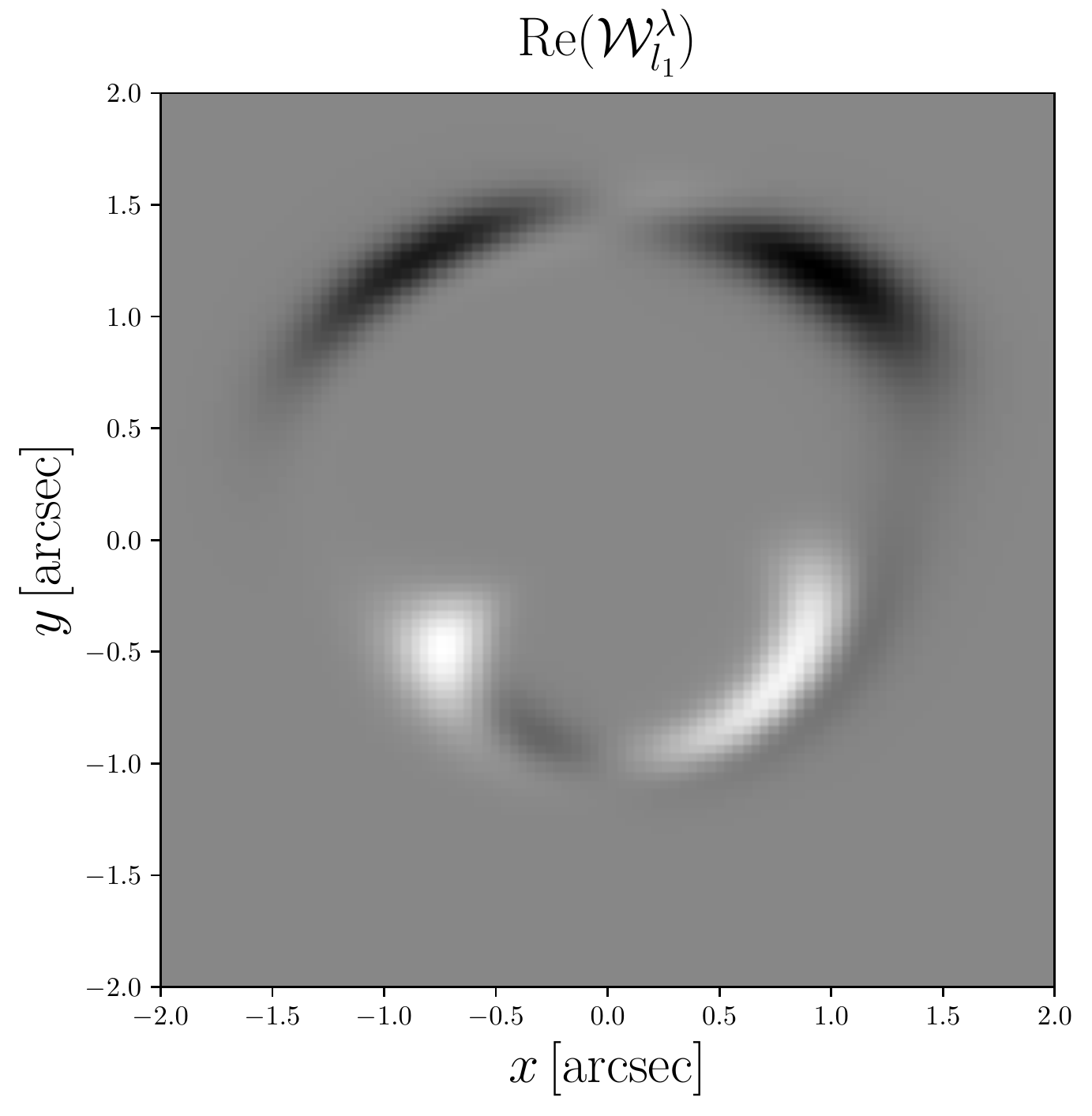}
\includegraphics[width=0.245\textwidth]{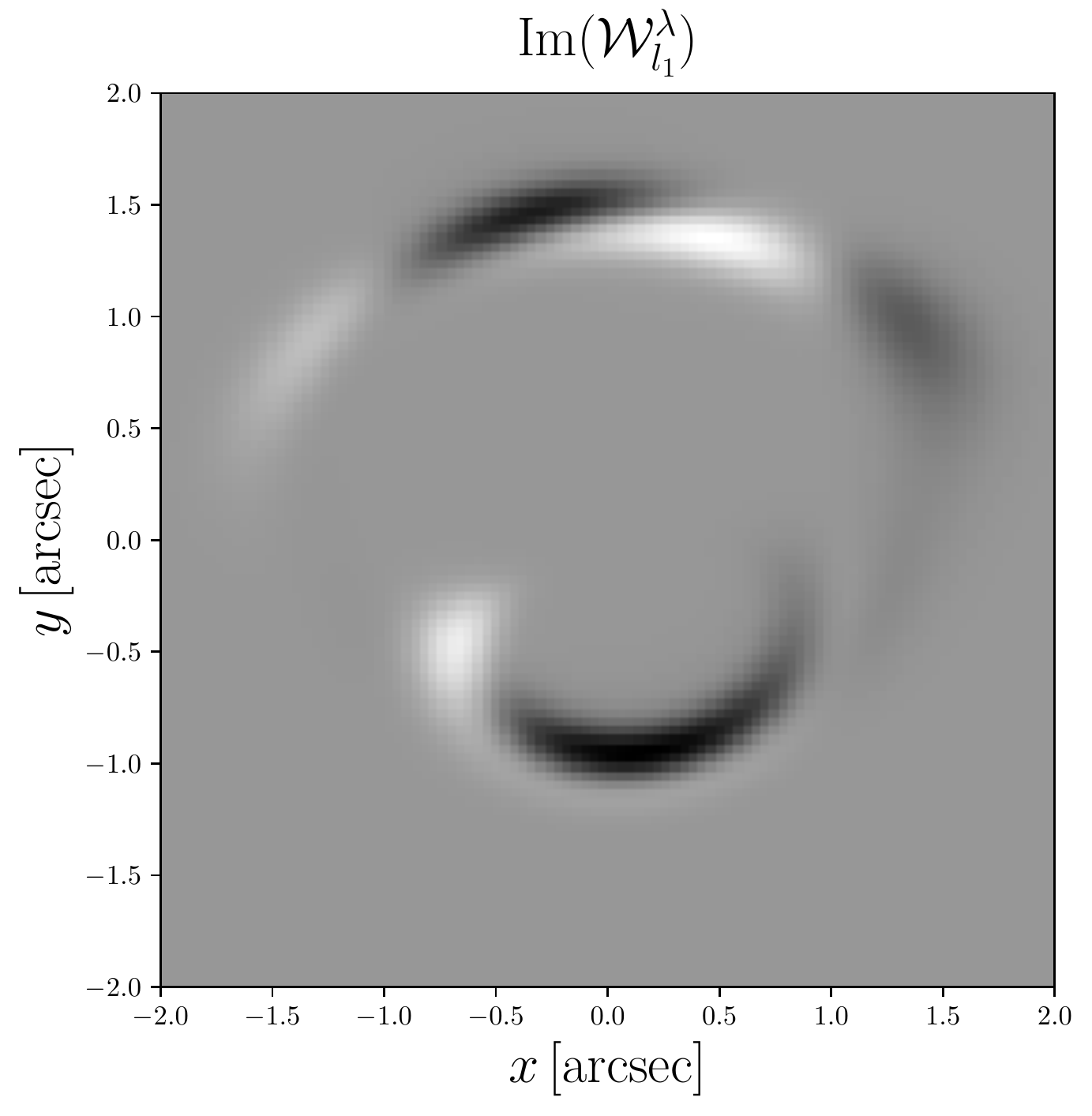}
\includegraphics[width=0.245\textwidth]{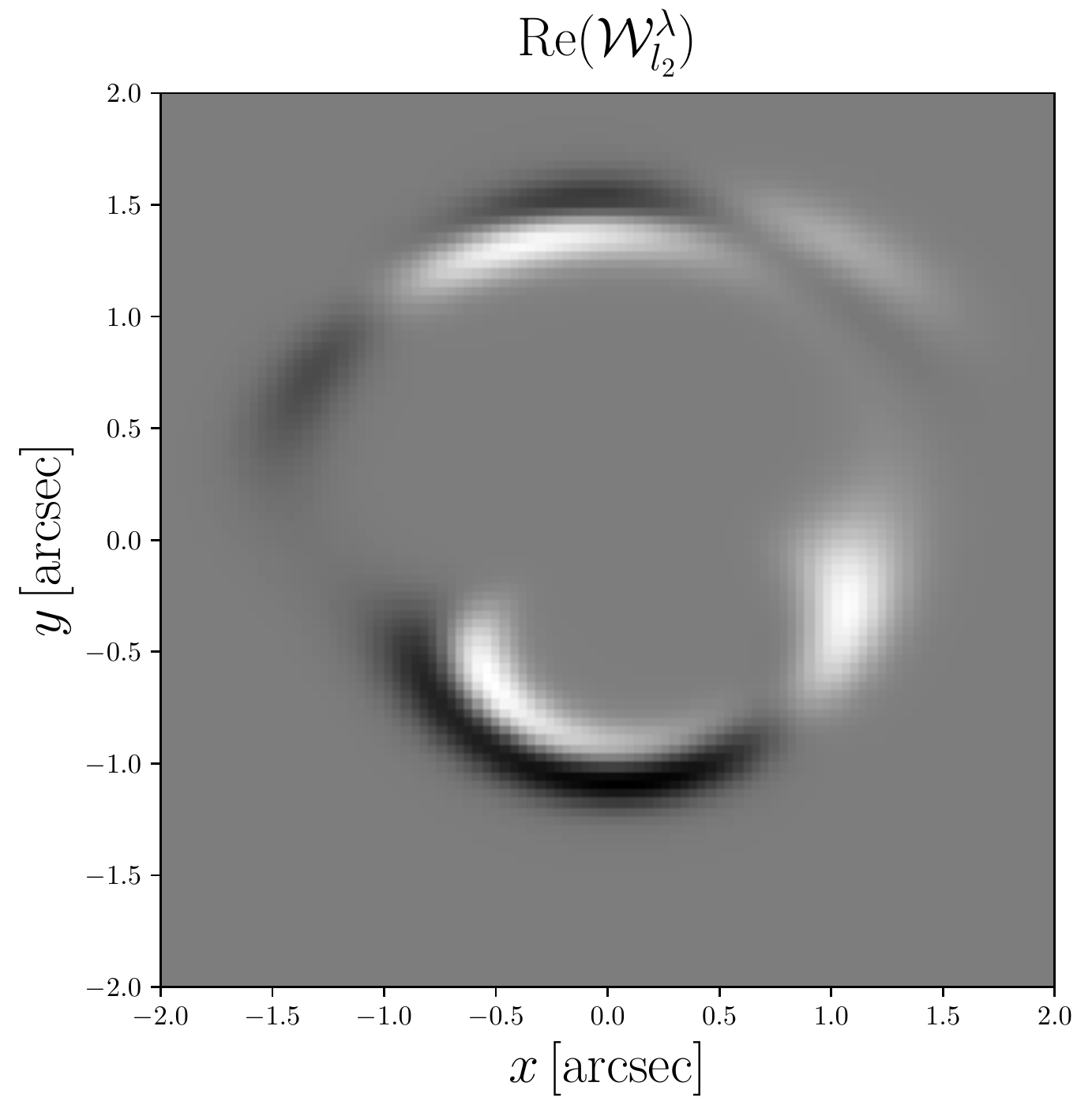}
\includegraphics[width=0.245\textwidth]{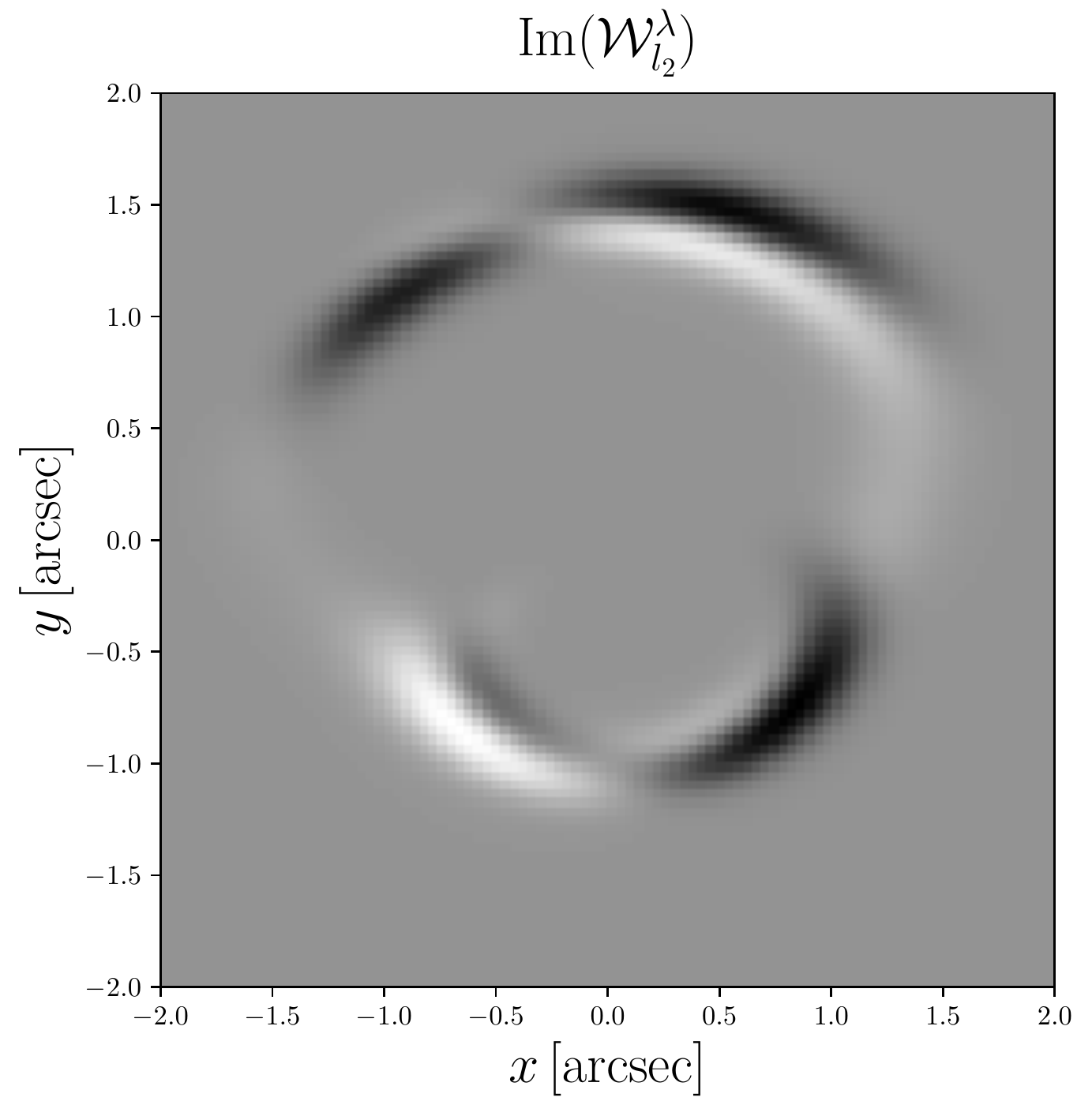}\\
\includegraphics[width=0.245\textwidth]{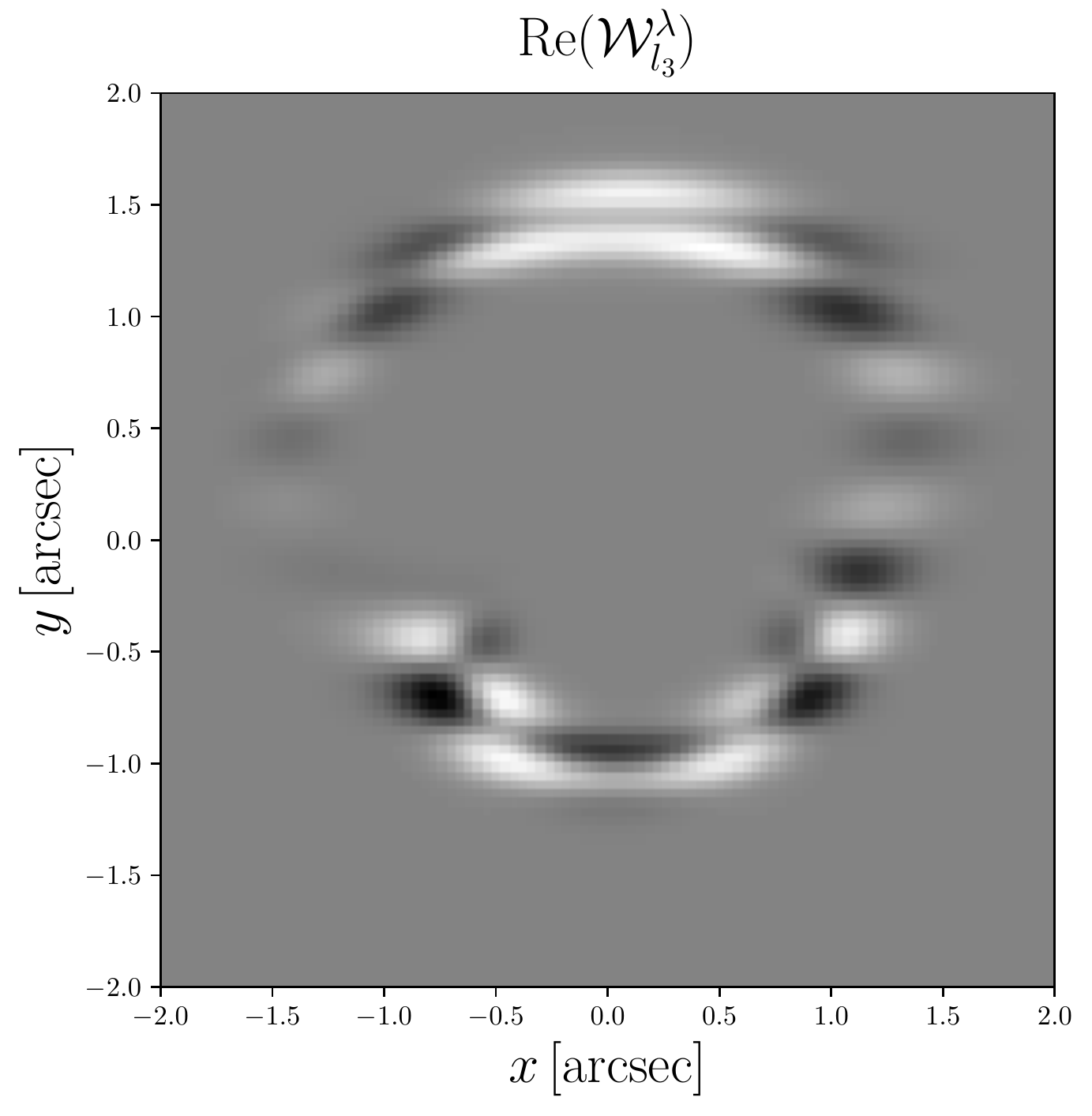}
\includegraphics[width=0.245\textwidth]{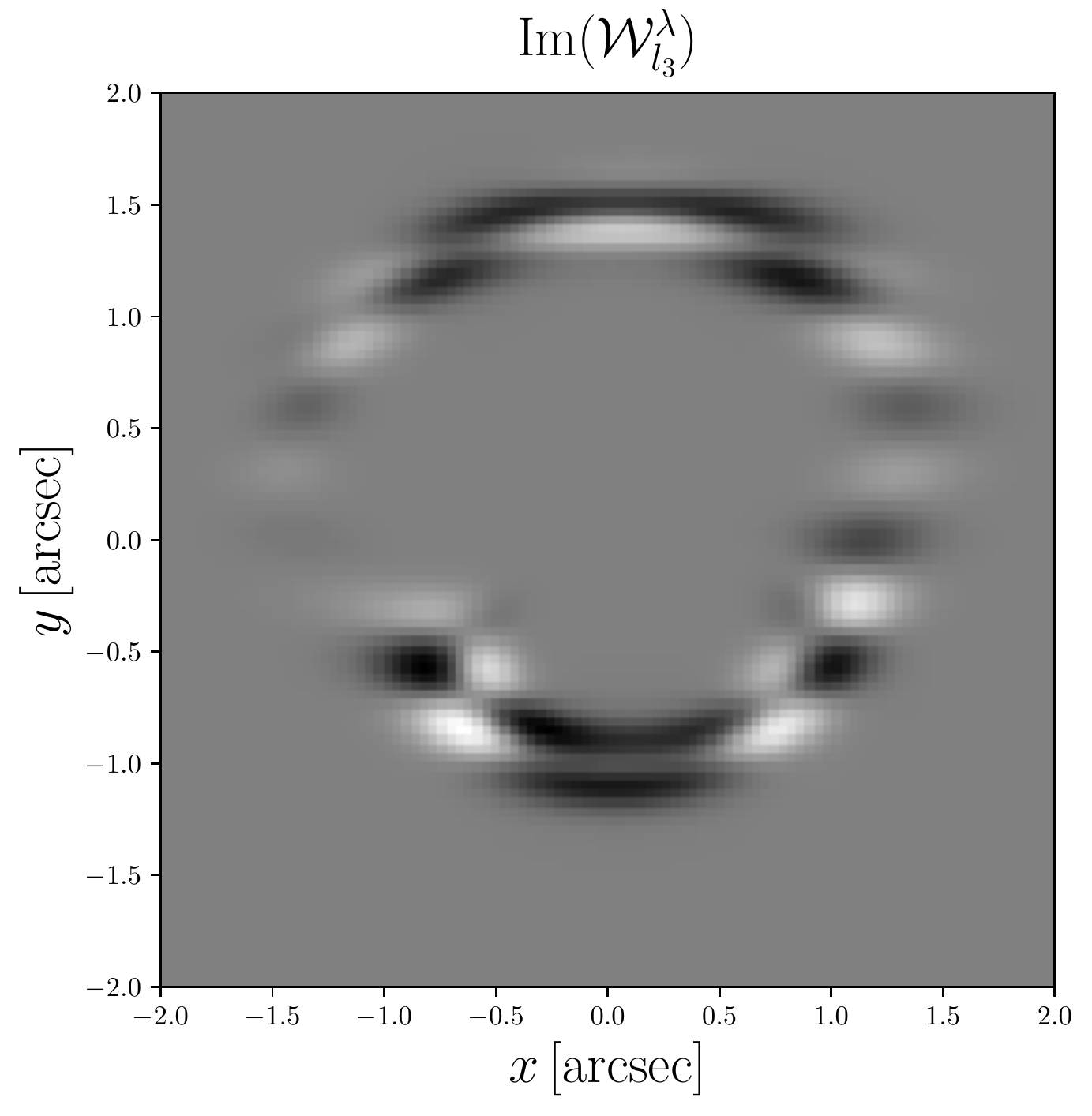}
\includegraphics[width=0.245\textwidth]{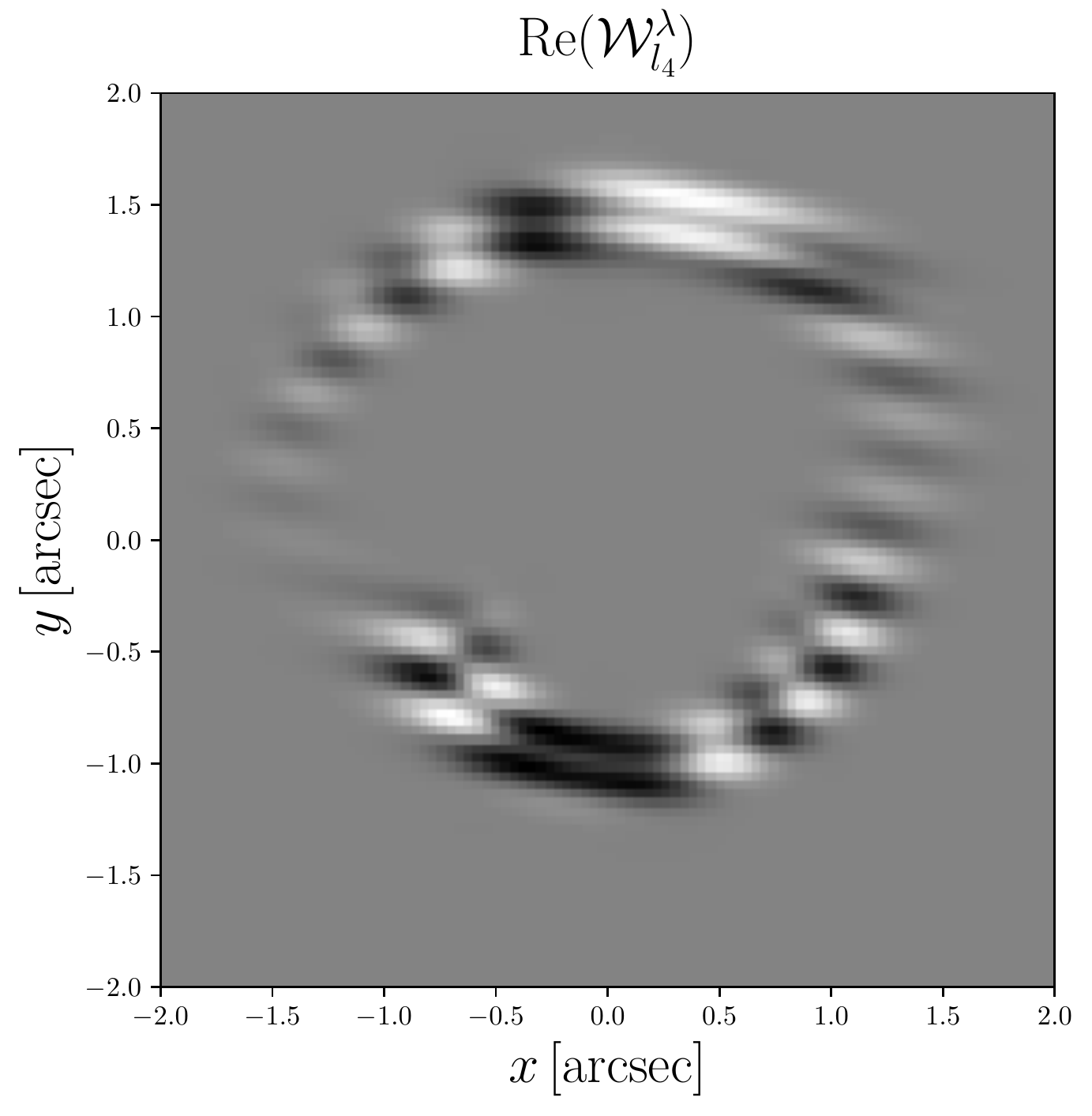}
\includegraphics[width=0.245\textwidth]{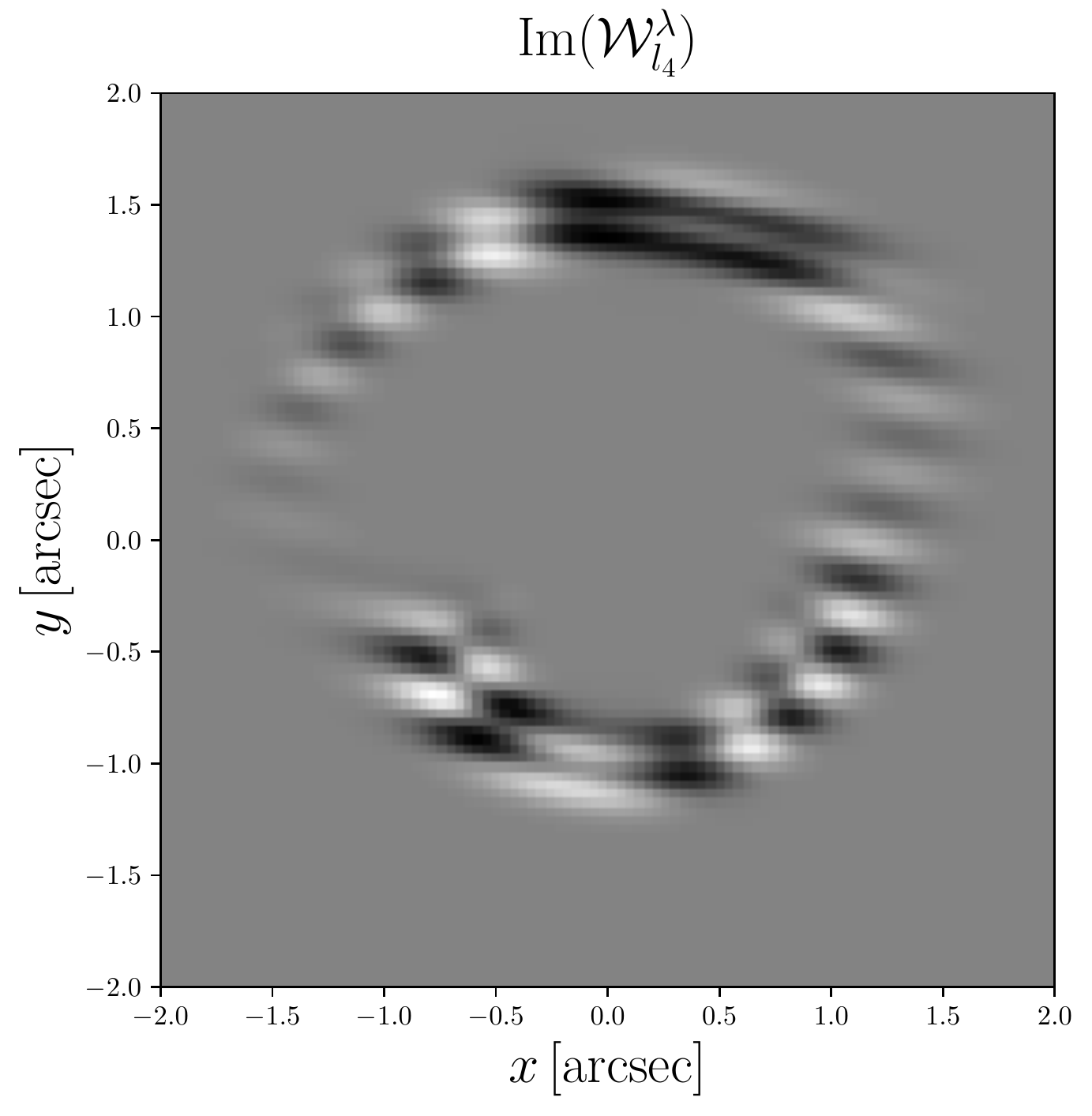}
\caption{Real and imaginary parts of the $\mathcal{W}_l^\lambda$ kernel for four different Fourier modes. The kernels are ordered from long wavelength modes (top left corner) to short wavelength modes (lower right corner) with $\{k_{l_1},k_{l_2},k_{l_3},k_{l_4}\} = \{1.57,2.22, 11.0,19.1\}$ arscec$^{-1}$. The source and lens model used here are the same as in Fig.~\ref{fig:residuals_example}.}\label{fig:kernel_example}
\end{figure*}

In practice, the size of the source and the finite width of the PSF put an upper bound on the largest k mode that can be measured from a given lensed image, and it is therefore sometime unnecessary to consider the full range of Fourier modes given in Eq.~\eqref{eq:full_Fourier_mode}. We illustrate examples of the $\mathcal{W}_{l}^\lambda$ kernel in Fig.~\ref{fig:kernel_example} for four different choices of Fourier modes. The source and lens model used to generate these are the same as in Fig.~\ref{fig:residuals_example}. Not surprisingly, we see that the various kernels pick out different features of the image residuals, with the lower wave numbers picking structures stretching across the image, and the higher wave numbers selecting small-scale brightness fluctuations along the lensed arc.

The Fourier transform of the mode functions given in Eq.~\eqref{eq:mode_Fourier} is
\be
\tilde{\varphi}_l(\kk) =  \frac{A_{\rm img}}{k_l}j_0(\frac{1}{2}(k_xR-2\pi l_x))j_0(\frac{1}{2}(k_yR-2\pi l_y)),
\ee
where $j_0(x)$ is the spherical Bessel function of order zero, and where we have written $\kk = (k_x,k_y)$. 
Since we are focusing our attention here on the monopole of the substructure convergence power spectrum [Eq.~\eqref{eq:monopople_power_spectrum}], it is useful to define the following window function
\be
\Phi_{ll'}(k) = \frac{k_l k_{l'}}{(2\pi)^2A_{\rm img}}\int_0^{2\pi} d\theta_k\, \tilde{\varphi}_l(\kk) \tilde{\varphi}^*_{l'}(\kk),
\ee
which has the convenient normalization
\be
\int_0^\infty dk\, k\,  \Phi_{ll}(k) =1.
\ee
This window function encodes the fact that we have access to only a small region of the sky, and thus can measure a limited number of Fourier modes. Now, compared to large-scale structure surveys, the window function plays a subdued role here since the gradient of the source appearing in $\mathcal{W}_l^\lambda$ kernel (Eq.~\eqref{eq:W_kernel_Fourier}) already limits the sensitivity of the data to Fourier modes with wavelength on the order of the lens' Einstein radius or smaller, independently of the size of $A_{\rm img}$. Furthermore, for the modes given in Eq.~\eqref{eq:ortho_modes}, $\Phi_{ll}(k)$ is strongly peaked at $k=k_l$ while $\Phi_{ll'}(k)$ is oscillatory for $l\neq l'$, hence leading to strong cancellation\footnote{For instance, we find that $\int_0^\infty dk\, k\, \Phi_{l,l+1} \sim 10^{-3}$.} of the off-diagonal elements. We can thus approximate the window function as
\be
\Phi_{ll'}(k)\approx \frac{\de(k-k_l)}{k_l} \de_{ll'},
\ee
which yields a ${\bf C}_{\rm sub}$ covariance matrix of the form
\begin{align}\label{eq:final_cov_fourier}
({\bf C}_{\rm sub})_{ll'} &= \frac{4}{A_{\rm img} k_l k_{l'}} \int dk\, k\, P_{\rm sub}^{(0)}(k) \Phi_{ll'}(k)\en
&\approx \frac{4 P_{\rm sub}^{(0)}(k_l)}{ A_{\rm img} k_l^2} \de_{ll'}.
\end{align}
We note that for a constant $P_{\rm sub}^{(0)}(k)$ (as in the case of a population of point masses), Eq.~\eqref{eq:final_cov_fourier} becomes exact for the diagonal elements of ${\bf C}_{\rm sub}$. In general, as long as the value of the convergence power spectrum does not vary rapidly over the width of the window function, we find Eq.~\eqref{eq:final_cov_fourier} to be an excellent approximation. For the remainder of this paper, we adopt for simplicity the approximation given in Eq.~\eqref{eq:final_cov_fourier} for the substructure covariance matrix, but note that it is straightforward to generalize our calculation to also include off-diagonal elements of ${\bf C}_{\rm sub}$. 

\subsection{Numerical implementation}\label{sec:num_impl}
To implement and test the likelihood presented in Secs.~\ref{sec:single_obs} and \ref{sec:multi_obs} in the Fourier basis, we have developed the software package \texttt{PkLens}\footnote{\texttt{PkLens} will be made publicly available upon publication of this manuscript.}. Written in pure \texttt{Python} 3, \texttt{PkLens} uses just-in-time compilation and automatic parallelization from the \texttt{numba} \cite{numba} package to accelerate key parts of the computation. 

The reality condition $\mathcal{W}^\lambda_{-l} = \mathcal{W}_l^{\lambda*}$ implies that the ${\bf G}$ matrix defined in Eq.~\eqref{eq:G_matrix} can be written in the following block structure
\be
{\bf G} = 
\begin{pmatrix}
{\bf X} & {\bf Y}\\
{\bf Y}^* & {\bf X}^*
\end{pmatrix},
\ee
where ${\bf X} = {\bf X}^\dagger$ is an Hermitian block and ${\bf Y} = {\bf Y}^{\rm T}$ is a symmetric block, both of size $N_{\rm modes, ind}\times N_{\rm modes, ind}$. We thus need to compute only half the elements of ${\bf X}$ and half that of ${\bf Y}$ (for a total of $N_{\rm modes, ind}$ entries) to fully characterize the matrix ${\bf G}$. This structure of the {\bf G} matrix allows us to use blockwise inversion in order to compute the matrix ${\bf D}^{-1}$ appearing in the likelihood given in Eq.~\eqref{eq:like_one_image}, hence significantly speeding up the linear algebra. Similarly, only half of the ${\bf g}_l$ vector entries need to be computed since ${\bf g}_{-l} = {\bf g}_l^*$.

\section{Fisher analysis}\label{sec:Fisher}
To develop some intuition about the sensitivity of different lens configurations and observational scenarios to the substructure convergence power spectrum, it is instructive to first carry out a simple Fisher analysis of the likelihood given in Eq.~\eqref{eq:like_multi_image}. We adopt a binned substructure convergence power spectrum as our fitting model, and the relevant parameters here are thus the logarithm of the amplitude of $P_{\rm sub}(k) $ within each bin, $\qsub = \{\ln{P_{{\rm sub},i}} \}_{i=1,\ldots,N_{\rm bins}}$. For the analysis shown in this section, we divide the range of scales probed by a given lensed image into four wave number bins that are evenly spaced in $\log_{10}(k)$. In the following, for each filter centered at wavelength $\lambda$, we assume that we have $N^{\lambda}_{\rm obs}$ observations of the same lens.
\subsection{Fisher matrix and sensitivity function}\label{sec:Fish_mat_and_sens}
The Fisher matrix for the binned log amplitude of the power spectrum takes the form
\begin{align}\label{eq:Fisher_mat}
F_{ij}  & \equiv -  \Big\langle \frac{\pa^2 \ln \mathcal{L}}{\pa\ln P_{{\rm sub},i} \pa \ln P_{{\rm sub},j}}\Big\rangle\en
&=\frac{P_{{\rm sub},i}P_{{\rm sub},j}}{2} {\rm Tr} \left[ {\bf \Gamma} \frac{\pa {\bf C}_{\rm sub}}{\pa P_{{\rm sub},i}}{\bf \Gamma}  \frac{\pa {\bf C}_{\rm sub}}{\pa P_{{\rm sub},j}}\right],
\end{align}
where
\be\label{eq:gamma_for_extended}
{\bf \Gamma} \equiv ({\bf G}^{-1} + \Cmat_\sub )^{-1} = {\bf G} -{\bf G}{\bf D}^{-1}{\bf G}.
\ee
To understand how the Fisher matrix scales with the observational parameters, it is instructive to consider a simple example where we neglect the off-diagonal entries of the ${\bf G}$ matrix. In this case, the diagonal entries of the Fisher matrix admit the form
\be\label{eq:Fisher_mat_approx}
F_{ii}  = \frac{1}{2} \sum_{l\in i} \frac{(\mathcal{S}_lP_{{\rm sub},i})^2}{\left(1+\mathcal{S}_l P_{{\rm sub},i}\right)^2},
\ee
where the sum runs over all Fourier modes whose magnitude falls within the range of the $i$th bin. We have assumed here that the noise for each observation is Poissonian with ${\bf C}_{{N_\lambda},ij}=\de_{ij}\sigma_1O_\lambda(\xx_i)$ [where $O_\lambda(\xx_i)$ is given in Eq.~\eqref{eq:brightness_to_count} and $\sigma_1=1$ for pure Poisson noise]. In Eq.~\eqref{eq:Fisher_mat_approx}, we have introduced the sensitivity $\mathcal{S}_l$ of a given gravitational lens observation to the $l$th mode of the substructure convergence field. It is defined as the product of a mode-independent prefactor $Q^\lambda_{\rm obs}$ that depends on the depth and quality of the observation and of a mode-dependent function $\Lambda^\lambda_l$ that only depends on the spatial structure of the macro lens, source, and PSF,
\be
\mathcal{S}_l = \sum_\lambda Q_{\rm obs}^\lambda \times \Lambda^\lambda_l,
\ee
where
\be\label{eq:Q_fac}
 Q_{\rm obs}^\lambda \equiv  N^\lambda_{\rm obs} \frac{ T_{\rm exp} \mathcal{F}_\lambda}{\sigma_1  \mathcal{S}_{\rm inv}^{(\lambda)}},
 \ee
and
\be\label{eq:l_sens}
\Lambda_l^\lambda \equiv \frac{4}{k_l^2 N_{\rm pix}}\sum_m \frac{|(W_\lambda * \nabla\varphi_l\cdot\nabla_{\uu}\hat{S}_\lambda)(\xx_m)|^2}{W_\lambda * \hat{S}_\lambda (\xx_m) },
\ee
where the sum runs over all the pixels in the image, and the ``$*$'' symbol stands for the convolution operation. Note that we have written the source surface brightness as $S_\lambda(\uu) = \mathcal{F}_\lambda \hat{S}_\lambda(\uu)$, where $ \mathcal{F}_\lambda$ is the total source flux within the bandpass of the filter and $\int d^2\uu\, \hat{S}_\lambda(\uu)=1$. For simplicity, we have omitted the foreground contribution when writing Eq.~\eqref{eq:l_sens}. 

Since $\Lambda_l^\lambda$ describes the intrinsic sensitivity of a given lens configuration to the $l$th mode of the substructure density field, it is a useful figure of merit to rapidly assess whether a given lens can provide competitive constraints on the substructure convergence power spectrum. The dimensionless prefactor $Q^\lambda_{\rm obs}$ simply captures how the sensitivity $S_l$ scales with exposure time, number of observations, source flux, noise level, and detector sensitivity. Not surprisingly, the sensitivity is improved for a longer total exposure, a brighter source, a lower noise level, and by lowering the value of $\mathcal{S}_{\rm inv}^{(\lambda)}$ (which could be done by using a larger telescope and/or a more sensitive camera). 

For very large value of the sensitivity, $P_{{\rm sub},i}S_l\gg 1$, Eq.~\eqref{eq:Fisher_mat_approx} implies that the measurement uncertainty $\de \ln P_{{\rm sub},i} = \sqrt{(F_{ii}^{-1})}$ on the amplitude of the binned power spectrum becomes sample variance dominated with
\be
\de \ln P_{{\rm sub},i} \simeq \frac{\sqrt{2} }{\sqrt{N_i}},\qquad (P_{{\rm sub},i}S_l\gg 1)
\ee
where $N_i$ is the number of modes within the $i$th bin. This is a familiar result that arises for instance in the study of cosmological large-scale structure. On the other hand, for low sensitivity $P_{{\rm sub},i}S_l\ll 1$, we can Taylor expand Eq.~\eqref{eq:Fisher_mat_approx} to obtain
\be\label{eq:noise_dom}
\de \ln P_{{\rm sub},i} \simeq  \frac{\sqrt{2} }{P_{{\rm sub},i} \sqrt{\sum_{l\in i} \mathcal{S}_l^2}},\qquad (P_{{\rm sub},i}S_l\ll 1)
\ee
from which we obtain $\de P_{{\rm sub},i} \propto 1/(N_{\rm obs} T_{\rm exp})$. 

\begin{figure}[t!]
\centering
\includegraphics[width=0.49\textwidth]{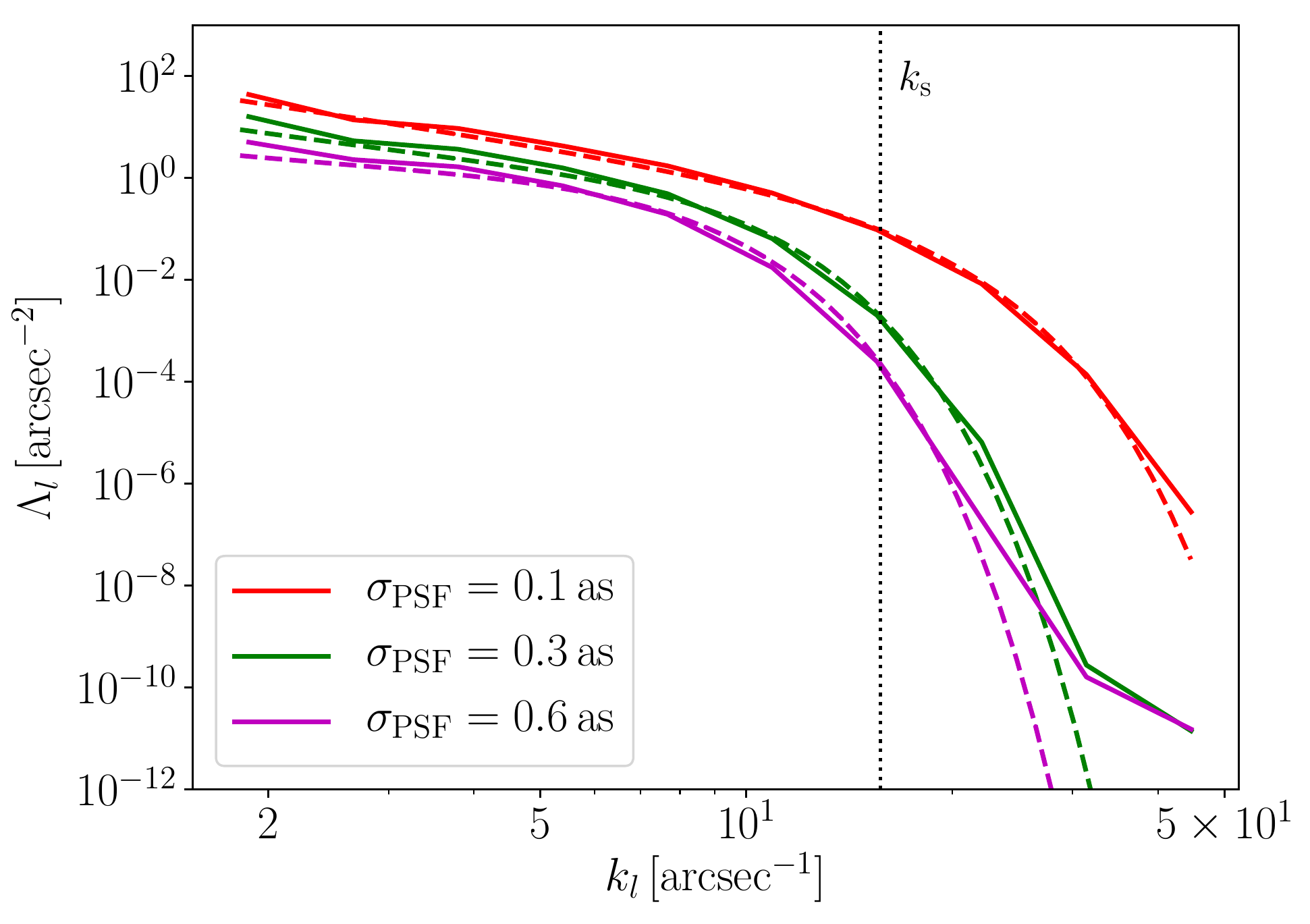}
\caption{Sensitivity function $\Lambda_l$ for a Gaussian source of size $\sigma_{\rm s}=0.2$ arcsec lensed by a singular isothermal sphere lens model with Einstein radius $b_{\rm ein}=1.2$ arcsec. The solid lines show the results of exact numerical computations for three different sizes of Gaussian PSF, while the dashed lines illustrate the approximate expression given in Eq.~\eqref{eq:sensitivity_approx}. The vertical dotted line shows the approximate wave number corresponding to the size of the source $k_{\rm s} = \pi/\sigma_{\rm s}$. Note that higher values of $\Lambda_l$ means that the mock lens has greater sensitivity to the substructure convergence power spectrum for wave number $k_l$. }\label{fig:sensitivity}
\end{figure}

To gain some intuition about the structure of the sensitivity function as a function of wave number $k_l$, let us consider a Gaussian source of width $\sigma_{\rm s}$ lensed by a singular isothermal sphere lens model into an Einstein ring with radius $b_{\rm ein}$. Also, let us consider a series of observations with a single filter and a Gaussian PSF of size $\sigma_{\rm PSF}$ (we shall drop the wavelength index $\lambda$ in the following). For this simple system, Eq.~\eqref{eq:l_sens} admits as leading behavior in the limit that $\sigma_{\rm PSF}, \sigma_{\rm s} < b_{\rm ein}$
\be\label{eq:sensitivity_approx}
\Lambda_l \propto b_{\rm ein} \frac{\sigma_{\rm PSF}^2 +\sigma_{\rm s}^2 + k_l^2\sigma_{\rm PSF}^4}{k_l^2(\sigma_{\rm PSF}^2 +\sigma_{\rm s}^2 )^{5/2}A_{\rm img}}e^{-\frac{1}{2}k_l^2 \sigma_{\rm eff}^2 },
\ee
where 
\be
\sigma_{\rm eff}  = \sqrt{2}\frac{\sigma_{\rm PSF}\sigma_{\rm s} }{\sqrt{\sigma_{\rm PSF}^2+\sigma_{\rm s}^2 }}.
\ee
We compare this analytical estimate of $\Lambda_l$ to exact numerical computations for three different choices of PSF size in Fig.~\ref{fig:sensitivity}. There, the solid lines show the numerical results, while the dashed lines display the approximate expression given in Eq.~\eqref{eq:sensitivity_approx}. The dotted vertical line shows the wave number $k_{\rm s}$ corresponding to the size of the Gaussian source. 

On scales larger than $\sigma_{\rm eff}$ ($k_l \lesssim \pi/\sigma_{\rm eff}$), the sensitivity decays as $\Lambda_l\propto k_l^{-2}$. This scaling is simply the result of the Poisson equation linking the deflection field probed by the lensing observations to the substructure convergence whose power spectrum we are trying to measure, that is, $|\tilde{\alpha}_{\rm sub}(k)|^2\sim 4 |\tilde{\kappa}_{\rm sub}(k)|^2/k^2$.  Since the substructure deflection field couples to the gradient of the source's surface brightness profile convolved with the PSF (see Eq.~\eqref{eq:delta_O}), the sensitivity becomes strongly suppressed for Fourier modes probing scales smaller than either the size of the source or the PSF, whichever is smallest. This can be seen in Fig.~\ref{fig:sensitivity} where the examples with $\sigma_{\rm PSF} > \sigma_{\rm s}$ (purple and green lines) have rapidly decaying sensitivity for $k _l\gtrsim k_{\rm s}$, whereas the example with  $\sigma_{\rm PSF} < \sigma_{\rm s}$ (red lines) roughly retains the $k_l^{-2}$ scaling until wave numbers corresponding to the PSF size. While this sensitivity cutoff is exponential for our example with a Gaussian source and PSF,  we generally expect it to be milder for more realistic choices of source and PSF models.

In addition to the sensitivity cutoff for $k_l\gtrsim \pi/\sigma_{\rm eff}$, Eq.~\eqref{eq:sensitivity_approx} admits the general scaling $\Lambda_l\propto b_{\rm ein}/\sigma_{\rm PSF}^3$ for PSF size larger than the source. The leading factor of $b_{\rm ein}/\sigma_{\rm PSF}$ essentially counts the number of independent sections of the Einstein ring that are available for the analysis. This is quite intuitive: at fixed PSF and image size, a larger Einstein ring contains more information about the substructure density field than a smaller one. The further $1/\sigma_{\rm PSF}^2$ factor stems from the reduced signal to noise per pixel as the PSF size increases. Indeed, for larger PSFs, the light from a given part of the source spreads to a larger area of the focal plane, resulting in a smaller photon count in each pixel and thus to a reduced sensitivity. Of course, this scaling breaks down once the PSF size becomes smaller than the typical size of the source, as can be seen in Eq.~\eqref{eq:sensitivity_approx}.  As a general rule of thumb, high-resolution images will always provide better constraining power on the substructure power spectrum than low-resolution images, and lenses with larger Einstein radii (or more complete Einstein rings) will generally have display greater sensitivity to the effect of substructure. 

\subsection{Simple Fisher forecast}\label{sec:fisher_simple}
\begin{figure}[t!]
\centering
\includegraphics[width=0.49\textwidth]{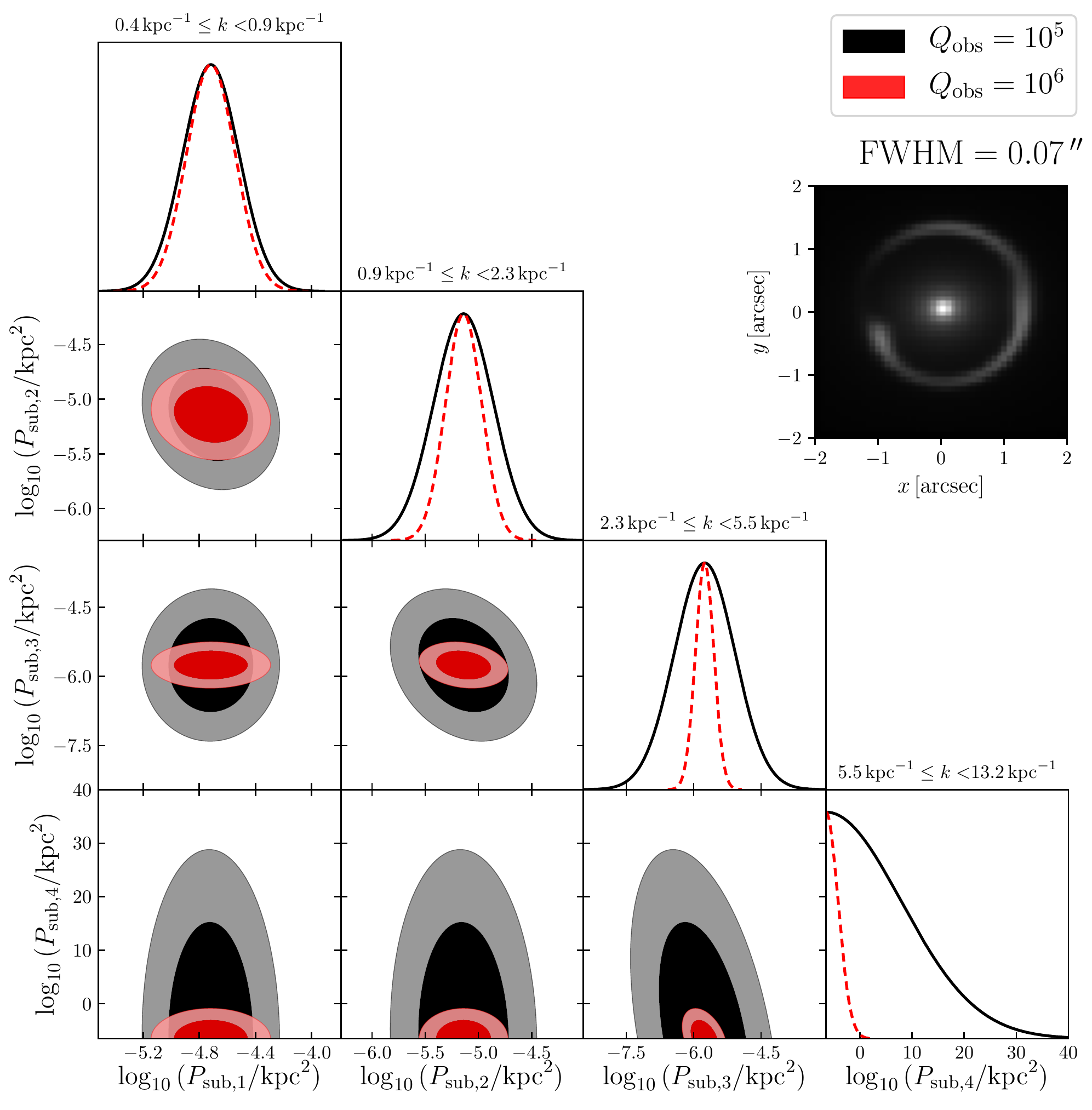}\\
\includegraphics[width=0.49\textwidth]{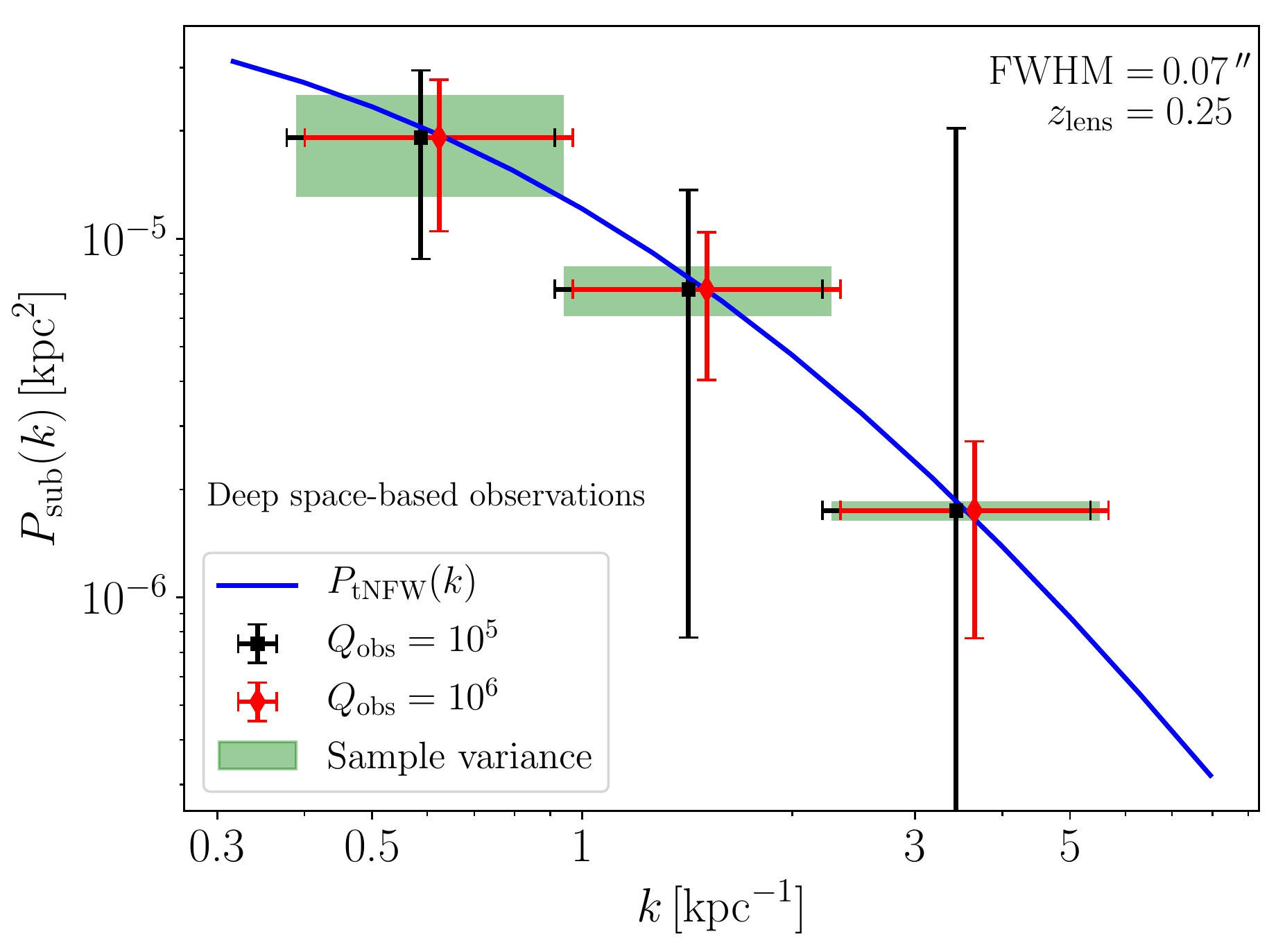}
\caption{Fisher forecast for the substructure convergence power spectrum in four logarithmic wave number bins. The top panel shows the different projection of the inverse Fisher matrix for an HST-like observations with a PSF FWHM of $0.07$ arcsec for two values of the quality factor $Q_{\rm obs}$ [defined in Eq.~\eqref{eq:Q_fac}]. The darkly and lightly shaded areas show the $68\%$ and $95\%$ confidence regions, respectively. The inset shows the lens configuration used. The lower panel shows the resulting Fisher error bars on $P_{\rm sub}(k)$ for the three lowest wave number bins. The blue solid line shows the fiducial power spectrum model used in the forecast, which corresponds to the truncated NFW model shown in Fig.~\ref{fig:Psub_example}. The error bars show the $1$-$\sigma$ regions, while the green rectangles display the sample variance contribution within each bin. For clarity, the wavenumber bin center for each observational scenario shown has been offset by $6\%$, with the green rectangle showing the true wave number bin used in the analysis. }\label{fig:pk_Fisher_space}
\end{figure}
We now use the Fisher matrix from Eq.~\eqref{eq:Fisher_mat} to quantitatively estimate the error on the binned substructure convergence power spectrum for two different observational scenarios. As usual, these Fisher forecasts should be interpreted with caution, especially since we neglect here possible covariances between the effect of substructures and changes to the macro lens and source parameters. These possible degeneracies will be explored in the next section once we perform complete Markov Chain Monte Carlo analyses of lensed images.The results shown here should thus be taken as illustrative of the best-case sensitivity to the substructure power spectrum that could be achieved within the observational scenarios we consider below. In the following, we consider images on a $50\times50$ pixel grid, and adopt the truncated NFW substructure power spectrum shown in Fig.~\ref{fig:Psub_example} as our fiducial model for the Fisher forecast.

\begin{figure}[t]
\centering
\includegraphics[width=0.49\textwidth]{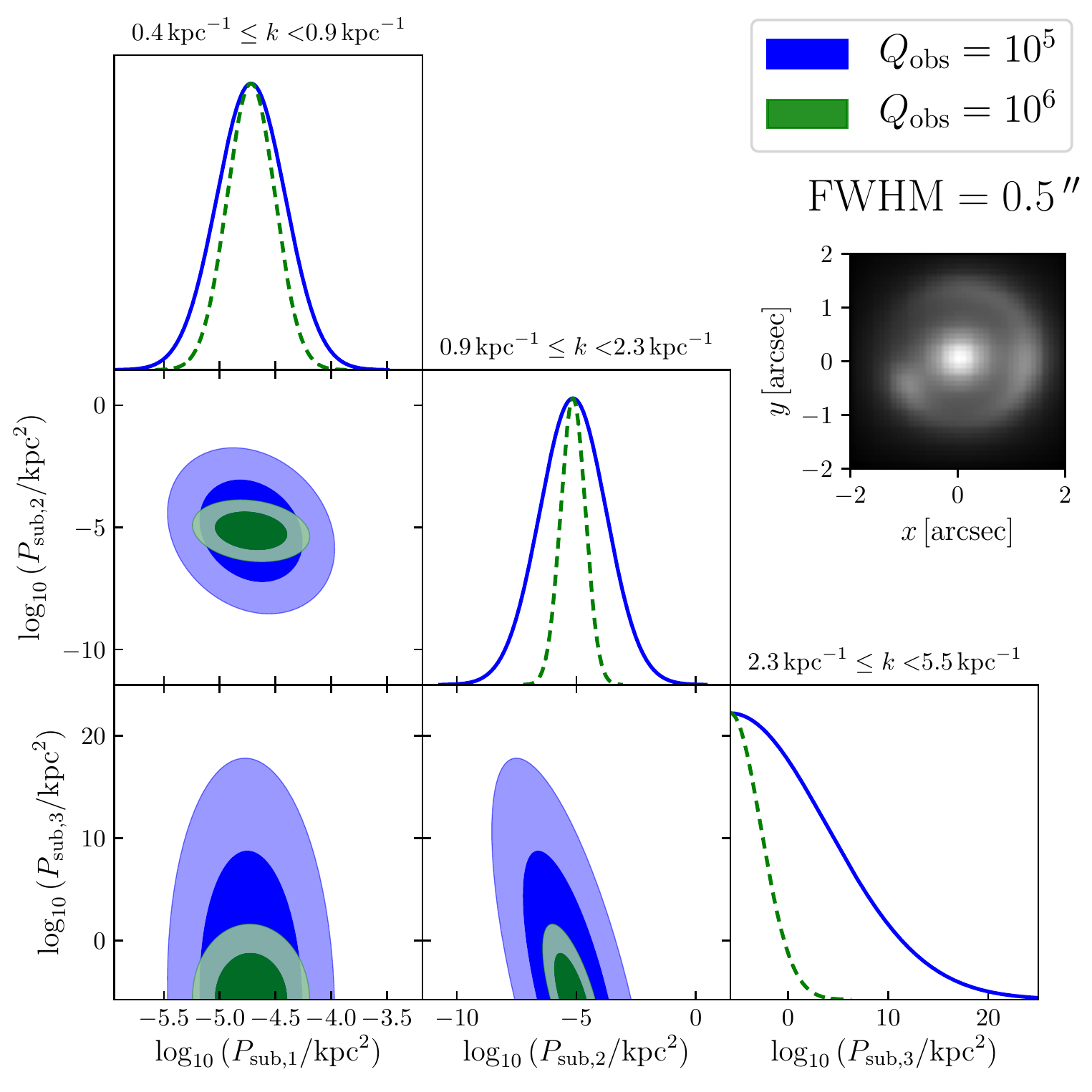}
\caption{Same as the upper panel of Fig.~\ref{fig:pk_Fisher_space} but for seeing-limited ground-based observations with FWHM $=0.5$ arcsec (modeled here as a Moffat profile). Here, we only display the results for the wave number range $0.4{\rm kpc}^{-1}\leq k\leq 5.5 {\rm kpc}^{-1}$ since there is very little sensitivity to higher wave numbers given the poor image resolution.  Note the larger range power spectrum value shown on the axes as compared to Fig.~\ref{fig:pk_Fisher_space} due to the much lower sensitivity here. The inset shows the lens configuration used. }\label{fig:pk_Fisher_ground}
\end{figure}

Figure~\ref{fig:pk_Fisher_space} shows the Fisher forecast for a high-resolution observations with a PSF full width at half maximum (FWHM) of  $0.07$ arcsec, similar to what is achievable with the Hubble Space Telescope (HST) for optical wavelengths. We display Fisher estimates of the error bars for two choices of the observation quality factor $Q_{\rm obs}$. As a concrete example, $Q_{\rm obs}=10^5$ could be achieved by combining 10 observations of a source with total flux $\mathcal{F}_\lambda = 10^{-18}$ erg/cm$^2$/s/\AA\,(approximately corresponding to an unlensed AB magnitude of 24), each observed through the FW555 filter for $T_{\rm exp}=2000$ sec with the UVIS detector on Wide-Field Camera 3 (WFC3) aboard HST.  

For $Q_{\rm obs}=10^6$, the substructure convergence power spectrum error bars approach the sample-variance limit within the lowest wave number bin, as shown in the lower panel of Fig.~\ref{fig:pk_Fisher_space}. For the second and third wave number bins, the error bars grow modestly according to Eq.~\eqref{eq:Fisher_mat_approx} for both image depths shown. There is then a significant decrease in sensitivity within the highest wave number bin due to the limits imposed by the source size and PSF (corresponding here to $k_{\rm s} \simeq 8\,{\rm kpc}^{-1}$ and $k_{\rm FWHM} \simeq 11\,{\rm kpc}^{-1}$, respectively), in accordance with our discussion above. For the wave number range $2.3\,{\rm kpc}^{-1}\leq k\leq 13.2\, {\rm kpc}^{-1}$, the deeper observations ($Q_{\rm obs}=10^6$) lead to a significant gain in sensitivity as compared to the shallower one ($Q_{\rm obs}=10^5$) due to these modes being in the noise-dominated regime described by Eq.~\eqref{eq:noise_dom}.  Within the highest wave number bin, only coarse upper limits appear possible in either case, which is why we do not illustrate them in the lower panel of Fig.~\ref{fig:pk_Fisher_space}.

 Figure~\ref{fig:pk_Fisher_ground} displays Fisher forecast for seeing-limited ground-based observations with PSF FWHM $=0.5$ arcsec for the same two values of $Q_{\rm obs}$. The error bars are significantly larger in this case, with the two highest wave number bins only yielding very coarse limits. While a measurement of the power spectrum amplitude for the lowest wave number bin appears possible, we caution that the degeneracies between the effects of substructure and the structure of the lens, source, and foregrounds are likely to be quite severe for the low-resolution ground-based data. This will likely degrade the constraining power of the low-resolution data compared to the simple Fisher forecast illustrated in Fig.~\ref{fig:pk_Fisher_ground}, unless a high-resolution image is also available to help break the various degeneracies. 

\begin{figure}[t]
\centering
\includegraphics[width=0.49\textwidth]{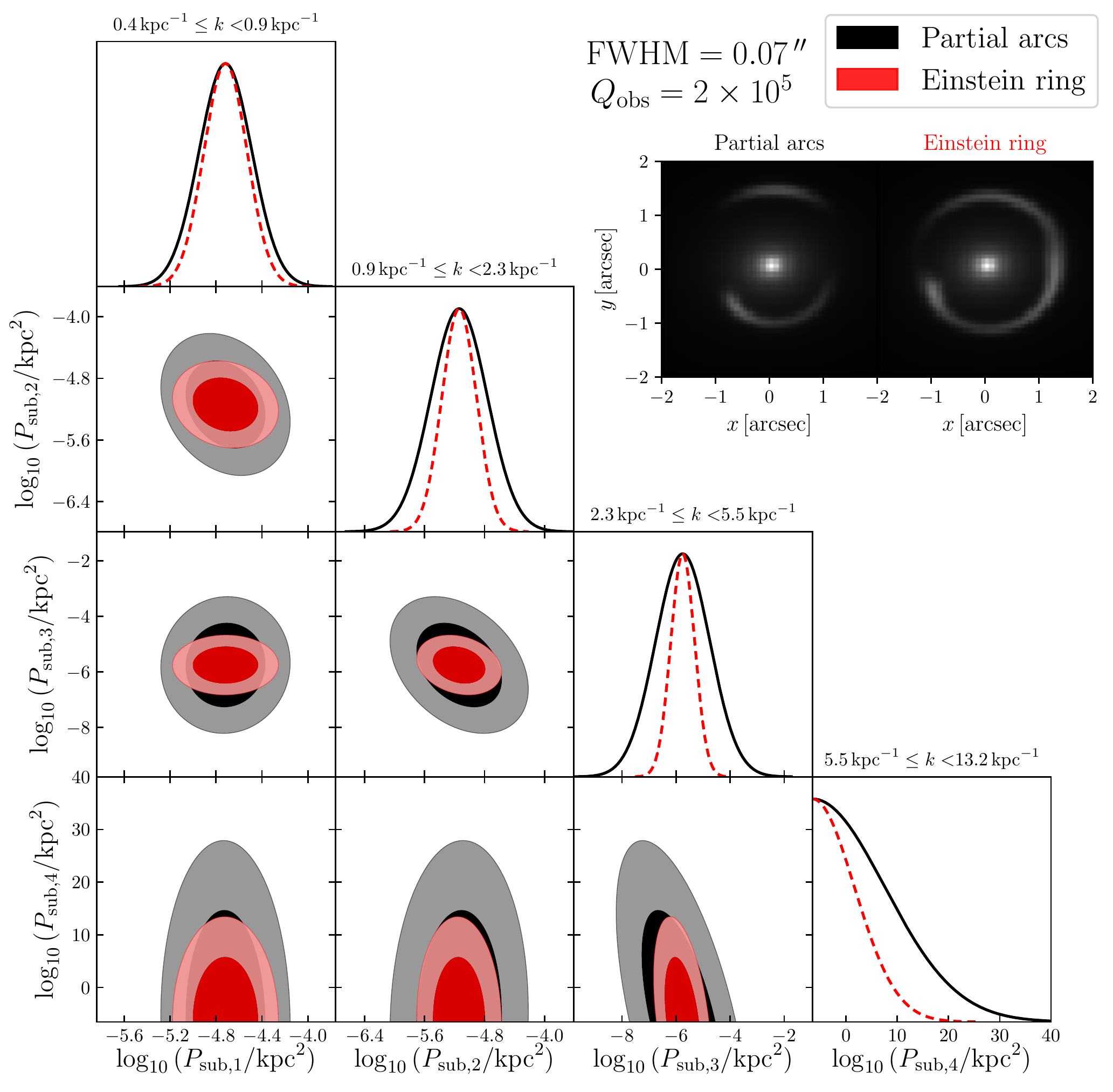}
\caption{Similar to the upper panel of Fig.~\ref{fig:pk_Fisher_space} but comparing the Fisher forecasts for two different lens configurations (shown in the inset). Here, the structure of the source, the macro lens, its environment and the foregrounds are all kept the same, and only the source position is changed. The nearly complete Einstein ring displays greater sensitivity to the substructure power spectrum than the partial arcs.}\label{fig:Fisher_compare_lens_config}
\end{figure}

It is important to realize that, at equal value of the observational quality factor, distinct lens configurations will display different sensitivity to the substructure power spectrum. An example of this is illustrated in Fig.~\ref{fig:Fisher_compare_lens_config} where we show the figure forecast for two different lens configurations: a set of partial arcs and a nearly complete Einstein ring (the latter being the configuration used in the forecast above). For these images, the structure of the source, macro lens, its environment and foregrounds are all kept fixed and only the position of the source is modified. Due to its greater coverage of the image plane that allows it to probe more substructure modes, the Einstein ring displays greater sensitivity to the substructure power spectrum is all wave number bins shown. As discussed in Sec.~\ref{sec:Fish_mat_and_sens}, at a given image resolution and depth, lensed images covering a larger area of the image plane will generally display greater sensitivity to the power spectrum. 

\section{Analysis of Simulated Images}\label{sec:numerical_results}

Having developed some intuition about the various factors affecting the sensitivity of a gravitational lens observation to the substructure convergence power spectrum, we now turn our attention to more realistic analyses of mock images. We explore in this section how the posterior distribution of the binned substructure power spectrum amplitudes is affected by degeneracies with the macro lens, source, foreground, and noise parameters for a few representative observational scenarios. Details on how we generate our mock observations of gravitational lenses are given in Appendix \ref{sec:img_sims}. Null tests to confirm the accuracy of our numerical implementation, and a study of possible degeneracies between the substructure power spectrum amplitudes and the noise parameters are presented in Appendix \ref{app:null_tests}. We present here the results of our Markov Chain Monte Carlo (MCMC) analyses exploring the complete degeneracies between lens, source, foreground, and noise parameters on the one hand, and substructure power spectrum on the other.

\subsection{Substructure power spectrum inference}
\begin{figure}[t!]
\centering
\includegraphics[width=0.49\textwidth]{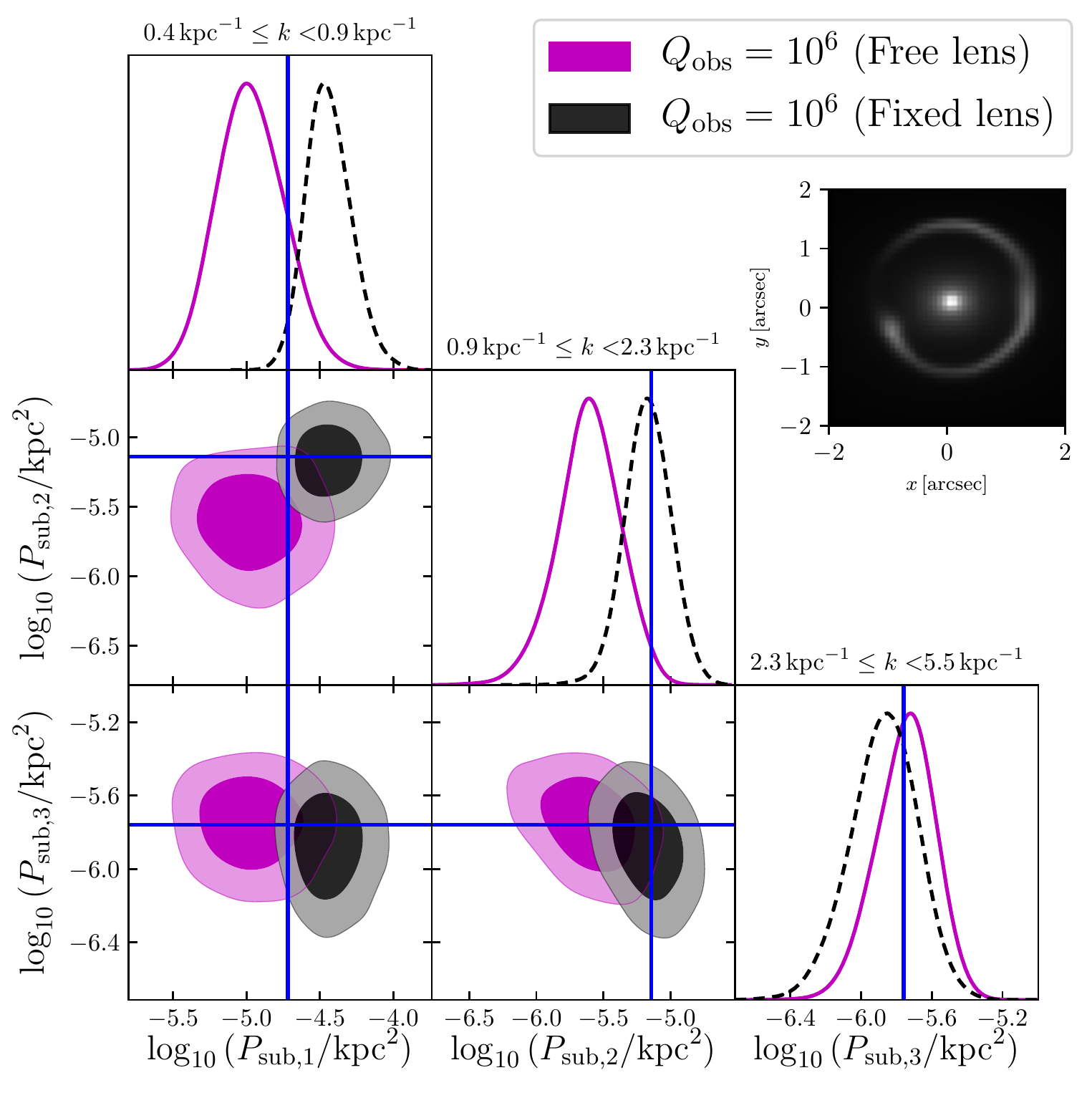}\\
\includegraphics[width=0.49\textwidth]{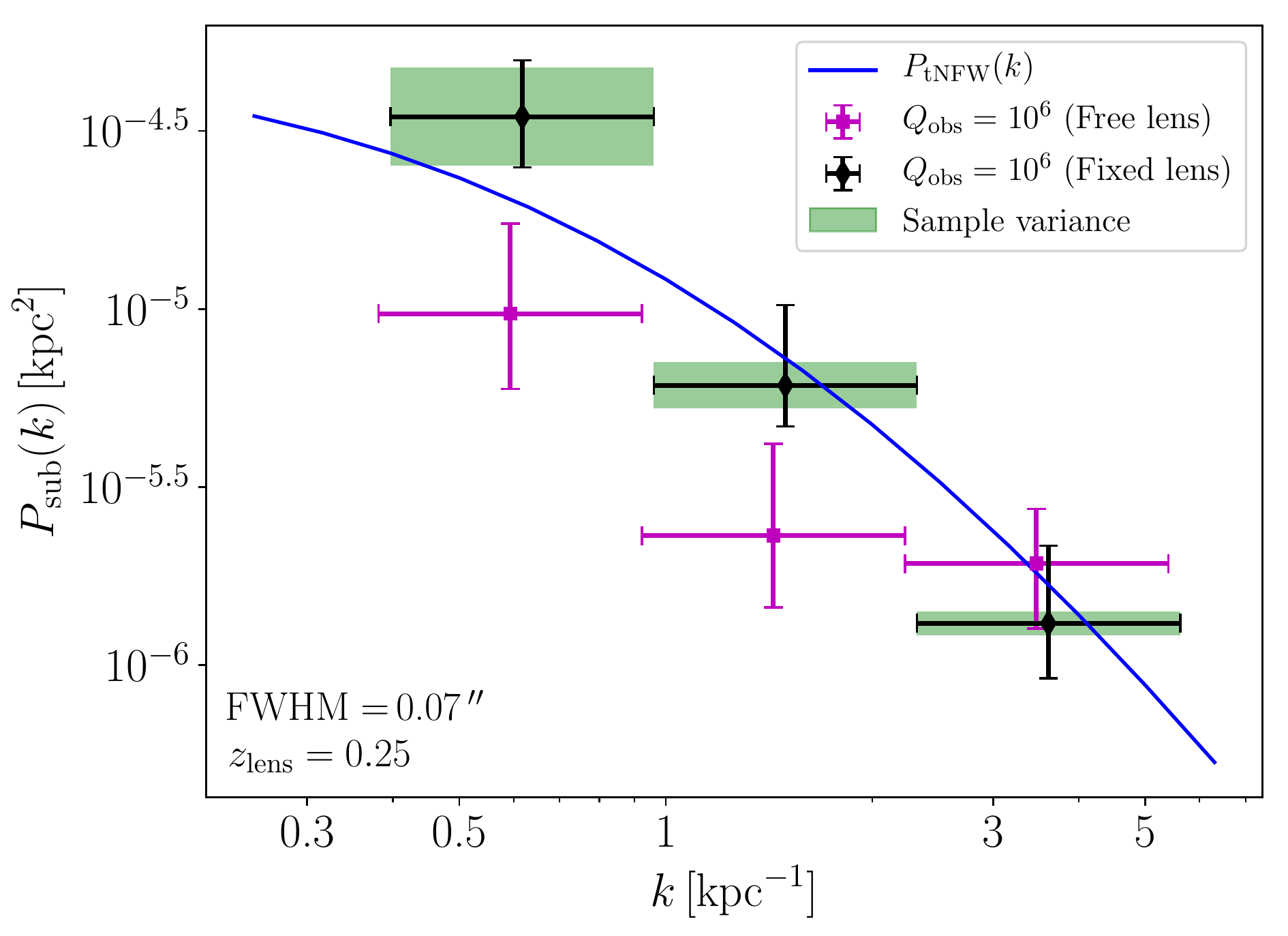}
\caption{Posterior summary for the binned substructure convergence power spectrum amplitudes for the case of a nearly complete Einstein ring observed with quality factor $Q_{\rm obs}=10^6$ and an HST-like PSF with a FWHM of $0.07$ arcsecond. The top panel shows the marginalized confidence intervals when the macro lens, source, foreground, and noise parameters are either held fixed (dashed black) or allowed to vary freely in the MCMC chains (magenta). The dark (light) contours show the $68\%$ ($95\%$) confidence intervals. The solid blue lines show the average values of the substructure power spectrum used to generate the mock image within each bin. The inset shows the mock image configuration used in the analysis. The lower panel shows the corresponding error bars in the substructure power spectrum space. The blue solid line shows the substructure power spectrum (corresponding to the truncated NFW model shown in Fig.~\ref{fig:Psub_example}) used to generate the mock image. The two-sided error bars show the $68\%$ highest posterior density intervals, and the green rectangles show the sample contribution to the error bars within each bin. }\label{fig:pk_ex_full_int}
\end{figure}

As in the case of our Fisher analysis, we adopt the logarithm of the binned substructure convergence power spectrum amplitudes as our fitting model. We divide the range of scales probed by a given lensed image into three wave number bins that are evenly spaced in $\log_{10}(k)$ within the range\footnote{Since we are considering lenses at redshift $z_{\rm lens}=0.25$ with a Planck 2015 Cosmology \cite{Ade:2015xua}, this interval corresponds to angular scales in the range $1.6\, {\rm arcsec}^{-1}\leq k\leq 22.2\,{\rm arcsec}^{-1}$.} $0.4\, {\rm kpc}^{-1}\leq k\leq 5.5\,{\rm kpc}^{-1}$. We note that the choice of binning presented in this work is arbitrary and driven only by convenience and simplicity. We leave a thorough study of the optimal binning strategy to future work, but note that it is very likely to depend on the specifics of each dataset. We adopt a broad scale free (log-uniform) prior on the amplitude within each bin, $\log_{10}(P_{{\rm sub},i}/{\rm arcsec}^2)\in [-10,-1]$. 

We use the affine invariant sampler \texttt{emcee} \cite{2013PASP..125..306F} to sample the likelihood given in Eq.~\eqref{eq:like_one_image}, allowing the macro lens, source, foreground, and noise parameters to vary freely within broad flat priors (see Appendix \ref{sec:img_sims} for details on the different components entering the mock lensed images). To determine how much the macromodel, source, noise, and foregrounds can somewhat reabsorb the effects of the substructure, we also run MCMC chains that keep the parameters of these latter components fixed to their true values. For each mock image and case considered, we use twice as many \texttt{emcee} walkers as the number of free parameters in the computation. As illustrative examples, we perform the substructure power spectrum inference for the two lens configurations used for the Fisher forecast shown in Fig.~\ref{fig:Fisher_compare_lens_config}: the nearly complete Einstein ring, and the two partial lensed arcs, each observed with a quality factor of $Q_{\rm obs}=10^6$ and an HST-like PSF with FWHM $= 0.07$ arcsec. 

The resulting posterior distribution summary for the case of the nearly complete Einstein ring is shown in Fig.~\ref{fig:pk_ex_full_int}. The top panel shows the marginalized confidence intervals of the substructure power spectrum for the three wave number bins. The black lines and contours denote the confidence intervals when the macrolens, source, foreground, and noise parameters are held fixed at their true values, while the magenta contours and lines show the intervals when all parameters all allowed to freely vary.  The solid blue lines denote the average values of the substructure convergence power spectrum within each wave number bin of the model used to generate the mock images. The lower panel shows the resulting $68\%$ highest posterior density (HPD) intervals for each bin, on top of the truncated NFW power spectrum used to general the mock data. The green rectangles illustrate the contribution from sample variance within each bin. 

We first note that even in the case of a fixed macro lens, source, and foregrounds (black contours), there are fluctuations of the highest posterior values with respect to the true input values, with the scatter being larger at low wave numbers. This is expected given that we are looking at a single realization of the substructure convergence field. The relatively small number of substructure modes within the lower wave number bins gives rises to a significant lens-to-lens variation on the inferred value of the power spectrum at larger scales. However, our test shows that averaging over a sufficient number of substructure realizations leads to an unbiased estimate of the substructure power spectrum amplitude within each bin. Of course, while this is easy to do for mock data, performing this average over substructure realizations with real data will require carefully combining the measurements from different lens systems, taking into account how their respective substructure population depends on the lens galaxy's properties. 
\begin{figure}[t!]
\centering
\includegraphics[width=0.49\textwidth]{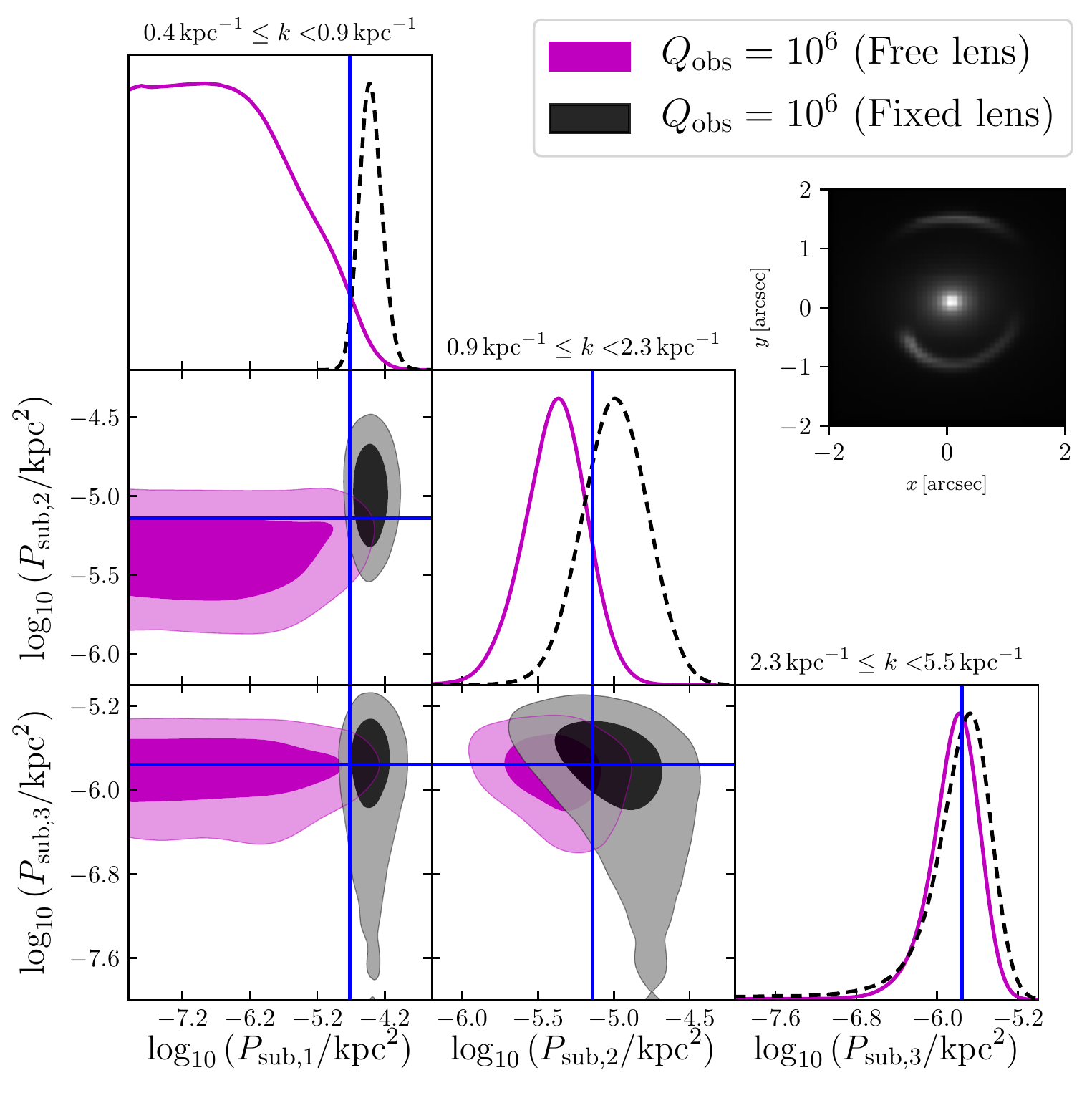}\\
\includegraphics[width=0.49\textwidth]{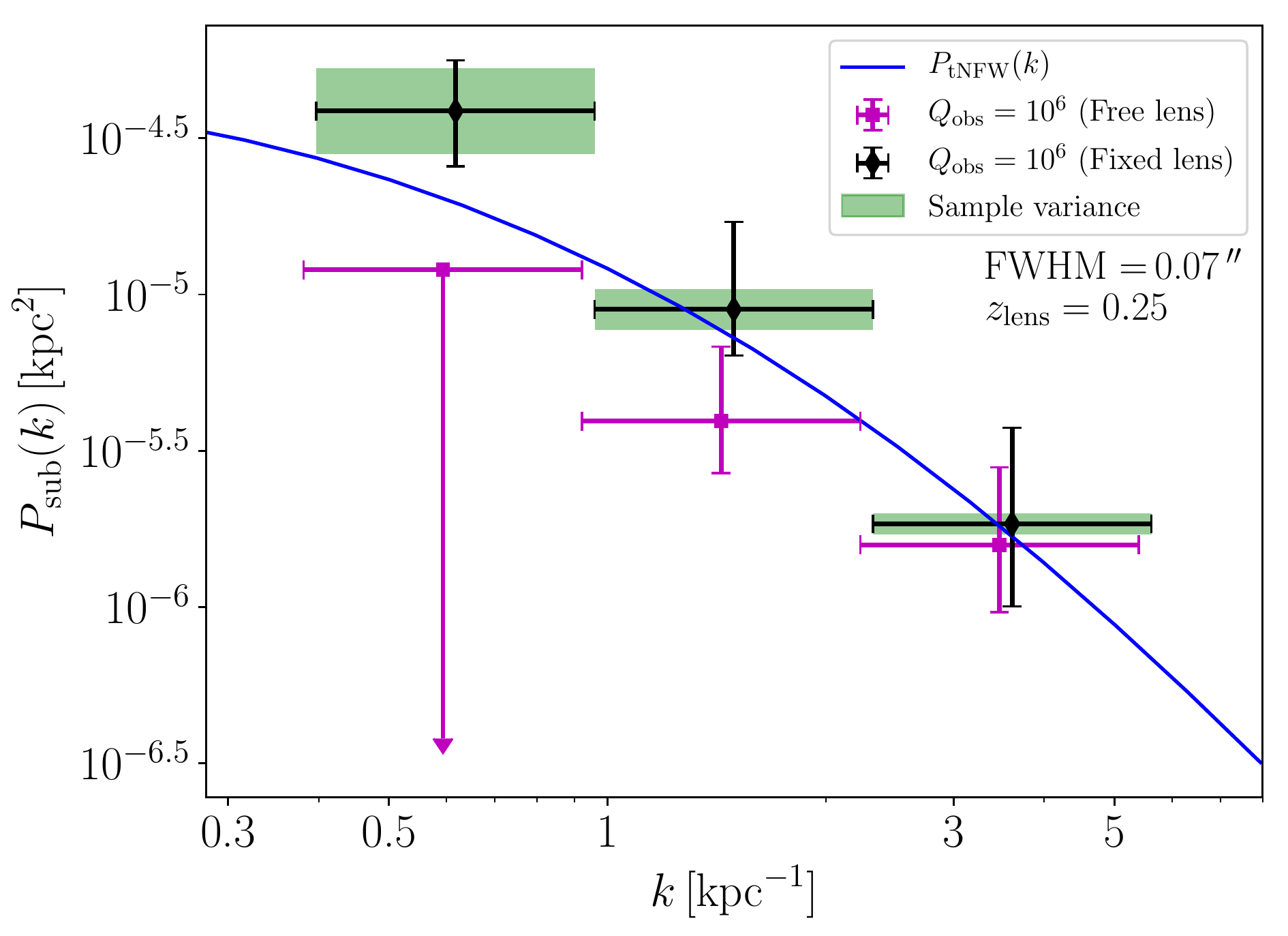}
\caption{Similar to Fig.~\ref{fig:pk_ex_full_int} but for a different lens configuration made of two partial lensed arcs (see inset in top panel). In the lower panel, the black diamond error bars show the highest posterior density (HPD) intervals for $Q_{\rm obs}=10^6$ while keeping the macro lens, source, foreground, and noise parameters fixed. The magenta square error bars show the HPD intervals for $Q_{\rm obs}=10^6$ allowing all macro lens, source, foreground, and noise parameters to vary within the MCMC chains.  The one-sided error bars show the $95\%$ HDP upper limits.
}\label{fig:pk_ex_full_int_2}
\end{figure}

We see in Fig.~\ref{fig:pk_ex_full_int} that even after the macro lens, source, and foregrounds are allowed to reabsorb some of the surface brightness features caused by the substructure, the mock data still retain significant sensitivity to the underlying convergence power spectrum. To some extent, the results shown here are still somewhat optimistic since the model used to fit the data is the same as the model used to generate the mock data. Nevertheless, it is encouraging that the inferred confidence regions on the substructure convergence power spectrum do not change dramatically in size once the possible degeneracies with the macro lens, source, and foregrounds are fully taken into account. The highest posterior values of the binned power spectrum amplitudes do shift with respect to the true input values, implying that the macro lens, source, and foregrounds are absorbing some of the substructure effects. This reabsorption is more important on the larger scales (low wave number) where it is easier for these components to compensate for small surface brightness fluctuations caused by the substructure. In any case, the resulting biases are on the binned power spectrum amplitude are modest, and the true input values are recovered within the $95\%$ confidence intervals. Overall, we observe little covariances among the different binned power spectrum amplitudes, in agreement with our Fisher estimates presented in Sec.~\ref{sec:fisher_simple}. 

Figure \ref{fig:pk_ex_full_int_2} shows the corresponding posterior summary for the case of the partial lensed arcs. Much of the same discussion as above applies here except that in this case, the macro lens, source, and foregrounds can reabsorb a significant fraction of the substructure effect within the lowest wave number bin. This is caused by the less redundant nature of this lens configuration (which essentially shows two distinct images of the extended source) as compared to the Einstein ring, which makes it more difficult to break the degeneracy between the macro lens, source, and foregrounds on the one hand, and the effects of the substructure on the other. 

\section{Generalization to compact time-dependent sources}\label{sec:likelihood_comp_src}
We now turn our attention to time-dependent compact (unresolved) sources, such as quasars, and consider how they can be used to probe the substructure within lens galaxies.  As outlined in Sec.~\ref{sec:valid_first_order}, the first-order perturbative approach used to compute the lensing residuals for extended sources fails in the case of compact sources since the surface brightness gradient of such object can become very large, hence leading to large second-order corrections to the lensing residuals. A different approach is thus required to treat these compact sources, which we summarize here.
\subsection{Lensing of time-dependent compact sources}\label{sec:TD_compact_source}
We assume that the compact source admits the following structure in the source plane
\be\label{eq:comp_src_1}
S_{\lambda}(\uu,t_{\rm s}) =  s_{\lambda}(t_{\rm s})\de(\uu - \uu_{\rm c}),
\ee
 where $s_{\lambda}(t_{\rm s})$ is the luminosity of the compact source at source-plane time $t_{\rm s}$ and wavelength $\lambda/(1+z_{\rm src})$, and $\uu_{\rm c}$ is the position of the compact source. If the compact source is not exactly pointlike, the delta function in Eq.~\eqref{eq:comp_src_1} could be replaced with the actual source profile, with only minor modifications to the remainder of the calculation. We can substitute the above into Eq.~\eqref{eq:full_gen_sol} to obtain the model for the lensed compact source
 \begin{align}\label{eq:gen_compact_src}
 \hat{O}_{\lambda}(\xx,t) & = \int d\yy\, W_{\lambda}(\xx-\yy,t) s_{\lambda}\left(t-\tau(\yy)\right) \en
 &\qquad\qquad \times\de \left(\yy - \nabla \phi_{\rm lens}(\yy)-\uu_{\rm c}\right) .
 \end{align}
 We recognize the argument of the Dirac delta function as the lens equation for the compact source, which allows us to immediately write
 \be
\de\left( \yy-\vec{\nabla}\phi_{\rm lens}(\yy) -\uu_{\rm c}\right) = \sum_{j=1}^{N_{\rm img}} \mu_j\de(\yy-\yy_j),
\ee
where $\yy_j$ and $\mu_j =  |{\rm det}\,{\bf M}(\yy_j)|$ are the location and the magnification of the $j$th image, respectively. Using the delta functions to perform the $\yy$ integration in Eq.~\eqref{eq:gen_compact_src}, our model for the lensed image of the time-dependent compact source takes the form
\be\label{eq:model_compact_imgs}
\hat{O}_{\lambda}(\xx,t) = \sum_{j=1}^{N_{\rm img}} \mu_j\,s_{\lambda}\left(t-\tau(\yy_j)\right)\,W_{\lambda}(\xx-\yy_j,t).
\ee
The substructure potential $\psub$ enters Eq.~\eqref{eq:model_compact_imgs} through the magnification $\mu_j$, the image position $\yy_j$, and the arrival time $\tau(\yy_j)$. In order to write down an explicit expression for Eq.~\eqref{eq:model_compact_imgs}, one would need to solve the full lens equation $ \yy-\vec{\nabla}\phi_{\rm lens}(\yy) -\uu_{\rm c}=0$ to determine the image positions for every possible choice of source position $\uu_{\rm c}$. This is a daunting task since the solution depends nonlinearly on the substructure potential $\psub$, whose statistics we are trying to probe in the first place. 

Our strategy to handle this difficulty is to treat the quantities $\left\{\mu_j,\yy_j, \Delta t_{j1}\equiv \tau(\yy_1)-\tau(\yy_j)\right\}$ as free nuisance parameters whose values are entirely driven by the data and an appropriate choice of priors. As we discuss below, the impact of the substructure on the image positions and arrival time delays can be \emph{exactly} captured by a series of constraint equations. It is important to note, however, that the impact of the substructure on the magnification of point images is often obscured by the presence of stellar microlensing and dust absorption.  While magnification information of point images has been used in the literature to constrain the presence of substructure within lens galaxies, we adopt here a conservative point of view and treat the $\mu_j$s as free parameters independent of $\psub$.

The remaining ingredient entering Eq.~\eqref{eq:model_compact_imgs} is the time dependence of the source $s_{\lambda}(t_{\rm s})$. Here, we choose to characterize the source variability as a Gaussian process.  Within this framework, the source function $s_{\lambda}\left(t-\tau(\yy_j)\right)$ is explicitly written in terms of a fluctuation vector $\de\vs_\lambda = \left\{ \{s_{\lambda,j}(t_k)-\hat{s}_\lambda\}_{j=1\ldots N_{\rm img}}\right\}_{k=1\ldots N^\lambda_{\rm obs}}$ describing the brightness of the source at each observation time $t_k$ and for each lensed image. Here, $\hat{s}_\lambda$ is the time-averaged compact source luminosity. Essentially, $\de\vs_\lambda$ contains the fluctuation light curve for each observed lensed image of the compact source. This vector is drawn from a Gaussian distribution given by 
\be\label{eq:DRW_PDF}
  \mathcal{P}_\src(\de\vs_\lambda) = \frac{1}{\sqrt{(2\pi)^{N_{\bf s}} |\Cmat_\src|}}
  e^{ -\frac{1}{2} \de\vs_\lambda^{\rm T}\, \Cmat_\src^{-1}\, \de\vs_\lambda },
\ee
where $N_{\bf s}$ is the total number of elements in the vector $\de\vs_\lambda$, which is $N_{\bf s} = N_\img \times N^\lambda_{\rm obs}$ because we reconstruct the source flux from every image at every epoch. The exact structure of the covariance matrix $\Cmat_\src$ depends on the type of Gaussian process used to describe the variability of the compact source. A few possible choices have been considered in the literature \cite{2009ApJ...698..895K,2011ApJ...728...26M,2011ApJ...743L..12M,Hojjati:2013aa,2013ApJ...765..106Z} including the damped random walk, the powered exponential, and the Mat\'ern covariance function. Our substructure analysis derived in the following sections is general enough to accommodate any choice of covariance function. We note that in general the covariance matrix $\Cmat_\src$ itself depends on the values of the time delays between images. 

Taking into account the conversion to counts within pixels (Eq.~\eqref{eq:brightness_to_count}), our model for the residuals of the lensed images of a time-dependent compact source observed at time $t_k$ is thus
\begin{align}\label{eq:from_src_to_image}
\de O_{\lambda}(\xx,t_k) &=  \frac{A_{\rm pix} T_{\rm exp}}{\mathcal{S}_{\rm inv}^{(\lambda)}}\sum_{j=1}^{N_{\rm img}} \mu_j\,\de s_{\lambda,j}(t_k)\,W_{\lambda}(\xx-\yy_j,t_k),
\end{align}
where $\de s_{\lambda,j}(t_k) \equiv s_{\lambda,j}(t_k)-\hat{s}_\lambda$.
%
\subsection{Constraint equations for compact sources}
As discussed in the previous section, the time-dependent source introduces nuisance parameters corresponding to the usual ``reduced observables'' (image position, magnification, and relative time delay) that are often used to characterize lensed images of quasars. The key point here is that these nuisance parameters depend on the substructure coefficients $\mathcal{A}_l$ through the lens equation. This leads to additional constraints on the relative amplitudes of the different substructure modes. To see this, we start by writing the lens equation for the compact source in the presence of mass substructures
\be\label{eq:constraint_1}
\uu_{\rm c} = \yy_j - \nabla\phi_0(\yy_j) - \sum_{l=1}^{N_{\rm modes}} \mathcal{A}_l\, \nabla\varphi_l(\yy_j),
\ee
where $\yy_j$ is the position of the $j$th image. We can use the $j=1$ image to eliminate the source position in Eq.~\eqref{eq:constraint_1} in order to obtain $2(N_{\rm img}-1)$ constraint equations for the $\mathcal{A}_l$ coefficients
\begin{align}\label{eq:A_l_constraints_pos}
&\Bigg\{\sum_{l=1}^{N_{\rm modes}}\mathcal{A}_l\left[\nabla\varphi_l(\yy_j)- \nabla\varphi_l(\yy_1)\right] \hspace{2cm}\\
&\qquad\qquad= \yy_j - \yy_1 +\nabla \phi_0(\yy_1)-\nabla\phi_0(\yy_j)\Bigg\}_{j=2,\ldots, N_{\rm img}}.\nonumber
\end{align}
 An additional set of constraint equations can be obtained by looking at the arrival time delay. Using Eq.~\eqref{eq:time_delay_def}, the arrival time delay for the $j$th image is
\begin{align}\label{eq:time_delay_first_const}
\tau_j &= t_0\left[\frac{1}{2}|\yy_j-\uu_{\rm c}|^2-\phi_0(\yy_j) - \sum_{l=1}^{N_{\rm modes}} \mathcal{A}_l\, \varphi_l(\yy_j)\right]\\
&=t_0\Bigg[\frac{1}{2}\left(|\uu_{\rm c}|^2-|\yy_j|^2\right)-\mathcal{D}\phi_0(\yy_j) \en
&\hspace{4cm}- \sum_{l=1}^{N_{\rm modes}} \mathcal{A}_l\,\mathcal{D}\varphi_l(\yy_j)\Bigg],\nonumber
\end{align}
where we have defined the projection operator 
\be
\mathcal{D} \equiv1 - \,\yy\cdot\nabla,
\ee
and where we have used Eq.~\eqref{eq:constraint_1} to eliminate the term $\uu_{\rm c}\cdot\yy_j$ in going from the first to the second line. We note that the operator $\mathcal{D}$ effectively ensures that only the gauge invariant part of the projected potential contributes to the arrival time delay.\footnote{Indeed, the term $\mathcal{D}\phi$ is invariant under the gauge transformation $\phi \rightarrow \phi + {\bf c}\cdot \xx$, where ${\bf c}$ is a constant vector.} As before, we can use the $j=1$ image to eliminate the source position from Eq.~\eqref{eq:time_delay_first_const} in order to obtain $N_{\rm img}-1$ additional constraints on the amplitudes $\mathcal{A}_l$
\begin{align}\label{eq:A_l_constraints_td}
\Bigg\{ \sum_{l=1}^{N_{\rm modes}}\mathcal{A}_l\mathcal{D}\left[\varphi_l(\yy_j)-\varphi_l(\yy_1)\right]= \frac{1}{2}\left(|\yy_1|^2-|\yy_j|^2\right)\\
\qquad+\mathcal{D}\phi_0(\yy_1)-\mathcal{D}\phi_0(\yy_j)-\frac{\Delta t_{1j}}{t_0}\Bigg\}_{j=2,\ldots, N_{\rm img}}.\nonumber
\end{align}
For each choice of nuisance parameters $\left\{\yy_1,\{\yy_j, \Delta t_{1j}\}_{j=2,\ldots,N_{\rm img}}\right\}$ describing the images of the compact source, Eqs.~\eqref{eq:A_l_constraints_pos} and \eqref{eq:A_l_constraints_td} form $3(N_{\rm img}-1)$ linear equations for $N_{\rm modes}$ unknown amplitudes $\mathcal{A}_l$. Schematically, this system takes the form
\be\label{eq:scheme_linear_a_syst}
{\bf L} \,{\bf a} = {\bf b},
\ee
where again ${\bf a} = \{\mathcal{A}_l\}$, and where the structure of the matrix ${\bf L}$ and vector ${\bf b}$ can be read off Eqs.~\eqref{eq:A_l_constraints_pos} and \eqref{eq:A_l_constraints_td}. It is important to emphasize that the constraints enforced by Eq.~\eqref{eq:scheme_linear_a_syst} are \emph{exact} and do not necessitate a perturbative expansion in the small substructure potential $\psub$. Essentially, the presence of a compact time-dependent source restricts the values of the $\{\mathcal{A}_l\}$ coefficients to lie on a hyperplane in the parameter space. In general, the number of constraints is much smaller than the number of measurable modes and the matrix ${\bf L}$ is therefore not invertible. In the following, we enforce this constraint by multiplying the likelihood by a Gaussian factor
\be
\mathcal{P}_{\rm c}({\bf a}|{\bf b}) \propto \frac{e^{-\frac{1}{2}({\bf L}\,{\bf a} -{\bf b})^\dagger{\bf \Sigma}^{-1}({\bf L}\,{\bf a} -{\bf b})}}{\sqrt{|{\bf \Sigma}|}},
\ee
and taking the limit ${\bf \Sigma}\to 0$ at the end of the calculation. Conveniently, this also allows us to relax the constraints by taking  ${\bf \Sigma}\to\infty$.
\subsection{Likelihood}
Having established the structure of these constraints, we can now write down the joint likelihood for the parameters $\qq$, $\qsub$, and the nuisance parameters $\pp{\rm Q} \equiv \left\{\{\mu_j,\yy_j, \Delta t_{1j}\}_{j=1,\ldots,N_{\rm img}}\right\}$ describing the properties of the lensed images of the compact source. As in Sec.~\ref{sec:single_obs}, it is useful to use a matrix and vector notation to simplify the derivation of the likelihood.  We denote the data vector as $\de{\bf O}_{\rm obs}$, which we take to contain the image residuals for all pixels from the $N_{\rm obs}$ observations. It thus has a total length of $N_{\rm d} \equiv N_{\rm pix}\times N_{\rm obs}$. As before, the image residuals caused by substructure are denoted as $\de {\bf O}_{\rm sub} = {\bf W}_{\rm E}\, {\bf a}$, but where the ${\bf W}_{\rm E}$ matrix now has dimensions of $N_{\rm d}\times N_{\rm modes}$. The contribution from the lensed compact source is denoted as $\de{\bf O}_{\rm C} = {\bf W}_{\rm C} \,\de{\bf s}$, where $\de{\bf s}$ is the compact source vector defined in Sec.~\ref{sec:TD_compact_source} of length $N_{\bf s} = N_{\rm img}\times N_{\rm obs}$, and ${\bf W}_{\rm C}$ is a matrix of size $N_{\rm d}\times N_{\bf s}$ which maps ${\bf s}$ to the observable space as given in Eq.~\eqref{eq:from_src_to_image}. 

With these definitions, our model for the observed image residuals takes the form 
\be
\de {\bf O}_{\rm obs} = {\bf W}_{\rm E}\, {\bf a} + {\bf W}_{\rm C} \,\de{\bf s} + {\bf N},
\ee
where ${\bf N}$ is a noise vector of length $N_{\rm d}$. As before, we are mainly interested in high signal-to-noise observations and we thus take the noise to have Gaussian statistical properties  specified by
\be
\langle {\bf N}{\bf N}\rangle_N = {\bf C}_N,
\ee
where ${\bf C}_N$ is now the $N_{\rm d}\times N_{\rm d}$ total noise covariance matrix for all observations. Since we expect the noise from different observations to be uncorrelated, the ${\bf C}_N$ matrix will generally have a block diagonal structure. The likelihood then takes the form 
\begin{align}
&\mathcal{L} \propto\int d{\bf a} d{\bf a}^\dagger \,d\de{\bf s}\,\mathcal{P}_{\rm c}({\bf a}|{\bf b})\mathcal{P}_{\rm sub}({\bf a}) \mathcal{P}_\src(\de\vs)\\
& \times  \frac{e^{-\frac{1}{2}(\de {\bf O}_{\rm obs}  - {\bf W}_{\rm E} {\bf a} - {\bf W}_{\rm C}\de{\bf s})^\dagger {\bf C}^{-1}_{N} (\de {\bf O}_{\rm obs}  - {\bf W}_{\rm E} {\bf a} - {\bf W}_{\rm C} \de{\bf s})}}{\sqrt{|{\bf C}_N| |{\bf C}_{\rm src}||{\bf C}_{\rm sub}||{\bf \Sigma}|}},\nonumber
\end{align}
where we have explicitly written down the marginalization over the substructure amplitudes specified by the vector ${\bf a}$, the source light curve fluctuations $\de{\bf s}$.  The constraints from Eq.~\eqref{eq:scheme_linear_a_syst} are implemented using the $\mathcal{P}_{\rm c}$ factor. Using the expressions for $\mathcal{P}_\src(\de\vs)$ and $\mathcal{P}_{\rm sub}({\bf a}|\qsub)$ given in Eqs.~\eqref{eq:DRW_PDF} and \eqref{eq:Gaussian_A_l}, respectively, and defining the vectors
\be
{\bf v} = \left[\begin{array}{c}
  {\bf a} \\
  \de{\bf s}
\end{array}\right],\qquad 
{\bf d} =  \left[\begin{array}{c}
  \de{\bf O}_{\rm obs} \\
 {\bf b}
\end{array}\right],
\ee
and the block matrices
\be
{\bf C}_{\rm s}^{-1} = \left[\begin{array}{cc}
  {\bf C}_{\rm sub}^{-1} & {\bf 0}  \\
  {\bf 0} & {\bf C}_{\rm src}^{-1}
\end{array}\right],
\ee
\be
{\bf A} = \left[\begin{array}{cc}
{\bf C}_N^{-1}& {\bf 0}  \\
  {\bf 0} & {\bf \Sigma}^{-1}
\end{array}\right],\quad
{\bf B} = \left[\begin{array}{cc}
{\bf W}_{\rm E}& {\bf W}_{\rm C}  \\
 {\bf L} & {\bf 0}
\end{array}\right],
\ee
we can write the likelihood as
\begin{align}\label{eq:L_tdep_tmp2}
\mathcal{L} \propto &\int d{\bf v} d{\bf v}^\dagger\times\\
&\frac{e^{-\frac{1}{2}\left( \left[\begin{array}{cc}
  {\bf d}^\dagger & {\bf v}^\dagger
\end{array}\right]\left[\begin{array}{cc}
{\bf A}& -{\bf AB} \\
-{\bf B}^\dagger{\bf A} & {\bf C}_{\rm s}^{-1} + {\bf B}^\dagger {\bf A}{\bf B}
\end{array}\right] \left[\begin{array}{c}
  {\bf d} \\
 {\bf v}
\end{array}\right]\right)}}{\sqrt{ |{\bf C}_N| |{\bf C}_{\rm sub}| |{\bf C}_{\rm src}||{\bf \Sigma}|}}.\nonumber
\end{align}
Using standard partial Gaussian integration, the marginalization over the ${\bf v}$ and ${\bf v}^\dagger$ vectors can be performed to yield
\be\label{eq:pixel_like_for_compact}
\mathcal{L} \propto \frac{e^{-\frac{1}{2}{\bf d}^\dagger {\bf M}^{-1}  {\bf d}}}{\sqrt{|{\bf M}|}},
\ee
where
\begin{align}
{\bf M} &\equiv \left({\bf A} -{\bf A} {\bf B}({\bf C}_{\rm s}^{-1}  + {\bf B}^\dagger {\bf A}{\bf B})^{-1}{\bf B}^\dagger {\bf A}\right)^{-1}\en
&={\bf A}^{-1} + {\bf B}{\bf C}_{\rm s} {\bf B}^\dagger,
\end{align}
where we have used the Woodbury matrix identity in the last line. Since only ${\bf \Sigma}$ (and not its inverse) appears within the covariance matrix ${\bf M}$ determining the likelihood given in Eq.~\eqref{eq:pixel_like_for_compact}, it is well defined to take the limit ${\bf \Sigma}\to 0$ to enforce the constraints from the image positions and time delays of the compact time-dependent source.

Equation \eqref{eq:pixel_like_for_compact} is the likelihood written in the ``pixel''  basis. As explained in Sec.~\ref{sec:single_obs}, it is often computationally advantageous to first perform the projection into the basis formed by the modes of the substructure deflection field since it reduces the dimensions of the matrices that need to be inverted. In this ``mode'' basis, the likelihood takes the form (after taking the limit ${\bf \Sigma}\to0$)
\be \label{eq:L_tdep_tmp4}
\mathcal{L} \propto 
\frac{e^{-\frac{1}{2}\left( \tilde\chi^2 - {\bf w}^\dagger{\bf U}^{-1}{\bf w} + ({\bf b}-\boldsymbol{\omega})^\dagger \boldsymbol{\Upsilon}^{-1} ({\bf b}-\boldsymbol{\omega}) \right)} }
{\sqrt{|{\bf C}_N| |{\bf C}_{\rm sub}| |{\bf C}_{\rm src}| |{\bf U}||{\boldsymbol{\Upsilon}}|}},
\ee
where
\be
{\bf U} = \left[\begin{array}{cc}
 \tilde {\bf D}& {\bf K}  \\
  {\bf K}^\dagger & {\bf F}
\end{array}\right],\quad
{\bf w} =  \left[\begin{array}{c}
 \tilde{\bf g} \\
 {\bf h}
\end{array}\right],
\ee
and where $\tilde{\chi}^2$, $\tilde{\bf g}$, and $\tilde{\bf D}$ are defined in Eqs.~\eqref{eq:chi_and_g_def} and \eqref{eq:D_def}. The other quantities appearing in Eq.~\eqref{eq:L_tdep_tmp4} are
\begin{align}
{\bf F} &= {\bf C}_{\rm src}^{-1} + {\bf W}_{\rm C}^\dagger \,{\bf C}^{-1}_{N}  {\bf W}_{\rm C},\\
{\bf K} &= {\bf W}_{\rm E}^\dagger\,{\bf C}^{-1}_{N}\, {\bf W}_{\rm C},\\
{\bf h} &= {\bf W}_{\rm C}^\dagger \,{\bf C}^{-1}_{N}\, \de {\bf O}_{\rm obs},
\end{align}
and
\be \label{eq:L_tdep}
\boldsymbol{\omega} = {\bf L}{\bf S}^{-1}\tilde{\bf g}
\quad\mbox{and}\quad
\boldsymbol{\Upsilon} = {\bf L}{\bf S}^{-1}{\bf L}^\dagger,
\ee
where ${\bf S}$ is the Schur complement of the matrix ${\bf F}$
\be
{\bf S} = \tilde{\bf D} - {\bf K}\,{\bf F}\,{\bf K}^\dagger.
\ee

The likelihood given in Eq.~\eqref{eq:L_tdep_tmp4} has a similar structure as that occurring for an extended source only (Eq.~\eqref{eq:like_multi_image}) except for the term in the exponent proportional to ${\bf \Upsilon}^{-1}$ which encodes the constraints from the compact source. Both $\boldsymbol{\omega}$ and ${\bf \Upsilon}$ have sizes determined by the number of linear compact source constraints, which for an $N_{\rm img}$ lens will be at most $3(N_{\rm img}-1)$, a small number indeed.

\begin{figure*}[t!]
\centering
\includegraphics[width=0.49\textwidth]{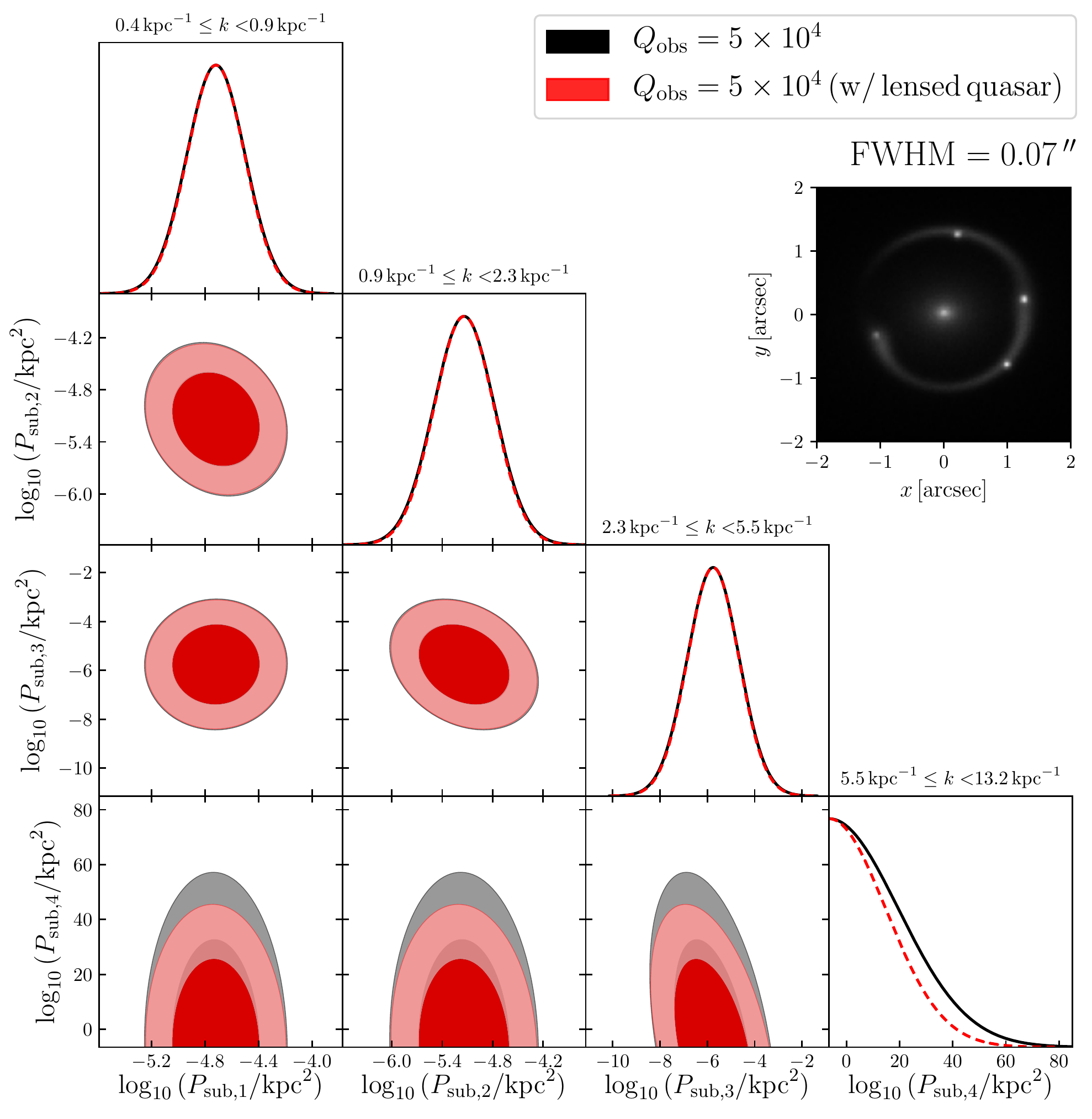}
\includegraphics[width=0.49\textwidth]{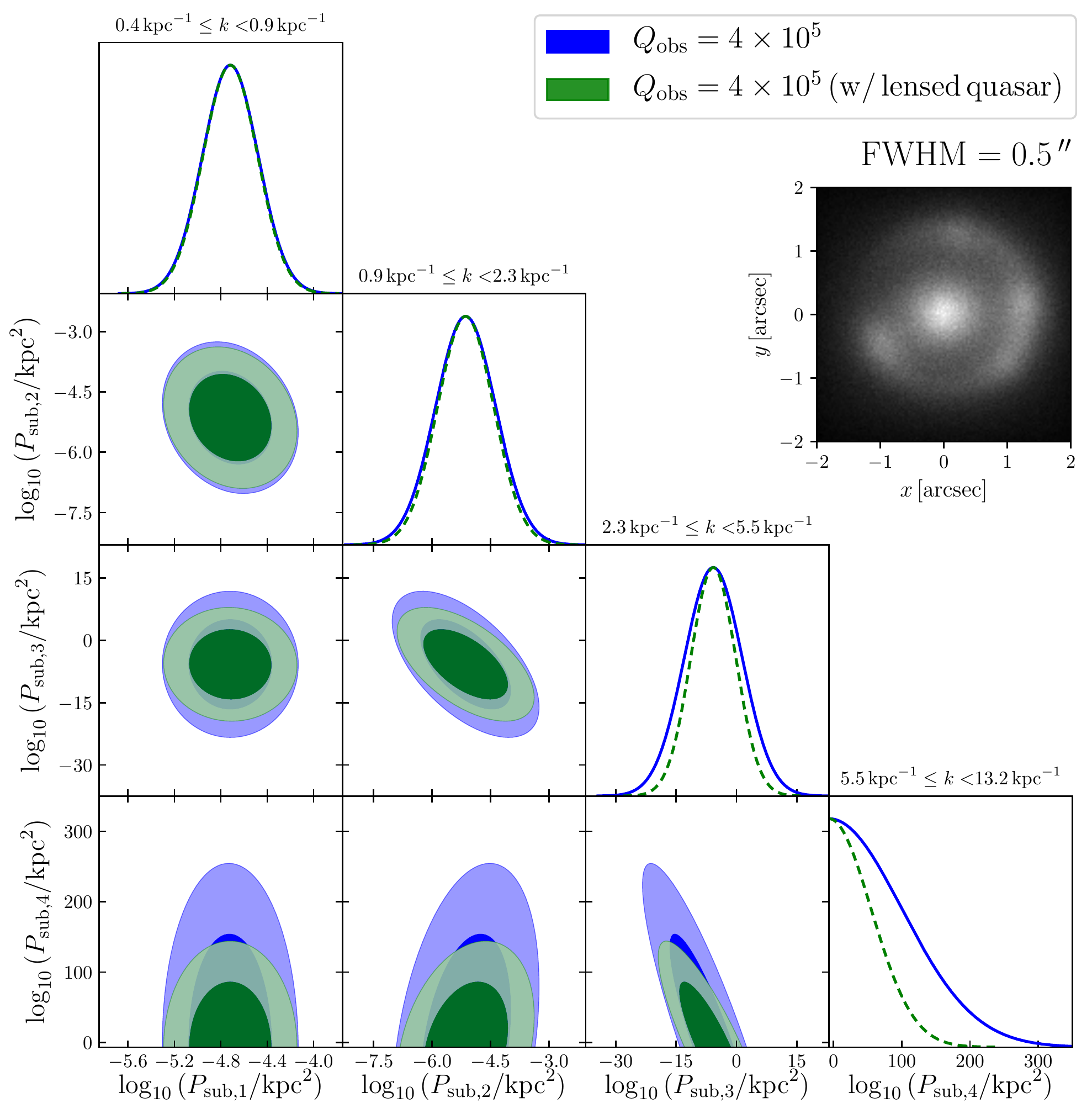}
\caption{Fisher forecast comparison for the binned substructure convergence power spectrum with and without the additional constraints from having a quadruply imaged lensed quasar within the image. Here, we use four wave number bins in the range $0.4\, {\rm kpc}^{-1} \leq k \leq 13.2\, {\rm kpc}^{-1}$, and only use the six constraints from the quasar image positions (no time delay constraints).  We adopt the truncated NFW substructure power spectrum shown in Fig.~\ref{fig:Psub_example} as our fiducial model. The left panel shows the comparison for a high-resolution image (FWHM $ = 0.07$ arcsec) with an observational quality factor of $Q_{\rm obs}=5\times10^4$. The right panel shows the Fisher forecast comparison for seeing-limited (FWHM $= 0.5$ arcsec) observations with a quality factor of $Q_{\rm obs}=4\times10^5$. The lensed image configuration used in each case are shown in the insets.}\label{fig:Fisher_w_compt_grnd}
\end{figure*}
%

\subsection{Compact sources and power spectrum sensitivity}
We wrap up this section by considering how the linear constraints imposed by the presence of a compact source can improve the constraints on the substructure convergence power spectrum. For this purpose, it is useful to isolate the part of the covariance matrix ${\bf M}$ that depends on ${\bf C}_{\rm sub}$ by writing it as
\be\label{eq:Isolate_Csub_in_M}
{\bf M} = {\bf M}_0 + {\bf Q}{\bf C}_{\rm sub}{\bf Q}^\dagger,
\ee
where 
\be
{\bf M}_0 \!=\! \left[\begin{array}{cc}
  {\bf C}_N \!+\! {\bf W}_{\rm C} {\bf C}_{\rm src} {\bf W}_{\rm C} ^\dagger & {\bf 0}\\
 {\bf 0} & {\bf \Sigma}
 \end{array}\right],\quad 
 {\bf Q} =  \left[\begin{array}{c}
  {\bf W}_{\rm E}\\
 {\bf L}
\end{array}\right].
\ee
Interestingly, we observe that the intrinsic variability of the compact source encoded in ${\bf C}_{\rm src}$ essentially acts as an additional source of noise for the pixels where its images form. On the other hand, the constraints on the modes of the substructure deflection field encoded in the projection operator ${\bf L}$ only appears in the second term of Eq.~\eqref{eq:Isolate_Csub_in_M}. Given the Gaussian structure of the likelihood given in Eq.~\eqref{eq:pixel_like_for_compact}, the Fisher matrix for the binned logarithmic amplitude of the substructure power spectrum takes the form
\begin{align}
 \Fmat_{ij} & = \frac{1}{2} \mbox{Tr}\left[ \Mmat^{-1} \frac{\partial\Mmat}{\partial \ln P_{\sub,i}} \Mmat^{-1} \frac{\partial\Mmat}{\partial \ln P_{\sub,j}} \right]\en
 &  = \frac{1}{2} \mbox{Tr}\left[ \Qmat^\dag \Mmat^{-1} \Qmat \frac{\partial\Cmat_\sub}{\partial \ln P_{\sub,i}} \Qmat^\dag \Mmat^{-1} \Qmat \frac{\partial\Cmat_\sub}{\partial\ln P_{\sub,j}} \right]\en
 & =  \frac{P_{\sub,i} P_{\sub,j}}{2} \mbox{Tr}\left[ \Gammat \frac{\partial\Cmat_\sub}{\partial P_{\sub,i}} \Gammat \frac{\partial\Cmat_\sub}{\partial P_{\sub,j}} \right],
\end{align}
where
\be\label{eq:Gammat}
  \Gammat = \left[ \left(\Qmat^\dag \Mmat_0^{-1} \Qmat\right)^{-1} + \Cmat_\sub \right]^{-1}.
\ee
We first note that Eq.~\eqref{eq:Gammat} exactly reduces to Eq.~\eqref{eq:gamma_for_extended} in the limit where the source has no compact component ($\Cmat_\src\to0$ and ${\bf \Sigma}\to\infty$). 

In the remainder of this section, we will focus our attention on the case where a compact source is present, but where we have only a single deep observation of the lens, which corresponds to the limit $\Cmat_\src\to0$ and ${\bf \Sigma}\to0$. In that case, the $\Gammat$ matrix takes the form
\be
 \Gammat = \left[ \tilde\Gmat^{-1} + \Cmat_\sub  -  \tilde\Gmat^{-1} {\bf L}^\da( {\bf L}  \tilde\Gmat^{-1} {\bf L}^\da)^{-1} {\bf L}   \tilde\Gmat^{-1}\right]^{-1}.
\ee
The two first terms in the square brackets are the same as in the purely extended source case, while the third term encodes the constraints from the presence of the compact source. 

We compare in Fig.~\ref{fig:Fisher_w_compt_grnd} the Fisher forecasts on the binned substructure power spectrum amplitudes with and without the additional constraints encoded in the ${\bf L}$ operator, for the case of a single deep observation of a lens (i.e.~no time delay constraints). We assume a cusplike lens configuration where four images of the lensed quasars are present in addition to a partial Einstein ring (as illustrated in the insets of Fig.~\ref{fig:Fisher_w_compt_grnd}). These highly idealized forecasts keep the structure of the source, foregrounds, the macro lens and its environment fixed to their true values.  As such, they should be cautiously interpreted as a best-case scenario for the additional sensitivity that lensed images of compact sources can bring to the substructure power spectrum measurements, in the absence of time delay measurements. 

The left panel displays the Fisher forecast comparison for a high-resolution image (FWHM $ = 0.07$ arcsec) where we observe that the added quasar constraints do not significantly modify the constraints, except in the highest wave number bin where the improvement is modest. The right panel shows the Fisher forecast comparison for a seeing-limited deep observation (FWHM $= 0.5$ arcsec) of the same lens. In this case, it is assumed that we have an independent measurement of the quasar image positions (from, e.g.,~a shallow HST observation) since the low-resolution nature of the image used for the power spectrum analysis likely precludes such a measurement. The improvement to the binned power spectrum constraints is slightly more significant in this case, with the largest gain in the two highest wave number bins. This reflects the fact that lensed images of compact sources are more sensitive to smaller-scale perturbations than the extended image studied in the rest of the paper. 

In general, the presence of a compact source can improve the bounds on the substructure convergence power spectrum at large wave numbers. The small relative number of constraints arising from a typical four-image quasar lens (six from image positions, with three additional constraints possible if time delays are known) compared to the total number of distinct substructure modes that can be probed (usually several hundreds) implies that the gain in sensitivity is modest. Nevertheless, adding the constraints from the quasar images could help break potential degeneracies between the macro lens and the effects of the substructure, so a further investigation is required to determine their full impact.

Here, we have treated the magnifications of the observed compact source images as free parameters. If microlensing or dust extinction (if relevant) could be properly understood and modeled, the substructure information contained in the brightness of the quasar images could also be harnessed by multiplying the likelihood given in Eq.~\eqref{eq:pixel_like_for_compact} by an extra (non-Gaussian) function describing the impact of substructure on the $\mu_j$ parameters. Given the highly nonlinear nature of the magnification perturbations caused by substructure \cite{Keeton:2009aa}, deriving such a function likely requires extensive forward simulations \cite{Dalal:2002aa,Fadely:2009aa,Gilman:2017voy}.

\section{Conclusions}\label{sec:conclusions}
In this paper, we have derived a mode function-based approach to extract statistical information about the projected substructure density field in proximity to strongly lensed images of high redshift sources. Focusing on two-point statistics, we have derived a likelihood for the substructure convergence power spectrum, given pixelated images of gravitationally lensed extended sources. We have implemented this likelihood within the software package \texttt{PkLens} and have performed simple Fisher forecasts to assess the sensitivity of different observational scenarios. Using simple lens, source, and foreground models, we have explored the possible degeneracies of these latter components and the collective effect of the substructure. We have finally generalized our power spectrum likelihood to take into account the presence of compact time-dependent lensed sources such as quasars within the observed images. 

Not too surprisingly, our results indicate that deep high-resolution images provide the best sensitivity to the substructure power spectrum, up to scales approximately corresponding to the smallest observable feature of the background source. At fixed image resolution and pixelization, lenses with larger Einstein radii or displaying a larger fraction of a complete Einstein ring provide better sensitivity to the collective effect of the substructure. We generally find that substructure perturbations with smaller wave numbers are more likely to be reabsorbed by changes in the macro lens and source models, compared to those with larger wave numbers. This can bias low the inferred amplitude of the substructure power spectrum on these scales. We leave to future work a detailed study of this potential bias and of possible techniques to address it. Finally, we find that for lenses containing both extended images and multiply-imaged quasars (such as RX J1131-1231 \cite{Sluse:2005km}), the extra constraints provided by the quasar image positions provide a modest improvement on the substructure power spectrum constraints, with most of the gain in sensitivity being concentrated at the smallest scales. While we have not considered the additional constraints coming from the relative time delays \cite{Keeton:2009ab}, it is likely that they could provide a slight improvement to the substructure sensitivity as compared to those shown here, while at the same time helping break some degeneracies in the lens modeling. 

Since one of the goals of this paper was to develop intuition about the different factors influencing the sensitivity to the substructure power spectrum, and not to perform detailed lens modeling (see, e.g., Ref.~\cite{Nightingale:2017cdh}), we have focused here on simple parametric lens and source models. The next step is to embed the numerical infrastructure developed here within a lens modeling framework that allows for more flexible source and foreground structures (see, e.g.~Refs.,~\cite{Warren:2003na,Suyu:2006fd,Birrer:2015rpa}), as well as dynamical PSF reconstruction. Given that the deflections caused by the substructure couple primarily to the gradient of the source, it could be argued that the forecasts based on simple sources presented here are conservative since more complex sources are likely to display enhanced sensitivity to substructure due to their greater spatial variability. It is however possible that this improvement is somewhat offset by the larger number of parameters needed to accurately describe the source. We study this tradeoff between source complexity and substructure constraints in an upcoming publication, in which we apply our formalism to actual observational data. 

In this work, we have used a binned substructure power spectrum as our fitting model, but the framework we developed here [see Eq.~\eqref{eq:C_sub_from_P_sub}] is completely general and can easily accommodate parametric models of $P_{\rm sub}(\kk)$ such as a power laws \cite{Bayer:2018vhy}, as well as anisotropic power spectra. While we have used the substructure power spectrum for a population of truncated NFW subhalos \cite{Rivero:2017mao} within the main lens as our fiducial model throughout this manuscript, we note that our derived likelihood is sensitive to the overall collective effect of the substructure between the lensed source and the observer, including line-of-sight structures either in front or behind the main lens, and baryonic structures such as globular clusters and giant molecular clouds. In order to eventually extract constraints on dark matter physics from substructure power spectrum measurements, robust predictions about the contribution from these latter objects to the power spectrum would have to be computed. Similarly, a more thorough study of the impact of non-Gaussianities on the inferred power spectrum would have to be performed. 

Eventually, deriving robust constraints on the Universe's small-scale structure from lensing power spectrum measurements will require combining the results from multiple lens systems. This will require a detailed study of how the inferred substructure power spectrum (including the line-of-sight contribution) depends on the properties (redshift, mass, concentration, environment, etc.) of the lens galaxy (see, e.g.~Ref.,~\cite{Mao:2015yua}). Alternatively, given the large number of gravitational lenses that will become known in the next decade \cite{Oguri:2010ns}, one could imagine building carefully selected samples that share similar properties and should thus have comparable substructure populations. Combining lens systems will allow the reduction of sample variance uncertainties on the substructure power spectrum at larger scales and, ultimately, provide a key test of the cold dark matter paradigm. 

The following software packages were used in this work: \texttt{Matplotib} \cite{doi:10.1109/MCSE.2007.55}, \texttt{Scipy} \cite{Scipy}, \texttt{Numpy} \cite{Numpy}, \texttt{astropy} \cite{astropy}, \texttt{numba} \cite{numba}, \texttt{emcee} \cite{2013PASP..125..306F}, \texttt{Getdist} \cite{getdist}, \texttt{h5py} \cite{h5py}, and \texttt{corner} \cite{corner}. 

\acknowledgments
F.-Y.~C.-R.~wishes to thank the Kavli Institute for Theoretical Physics for their hospitality during the completion of this work. We thank the Aspen Center for Physics, which is supported by National Science Foundation (NSF) Grant No.~PHY-1607611, for their hospitality during the initial stages of this work. This research was supported in part by the NSF under Grant No.~PHY-1748958. F.-Y. C.-R.~acknowledges the support of the National Aeronautical and Space Administration (NASA) ATP Grant No.~NNX16AI12G at Harvard University. C.~R.~K.~acknowledges the support of NSF Grant No.~AST-1716585. Part of the research described in this paper was carried out at the Jet Propulsion Laboratory, California Institute of Technology, under a contract with NASA.


\appendix
\section{Creating a random realization with a given power spectrum}\label{app:sec:psub_realization}
To create a realization of the convergence field with a given power spectrum, we first generate a map where each pixel is drawn from a standard normal distribution, that is,
\be\label{app:eq:variance_white_noise}
\langle \kappa_{ij} \kappa_{kl}\rangle = \de_{ik}\de_{jl},\qquad \langle\kappa_{ij}\rangle=0,
\ee
where $\kappa_{ij}$ is the value of the convergence at the pixel with indices $ij$, with the first index referring to the $x$ axis and the second referring to the $y$ axis. We note that this is a white noise map. The discrete Fourier transform of $\kappa_{kl}$ is given by
\be
\tilde{\kappa}_{mn} = A_{\rm pix} \sum_{k=0}^{N_x-1}\sum_{l=0}^{N_y-1} \kappa_{kl} e^{-2\pi i \frac{k\,m}{N_x}} e^{-2\pi i \frac{l\,n}{N_y}},
\ee 
where $A_{\rm pix}$ is the area of one pixel, $N_x$ is the number of pixels along the $x$ axis, and $N_y$ is the number of pixels along the $y$ axis. The power spectrum of $\tilde{\kappa}_{mn}$ is given by
\begin{align}
\langle \tilde{\kappa}_{mn}^* \tilde{\kappa}_{m'n'}  \rangle & = A_{\rm pix}^2 \sum_{k=0}^{N_x-1}\sum_{l=0}^{N_y-1} \sum_{k'=0}^{N_x-1}\sum_{l'=0}^{N_y-1} \langle \kappa_{kl} \kappa_{k'l'}\rangle\en
&\quad\times e^{2\pi i \frac{k\,m}{N_x}} e^{2\pi i \frac{l\,n}{N_y}} e^{-2\pi i \frac{k'\,m'}{N_x}} e^{-2\pi i \frac{l'\,n'}{N_y}}\en
&= A_{\rm pix}^2  \sum_{k=0}^{N_x-1}\sum_{l=0}^{N_y-1}  e^{2\pi i \frac{k\,(m-m')}{N_x}} e^{2\pi i \frac{l\,(n-n')}{N_y}}\en
& = A_{\rm pix}^2 N_x N_y\de_{mm'}\de_{nn'},
\end{align}
where we used Eq.~\eqref{app:eq:variance_white_noise} and the definition of the Kronecker delta in terms of Fourier series. Now, remember that we want to create a map of a convergence field with an input monopole power spectrum $P^{(0)}_{\rm sub}(k)$. The discrete Fourier transform of such a convergence field is given by
\be
\langle |\tilde{\kappa}^{\rm sub}_{mn}|^2\rangle =   A_{\rm pix} N_x N_y P^{(0)}_{\rm sub}(k_{mn}).
\ee
Thus, to convert from the white noise Fourier variables to the desired substructure $\tilde{\kappa}^{\rm sub}_{mn}$ Fourier mode, we write
\be
\tilde{\kappa}^{\rm sub}_{mn} = \sqrt{\frac{ P^{(0)}_{\rm sub}(k_{mn})}{A_{\rm pix} }} \tilde{\kappa}_{mn}.
\ee
Note that $P_{\rm sub}$ has units of area so the ratio $P^{(0)}_{\rm sub}(k_{mn})/A_{\rm pix}$ is dimensionless. Finally, to compute the actual convergence map, we perform the inverse discrete Fourier transform
\be\label{eq:inv_fourier_trans_kappa}
\kappa^{\rm sub}_{kl} = \frac{1}{N_x N_y A_{\rm pix}}\sum_{m=0}^{N_x-1}\sum_{n=0}^{N_y-1} \tilde{\kappa}^{\rm sub}_{mn} e^{2\pi i \frac{m\,k}{N_x}} e^{2\pi i \frac{n\,l}{N_y}},
\ee
where we note that the factor $N_x N_y A_{\rm pix}$ is nothing more than the area of the whole region where we are computing the convergence field. The substructure deflection field is generated in Fourier space as
\be
\tilde{\boldsymbol{\alpha}}_{\rm sub}(\kk) = \left( \frac{2i k_x \tilde{\kappa}_{\rm sub}(\kk)}{k^2}, \frac{2i k_y \tilde{\kappa}_{\rm sub}(\kk)}{k^2} \right).
\ee
 An inverse discrete Fourier transform similar to Eq.~\eqref{eq:inv_fourier_trans_kappa} is then performed to create the configuration space deflection field used to create the lensed image.

\section{Image Simulation}\label{sec:img_sims}
We explain here how we generate mock lensed images that are perturbed by a random realization of a substructure deflection field. Our procedure to generate a random realization of a substructure convergence and deflection field from a given input power spectrum is reviewed in Appendix \ref{app:sec:psub_realization}. To avoid periodicity effects due to the use of numerical fast Fourier transform (FFT), we generate a substructure deflection field spanning an area several times larger than the strong lensing region where we have sensitivity to substructures. Our actual mock images are generated by combining a model image gotten by applying Eq.~\eqref{eq:brightness_to_count} with a realistic noise realization. We summarize below the different ingredients entering our simulated images.
\begin{itemize}
\item {\bf Macro lens}:  We take the macro lens to consist of an isothermal ellipsoidal mass distribution. This model has five free parameters: the $(x,y)$ lens position, the Einstein radius, the ellipticity, and the direction of the latter in the plane of the sky.

\item {\bf Lens environment}: We model the lens environment as an external shear (two parameters).

\item {\bf Source}:  The source is taken to be an elliptical S\'ersic profile, which is described by six parameters: the 2D source position, the source flux, its half-light radius, the source ellipticity, and the angle of the semimajor axis in the plane of the sky.

\item {\bf Foregrounds}: We take the light from the lens galaxy to also be given by an elliptical $n=4$ S\'ersic profile (six parameters). We allow for a possible offset between the centroid of the lens light and that of its mass model (see e.g.~Ref.~\cite{2016ApJ...820...43S}). We also add a uniform sky background (one parameter) to take into account zodiacal light, Earth shine, etc. Unless otherwise mentioned, we take the sky background surface brightness to be $L_{\rm sky} = 4.5\times10^{-18}$ erg/cm$^2$/\AA/s/arcsec$^2$.

\item {\bf Point Spread Function (PSF)}: In most cases, we convolve our images with actual Hubble Space Telescope PSFs\footnote{See \url{http://www.stsci.edu/hst/wfc3/analysis/PSF}.} as measured by the UVIS detector of the Wide Field Camera 3 (WFC3) using the F555W filter. However, to generate ground-based seeing-limited observations, we adopt a simple Moffat profile with power-law index $2$ whose FWHM is a free parameter. PSF convolution is handled with standard FFT techniques.

\item {\bf Noise}: We assume uncorrelated noise across the pixel array, and adopt the following model for the noise within each pixel
\be\label{eq:noise_model}
{\bf C}_{{N_\lambda},ij} = \de_{ij}\left(\sigma_0^2 + \sigma_1 O_\lambda(\xx_i) + s(\xx_i)\right),
\ee
where the first term describes a constant noise contribution (one parameter) across each pixel (such as readout noise), the second term (one parameter) is proportional to the observed count in each pixel and mimic Poisson (shot) noise, while the last term is introduced to take into account bad pixels or cosmic rays and can be used to mask certain pixels by giving them large uncertainties. For simplicity, we set $s(\xx_i)=0$ here. When generating a mock image, we draw a random zero-mean Gaussian noise realization whose variance is given by Eq.~\eqref{eq:noise_model} and add it to the image. Here, we conservatively take $\sigma_0 = 1$ count/pixel, which is significantly larger than the read noise of the UVIS detector of WFC3, and take $\sigma_1=1$, which makes the second term of Eq.~\eqref{eq:noise_model} be exactly Poissonian.

\item {\bf Pixelization}:  Throughout this work, we consider images of size $4''\times4''$, which is large enough to capture the relevant features of most galaxy-scale lenses. We restrict ourselves to images with $50\times 50$ pixels, which ensures that the likelihood from Eq.~\eqref{eq:like_one_image} can be evaluated in a few seconds on a modest computer. This implies a linear pixel size of $0.08''$, which is about twice the pixel size of the UVIS detector of WFC3. When computing a model image using Eq.~\eqref{eq:brightness_to_count}, we use 64 light rays per pixel to estimate the photon count within each pixel. We have checked that this number is large enough to ensure that the photon count within the pixels is converged to better than $0.1\%$.
\end{itemize}

Throughout the simulated images used in this work, we use the quantity $Q_{\rm obs}$ defined in Eq.~\eqref{eq:Q_fac} to characterize the quality of the mock observations. As an example, for the UVIS detector of WFC3 with the F555W filter on (for which $\mathcal{S}_{\rm inv}^{(\lambda_i)}= 1.8\times10^{-19}$ erg/cm$^2$/\AA/count)\footnote{See the technical document \url{http://www.stsci.edu/hst/wfc3/documents/ISRs/WFC3-2017-14.pdf}.}, we obtain $Q_{\rm obs} = 4.4\times10^5$ for a combination of ten 45-minute exposures of an unlensed source of AB magnitude 23.

\section{Null test and degeneracy with noise parameters}\label{app:null_tests}
\begin{figure}[t!]
\centering
\includegraphics[width=0.49\textwidth]{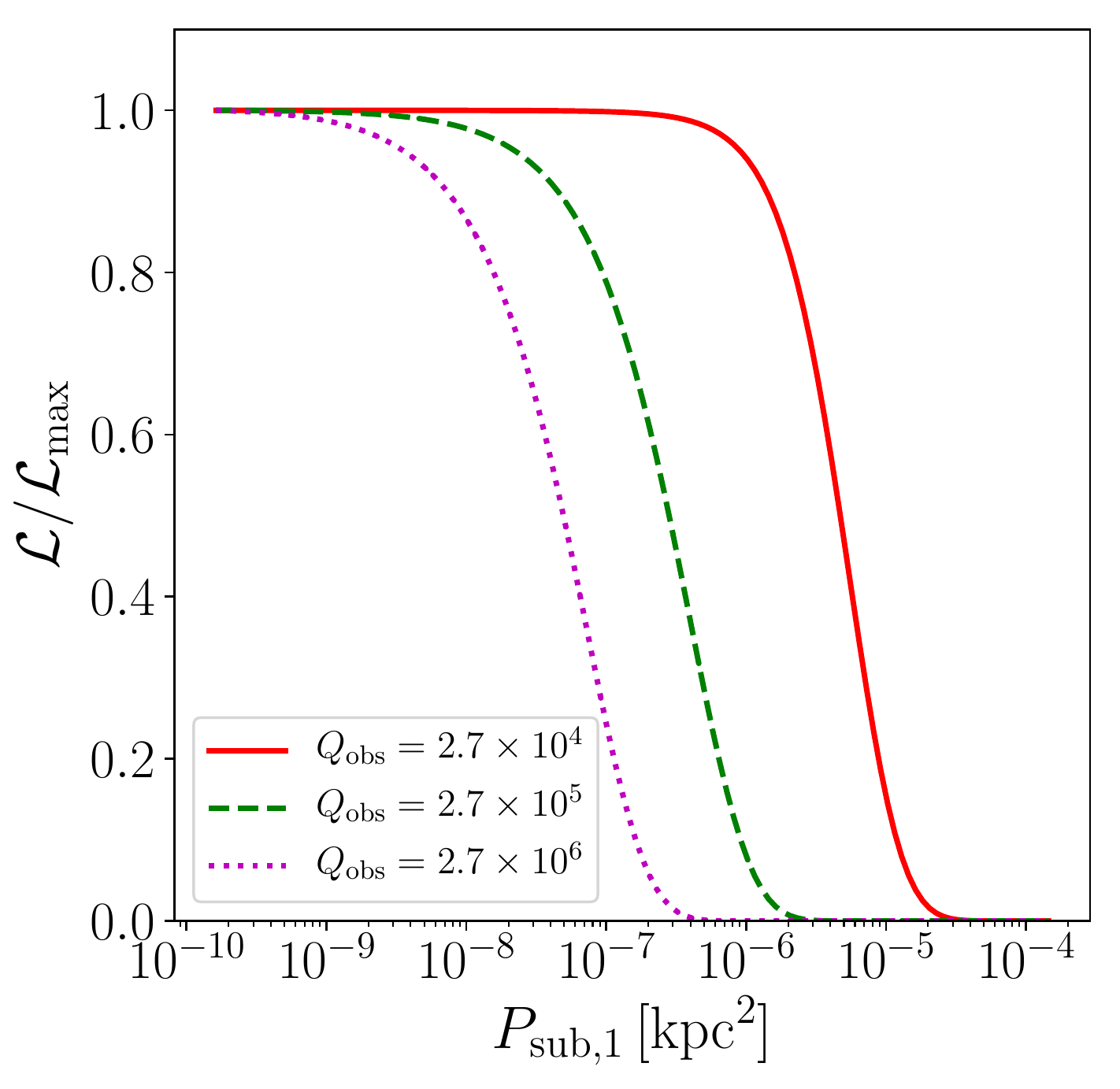}
\caption{Posterior distributions for the amplitude $P_{{\rm sub},1}$ of the substructure convergence power spectrum within the wave number range $0.4\, {\rm kpc}^{-1}\leq k \leq 5.5 \,{\rm kpc}^{-1}$. The mock lensed images used in this analysis (similar in configuration to that shown in Fig.~\ref{fig:residuals_example} with an HST-like PSF) do not contain substructure perturbations, and the resulting posterior distributions, which are consistent with a vanishing power spectrum amplitude, provide an important null test of our inference framework. The three different curves display different of the quality factor $Q_{\rm obs}$, with the upper bound on the substructure power spectrum amplitude getting more stringent for the deeper images, as expected.}\label{fig:null_test}
\end{figure}
\begin{figure}[t!]
\centering
\includegraphics[width=0.49\textwidth]{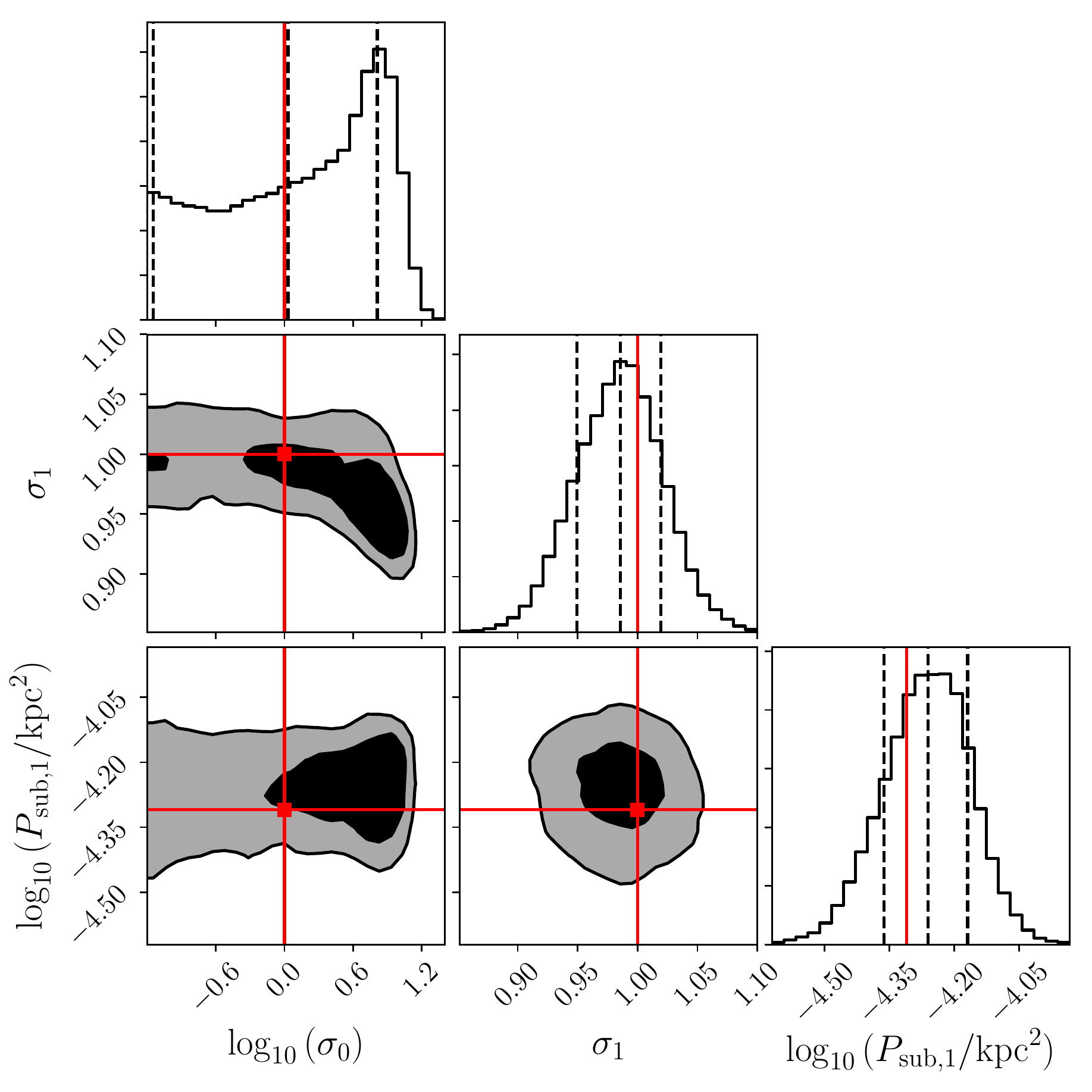}
\caption{Posterior distribution for the noise parameters given in Eq.~\eqref{eq:noise_model} and the amplitude of the substructure convergence power spectrum. The black and gray regions represent the $68\%$ and $95\%$ confidence regions, respectively. The dashed vertical lines shown in the one-dimensional posteriors shown along the diagonal correspond to the $16\%$, $50\%$, and $84\%$ quartiles. The solid red lines and squares represent the true values used to generate the mock data. Here, the fiducial data are generated assuming the point-mass substructure power spectrum shown in Fig.~\ref{fig:Psub_example}, with an image configuration similar to that shown in Fig.~\ref{fig:residuals_example} with a quality factor $Q_{\rm obs}=2.7\times10^5$ and an HST-like PSF.}\label{fig:Deg_w_noise}
\end{figure}

To test the sensitivity of gravitationally lensed extended sources to the substructure convergence power spectrum, we feed mock images generated using the procedure outlined in the Appendix \ref{sec:img_sims} to the likelihood given in Eq.~\eqref{eq:like_multi_image} and generate MCMC samples from the posterior using affine-invariant sampler \texttt{emcee} \cite{2013PASP..125..306F}. We adopt broad uniform priors for all macro lens, source, and foreground parameters, except for the surface brightness parameters for which log-uniform priors are taken. The constant noise contribution $\sigma_0$ also receives a log-uniform prior while we adopt a uniform prior on $\sigma_1$. This latter choice has very little impact on our results since the posterior for $\sigma_1$ is generally very sharply peaked around unity (consistent with Poisson noise).

As in the case of our Fisher analysis, we adopt the logarithm of the binned substructure convergence power spectrum amplitudes as our fitting model, and divide the range of scales probed by a given lensed image into one to four wave number bins that are evenly spaced in $\log_{10}(k)$. We adopt a broad log-uniform prior on the amplitude within each bin, $\log_{10}(P_{{\rm sub},i}/{\rm arcsec}^2)\in [-11,-1]$. 

We first test the validity of our framework by performing null tests in which we sample the likelihood given in Eq.~\eqref{eq:like_multi_image} for lensed images that have not been perturbed by substructure. For simplicity, we take the substructure fitting model to consist of a single power spectrum bin spanning the range $0.4\, {\rm kpc}^{-1}\leq k \leq 5.5 \,{\rm kpc}^{-1}$. Using a lensed image configuration similar to that shown in Fig.~\ref{fig:residuals_example} and considering three different values of the quality factor $Q_{\rm obs}$, we obtain the posterior distributions shown in Fig.~\ref{fig:null_test} for the convergence power spectrum amplitude $P_{{\rm sub},1}$ within that single bin. These posterior distributions are all consistent with a vanishing amplitude of the substructure perturbations, with the upper limits on the power spectrum amplitude improving for images with higher $Q_{\rm obs}$, as should be expected. This indicates that our formalism does not appear to interpret random noise features as spurious substructure within the lens galaxy. While this result is not surprising given the idealized nature of our mock images, it does provide an important consistency check that our numerical implementation of the likelihood given in Eq.~\eqref{eq:like_multi_image} is sound.

We then turn our attention to the potential degeneracy between image noise and the amplitude of the substructure power spectrum. For simplicity, we consider here mock lenses that are perturbed by a random realization of point-mass substructures with a power spectrum given by the dashed black line of Fig.~\ref{fig:Psub_example}.  While the point-mass power spectrum is likely unrealistic, it has the advantage and being described by a single parameter: its amplitude. We show the joint posterior distribution for the noise parameters and the amplitude of the substructure power spectrum (denoted $P_{{\rm sub},1}$ here) in Fig.~\ref{fig:Deg_w_noise}. To perform the inference, we use all Fourier modes spanning the range $0.4\, {\rm kpc}^{-1}\leq k \leq 5.5 \,{\rm kpc}^{-1}$ (for a total of 180 independent Fourier modes\footnote{Given the resolution of our mock images (0.08 arcsec pixel size), we could technically extend this range to $0.4\, {\rm kpc}^{-1}\leq k\lesssim 13.2$ kpc$^{-1}$. However, the small-scale modes  with $k > 5.5$ kpc$^{-1}$ are noise dominated for our choice of exposure, and adding them does not change the posterior shown in Fig.~\ref{fig:Deg_w_noise}.}).

We observe no degeneracy between the substructure power spectrum amplitude and the noise parameters, and note that we successfully recover the true amplitude of the input power spectrum (solid red line) to within less than one standard deviation. The marginalized posterior for the parameter $\log_{10}(\sigma_0)$ is quite broad and extends to significantly larger (and lower) values than that used to generate the mock image. This is a consequence of the Poisson term [that proportional to $\sigma_1$ in Eq.~\eqref{eq:noise_model}] dominating the noise budget in the mock image. Only when the contribution from $\sigma_0^2$ becomes comparable in magnitude to $\sigma_1 O_\lambda$ can the mock data start displaying sensitivity to $\sigma_0$, explaining the long flat tail of the posterior at low values of this noise parameter.

These findings are broadly consistent with those of Ref.~\cite{Hezaveh_2014} which found no degeneracy between uncorrelated image noise and the substructure power spectrum amplitude. Again, this is a consequence of the highly nonrandom structure of the image residuals caused by substructure [Eq.~\eqref{eq:delta_O}] which is not easily mimicked by pure random noise. In a realistic image where adjacent pixels may have correlated noise due, for example, to drizzling, it is possible however for the noise structure to be somewhat degenerate with the effect of the substructure.

\bibliography{../stochastic_lensing}

\end{document}